\documentclass[11pt]{article}
\usepackage{jheppub}
\usepackage{enumitem}
\usepackage{tcolorbox}

\usepackage{jheppub}
\usepackage{mathrsfs}
\usepackage{color}
\usepackage{hyperref}

\usepackage{slashed}
\usepackage{feynmp-auto}
\usepackage{simplewick}
\usepackage{cancel}

\usepackage{todonotes}

\usepackage{amsmath}
\usepackage{amsfonts}
\usepackage{graphicx}
\usepackage{amssymb}
\usepackage{xcolor}
\usepackage{empheq}
\usepackage{mathtools}

\newcommand{\eq}{\begin{equation}}

\newcommand{\eqe}{\end{equation}}
\newcommand{\eqa}{\begin{eqnarray}}
\newcommand{\eqae}{\end{eqnarray}}
\title{The geometry of optimal functionals}

\author{Yu-tin Huang$^{1,2}$}
\author{Wei Li$^{1}$}
\author{Guan-Lin Lin$^{1}$}

\affiliation{$^1$ Department of Physics and Astronomy, National Taiwan University, Taipei 10617, Taiwan}
\affiliation{$^2$ Physics Division, National Center for Theoretical Sciences, National Tsing-Hua University, No.101, Section 2, Kuang-Fu Road, Hsinchu, Taiwan}
\emailAdd{yutinyt@gmail.com}
\emailAdd{r07222072@ntu.edu.tw}
\emailAdd{klwalt861206@gmail.com}

\abstract{In this paper, we give a geometric interpretation of optimal functionals in the context of intersection of symmetry planes and cyclic polytopes. For 1D CFTs, we demonstrate that at given derivative order, the functional is given by a degenerate simplex of the cyclic polytope. More precisely the derivative functionals at $2N{+}1$-th order, is given by an unique $N$-dimensional simplex enclosing the origin. Taking the continuous limit, in the large $\Delta$ approximation this qualitatively agrees with that derived by Mazac et al. Remarkably similar construction applies to 2D CFT in the diagonal limit as well as the spin-less modular bootstrap. Finally we show that such geometric interpretation can be extended to functionals associated with bounds beyond the leading operator.}

\begin{document}
\begin{flushright}
\vspace{10pt} \hfill{NCTS-TH/1909} \vspace{20mm}
\end{flushright}
\maketitle

\section{Introduction}
The union of constraints arising from symmetries and unitarity, when projected onto physical observables, often result in a convex hull problem. This has been well appreciated in the modern revival of the conformal~\cite{Rattazzi:2008pe} and modular bootstrap~\cite{Hellerman:2009bu}\,. For the former, the observable is the four-point correlation function, while for the later it is the torus partition function. More recently, such an approach has been applied to EFT bootstrap where the relevant observable is the four-point S-matrix~\cite{EFT}. The convex hull defines the region in which the physical observable can take value, from which one extracts bounds on the quantum number of general consistent theories. Obviously the result depends crucially on the boundaries of the convex hull, which in general is not computable since the number of vertices for the hull are infinite. Techniques from linear programming, both in its original numerical approach as well as analytic extension, was precisely devised to tame this complexity and has brought spectacular progress over the past decade (See \cite{Rychkov:2016iqz, Simmons-Duffin:2016gjk, Poland:2016chs, Poland:2018epd} for reviews).

Recently, in the context of 1-dimensional CFT, it was shown that the convex hull of the conformal blocks is in fact a cyclic polytope~\cite{Arkani-Hamed:2018ign}. More precisely, when expanded around the self-dual point, the resulting Taylor coefficients form vectors whose ordered determinant is always positive. Here the ordering is with respect to conformal dimensions. Once this is established, then the resulting convex hull has the property that all boundaries are known before hand. Any solution to the bootstrap problem is then casted into whether this convex hull intersects a subspace required from crossing symmetry. For each solution one has a distinct cyclic polytope. Translated to the constraint on the spectrum, this turn into the requiring the existence of some fixed number of operators forming a simplex that incloses the origin. This immediately leads to the following question:
\begin{itemize}
  \item What is the finger print of the convex hull being a cyclic polytope, manifested on the spectrum? 
  \item  How does knowing the boundaries of the cyclic polytope, directly lead to boundaries of the theory space, i.e. the gap of each operator? 
\end{itemize}
To make the connection from the geometry to the gaps, we consider the linear functional point of view. The gap is determined from the zeros of the optimal functional, which are derivative or integral functionals acting on the conformal blocks. The relation between the boundaries of the cyclic polytope and the zeros of the optimal functional provides the potential link between the walls of the two spaces.  An exact functional for 1D CFT was proposed in~\cite{Mazac:2016qev, Mazac:2018mdx}, and similarly for modular bootstrap ~\cite{Hartman:2019pcd}. Discussions for general dimensions see~\cite{Mazac:2019shk}. The fact that the functional relevant for modular bootstrap is identical to that of the sphere packing problem further supports that such functionals must have a geometric interpretation behind it.  

In this paper, we initiate the exploration of the implications of the positive geometry on optimal functionals. In particular, we show that properties of cyclic polytopes immediately lead to the conclusion that the optimal functional for $2N{+}1$-th derivative order must be given as 
\eq\label{Answs}
N\in odd\;\;\langle \mathbf{X}, 0, i_{1},i_{1}{+}1,i_{2},i_{2}{+}1,\cdots,\Delta\rangle, \quad N\in even\;\;\langle \mathbf{X}, 0, i_{1},i_{1}{+}1,i_{2},i_{2}{+}1,\cdots,\infty,\Delta\rangle
\eqe
where $\langle \cdots\rangle$ represent taking the determinant while $\mathbf{X}$ is the crossing plane, $0$ the identity and $i$s are the block vectors with conformal dimension $\Delta_i$. Furthermore, the specific value of $i_{1},i_{2},\cdots,i_{gap}$ are exactly the vertices of a highly degenerate simplex, which for 1D CFT yields unique solutions. We've verify that eq.(\ref{Answs}) yields the correct result by matching to numerical analysis up to 23 order in derivatives, and demonstrate qualitative equivalence to the result given in~\cite{Mazac:2016qev, Mazac:2018mdx} in the continuous limit.  

The cyclic polytope property of the convex hull was also recently found in the diagonal limit of 2D CFT bootstrap~\cite{Sen:2019lec}, as well as the spin-less modular bootstrap~\cite{ShuHeng}. Not surprisingly, we find that the same approach also yields optimal functionals that agree with the numeric bootstrap for 2D CFT with the external dimension $\Delta_\phi>0.088$. Remarkably, when spin is included, we find that the the functional is once again given by a degenerate simplex involving spin-4 block vectors. In particular, the functional is given by:
\eq
N\in odd\;\;\langle \mathbf{X}, 0, i_{\ell=4},i_{\ell=4}{+}1,i_{2},i_{2}{+}1,\cdots,\Delta\rangle
\eqe 
where now it is $i_{\ell=4}, i_{1},i_{2},\cdots,$ that satisfy the co-plane condition. For these cases, unlike 1D CFT, the degenerate simplex is not unique. But we can still select out the one that corresponds to the optimal functional using other constraints.

Finally, we study the global constraints of the spectrum in 1D CFT where we consider bounds beyond the leading operators. We find that the geometric consideration of having a simplex inclosing the origin, allows us to derive bounds on the higher dimensional operators in a way depending on the lower dimension ones. The bounds and the boundaries  between the distinct sectors, once again are all given by geometric considerations, which we confirm matches with numerical bootstrap results. 

This paper is organized as follows: in the next section we begin with a review of 1D CFT bootstrap, both in the language of projective geometry, and in terms projective geometry and optimal functionals. In subsection~\ref{sec:Optimal}, we recast the optimal functional in terms of finding the degenerate simplex of the complex hull such that the origin is included. We explicitly verify this conjecture by comparing with results from numerical analysis, as well as large $\Delta$ approximation in the continuous limit. In section~\ref{2DCFT}, we extend the proposal to 2D CFT in the diagonal limit as well as modular bootstrap. Finally, in section~\ref{sec:TheorySpace}, we demonstrate that the geometric approach continues to yield the correct optimal functional beyond leading operator.  

\section{1D CFT} 
In this section we give a brief review of the conformal bootstrap in one dimensions, especially the set up of the problem in terms of projective geometry introduced in \cite{Arkani-Hamed:2018ign}. By one dimensional CFT we are really referring to the imposition of SL(2,$\mathbb{R}$) symmetry on correlation functions where the operators are positioned on a line. Thus the constraint obtained from the bootstrap equations must be obeyed in any dimensions.

Let's start with the 4-pt function $\langle\phi(x_1)\phi(x_2)\phi(x_3)\phi(x_4)\rangle$ of identical primary operator $\phi$ with scaling dimension $\Delta_\phi$. We will assume $x_1<x_2<x_3<x_4$ and define $SL(2,\mathbb{R})$ invariant cross-ratio $z$ as
\begin{equation}
z=\frac{x_{12}x_{34}}{x_{13}x_{24}}\in(0,1)
\end{equation}
where $x_{ij}=x_i-x_j$. The $SL(2,\mathbb{R})$ covariance of 4-pt function implies that up to an overall prefactor, it can be written as a function only of cross-ratio $z$
\begin{equation}\label{4pt}
\langle\phi(x_1)\phi(x_2)\phi(x_3)\phi(x_4)\rangle=\frac{\mathcal{G}(z)}{|x_{12}|^{2\Delta_\phi}|x_{34}|^{2\Delta_\phi}}\,.
\end{equation}
Using OPE between $\phi(x_1)$ and $\phi(x_2)$, $\phi(x_3)$ and $\phi(x_4)$, $\mathcal{G}(z)$ can be decomposed into a sum of 1D conformal blocks $G_\Delta(z)=z^\Delta {}_2F_1(\Delta,\Delta,2\Delta,z)$, 
\begin{equation}\label{schannel}
Unitarity:~\mathcal{G}(z)=\sum_{\Delta_O\in\phi\times\phi}c_{\Delta_O}^2G_{\Delta_O}(z)
\end{equation}
where $c_{\Delta_O}$ is 3-pt function coefficient of $\phi$, $\phi$ and $O$. Since the coefficients of the conformal blocks are squares of the three-point function, and $\mathcal{G}(z)$ is positively expandable on the blocks.  
Alternatively, we can use OPE between $\phi(x_1)$ and $\phi(x_4)$, $\phi(x_2)$ and $\phi(x_3)$ to express 4-pt function (\ref{4pt}) and (\ref{schannel}). Conformal symmetry tells us that the radius of convergence for the two channels overlap, and hence the four-point function written in one expansion is equivalent to the other. In other words, the crossing equation
\begin{equation}\label{crossing}
Crossing:~\mathcal{G}(z)=\left(\frac{z}{1-z}\right)^{2\Delta_\phi}\mathcal{G}(1-z)\,,
\end{equation}
is applicable when $\mathcal{G}(z)$ is written in terms of the block expansion. 
\subsection{Cyclic polytopes and crossing plane}
It is useful to analyze the two constraints, eq.(\ref{schannel}) and eq.(\ref{crossing}), in terms of derivative expansion around $z=1/2$. For conformal blocks, this defines a finite dimensional vectors,
\begin{align}\label{taylor1}
G_\Delta(z)\to\vec{\mathbf{G}}_\Delta=
\begin{pmatrix}
G_\Delta^0\\
G_\Delta^1\\
G_\Delta^2\\
\vdots\\
G_\Delta^n
\end{pmatrix}
\end{align}
where $G_\Delta^n=\frac{G_\Delta^{(n)}(1/2)}{n!}$. Note that it is useful to consider the $n{+}1$ dimensional geometry projectively, by normalizing the first entry to $1$ and absorbing the overall positive factor into $c_{\Delta}^2$. Doing the same for the four-point function $\mathcal{G}(z)$, eq.(\ref{schannel}) becomes a convex hull condition:
\eq
\mathcal{G}(z)=\sum_{\Delta}c_{\Delta}^2G_{\Delta}(z)\;\;\Rightarrow\;\;\vec{\mathcal{G}}=\sum_\Delta c_\Delta^2\vec{\mathbf{G}}_\Delta
\eqe
where, 
\eq
\mathcal{G}(z)\to\vec{\mathcal{G}}=
\begin{pmatrix}
\mathcal{G}^0\\
\mathcal{G}^1\\
\mathcal{G}^2\\
\vdots\\
\mathcal{G}^n
\end{pmatrix}\quad \mathcal{G}^n=\frac{\mathcal{G}^{(n)}(1/2)}{n!}\,.
\eqe
and we have $\sum_{\Delta}c_{\Delta}^2=1$. The convex hull is then in $\mathbb{P}^{n}$. Following \cite{Arkani-Hamed:2018ign} we take $n=2N{+}1$. For a given spectrum, the convex hull of the associated block vectors form a polytope in $\mathbb{P}^{2N{+}1}$,which we will refer to as the unitary polytope $\mathbf{U}_N$.   

Crossing symmetry eq.(\ref{crossing}), will impose linear relation on the different Taylor coefficients $\mathcal{G}^i$, and allows us to express $\mathcal{G}^{odd}$ in terms of $\mathcal{G}^{even}$
\begin{align}\label{taylor2}
&\mathcal{G}^1=4\Delta_\phi\mathcal{G}^0\notag\\
&\mathcal{G}^3=\frac{16}{3}(\Delta_\phi-4\Delta_\phi^3)\mathcal{G}^0+4\Delta_\phi\mathcal{G}^2\notag\\
&\mathcal{G}^5=\frac{64}{15}\Delta_\phi(32\Delta_\phi^4-20\Delta_\phi^2+3)\mathcal{G}^0-\frac{16}{3}\Delta_\phi(4\Delta_\phi^2-1)\mathcal{G}^2+4\Delta_\phi\mathcal{G}^4
\end{align}
Since for $n=2N{+}1$ there are $N{+}1$ unfixed $\mathcal{G}^i$, the ``crossing plane" on which the four-point function must live, is $N$-dimensional. In the rest of paper we will refer the crossing plane as $\mathbf{X}$

\begin{figure}
\begin{center}
\includegraphics[scale=0.4]{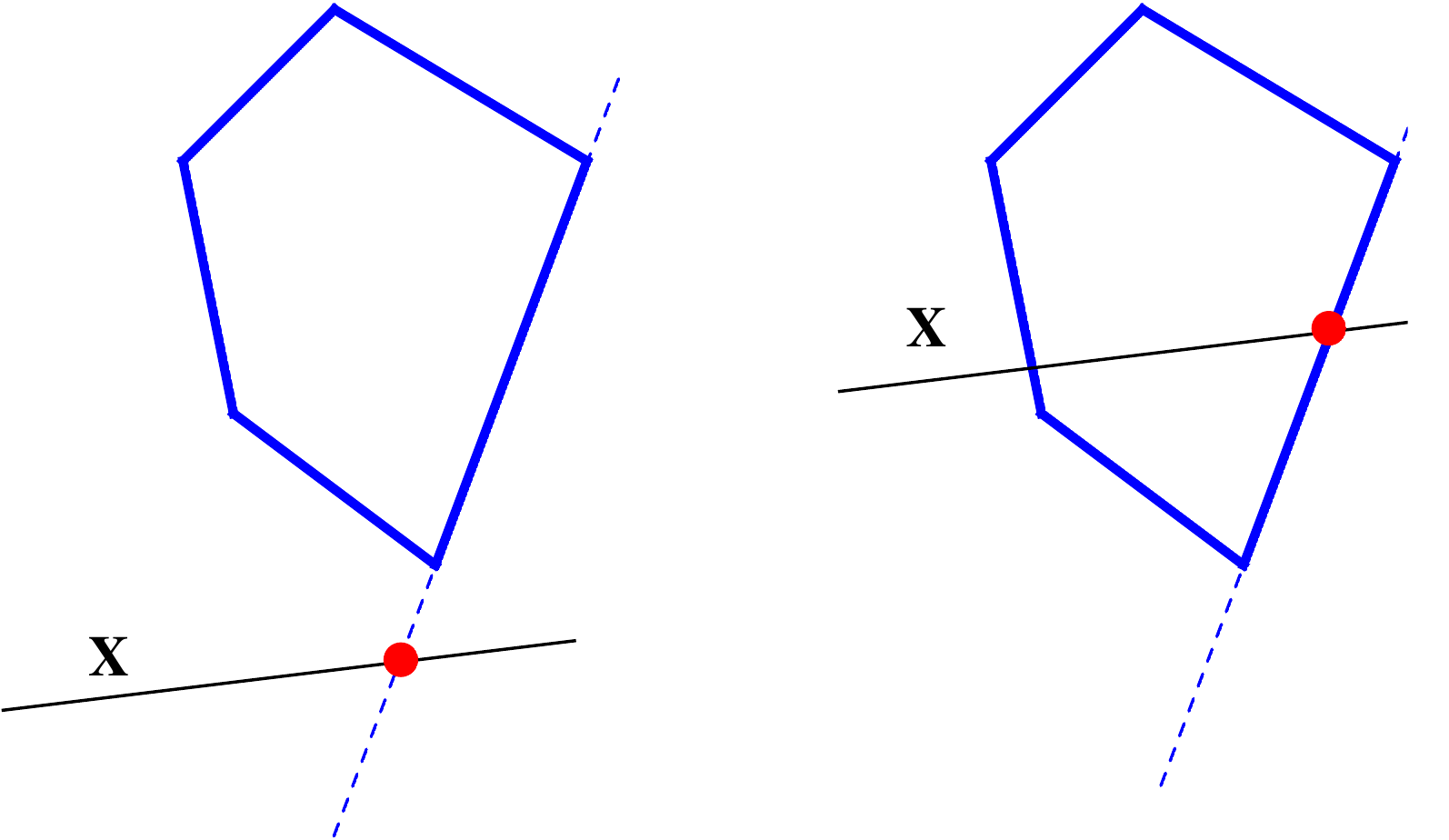}
\caption{The intersection of the ``crossing line" $\mathbf{X}$ with one of the boundary lines of the polygon. In 2 D generically two lines always intersects. The non-trivial constraint is that the intersecting point must be inside the polygon. }
\label{Intersection}
\end{center}
\end{figure}

After such discretization, a solution to the 1D bootstrap problem can be formulated as whether or not the unitary polytope $\mathbf{U}_N$ intersects with the crossing plane $\mathbf{X}$. As the polytope lives in $\mathbb{P}^{2N{+}1}$, if $\mathbf{X}$ intersects with $\mathbf{U}_N$, it will intersect at a point on one of its $N{+}1$-dimensional faces. Let's say that the face in question is comprised of operators $\{\vec{\mathbf{G}}_{\Delta_1},\vec{\mathbf{G}}_{\Delta_2},\cdots,\vec{\mathbf{G}}_{\Delta_{N{+}2}}\}$, the point $\vec{\mathbf{A}}$ for which it intersects $\mathbf{X}$ can be written projectively as 
\eqa\label{Intersect1}
\vec{\mathbf{A}}&=&\langle \mathbf{X},\Delta_1,\Delta_2,\cdots, \Delta_{N{+}1}\rangle  \vec{\mathbf{G}}_{\Delta_{N{+}2}}{+}(-)^{N{+}1}\langle \mathbf{X},\Delta_2,\Delta_3,\cdots, \Delta_{N{+}2}\rangle  \vec{\mathbf{G}}_{\Delta_1}{+}\cdots\\
&+&(-)^{N{+}1}\langle \mathbf{X},\Delta_{N{+}1},\Delta_1,\cdots, \Delta_{N}\rangle  \vec{\mathbf{G}}_{\Delta_{N{+}1}}
\eqae
where by $\langle \mathbf{X},\Delta_1,\Delta_2,\cdots, \Delta_{N{+}1}\rangle$ we are referring to the determinant of the $(2N{+}2)\times(2N{+}2)$ dimensional matrix comprise of the $N{+}1$ vectors of the crossing plane and the $N{+}1$ block vectors. For example, for N=1, we have
\eq
\langle \mathbf{X},\Delta_a,\Delta_b\rangle=\left(\begin{array}{cccc}1 & 0 & 1 & 1 \\ 4\Delta_\phi  & 0 & \frac{G^1_{\Delta_a}}{G^0_{\Delta_a}} &  \frac{G^1_{\Delta_b}}{G^0_{\Delta_b}} \\0 & 1 & \frac{G^2_{\Delta_a}}{G^0_{\Delta_a}} & \frac{G^2_{\Delta_b}}{G^0_{\Delta_b}} \\ \frac{16}{3}(\Delta_\phi-4\Delta^3_\phi) & 4\Delta_\phi  & \frac{G^3_{\Delta_a}}{G^0_{\Delta_a}} & \frac{G^3_{\Delta_b}}{G^0_{\Delta_b}}\end{array}\right)\,.
\eqe 
Now any generic $N{+}1$-dimensional plane will intersect with $\mathbf{X}$ on a point, the question is, whether such point lies inside the unitary polytope, see fig.\ref{Intersection}. To formalize this condition we need to characterize the boundaries of $\mathbf{U}_N$, i.e. its co-dimension one facets. Since these are $2N$-dimensional planes, they are given by $2N{+}1$ vertices, say $\{\Delta_1,\Delta_2,\cdots,\Delta_{N{+}1}\}$, and can be denoted in terms of dual vectors $\vec{\mathcal{W}}_{\{I\}}$ with 
\eq
\vec{\mathcal{W}}_{\{I\}}\equiv \langle ^*,\Delta_1,\Delta_2,\cdots,\Delta_{N{+}1}\rangle\,.
\eqe
The condition for which the intersection point lies inside the polytope is then simply
\eq
\vec{\mathcal{W}}_{\{I\}}\cdot \vec{\mathbf{A}}=\langle \vec{\mathbf{A}},\Delta_1,\Delta_2,\cdots,\Delta_{N{+}1}\rangle>0\,.
\eqe

From the above discussion, we see that to solve the constraints it is important that we know all the facets of $\mathbf{U}_N$, which is in tractable in general given the fact that generic CFT has infinite number of primary operators, and hence the vectors that form the convex hull are infinite. However, in~\cite{Arkani-Hamed:2018ign}, the authors found that the convex hull of 1D block vectors form a \textit{cyclic polytope}. More precisely, as the vectors satisfy
\eq\label{cyclicPoly}
\langle  \mathbf{G}_{\Delta_1},\mathbf{G}_{\Delta_2},\cdots,\mathbf{G}_{\Delta_n}\rangle >0,\quad \Delta_1<\Delta_2<\cdots<\Delta_n\,,
\eqe
i.e. its ordered determinant is definite positive. Similar property was also found in 2D CFT in the diagonal limit~\cite{Sen:2019lec}, which we will come back to in sec.\ref{2DCFT}. For cyclic polytopes, the boundaries are known before hand, and the facets are 
\begin{align}
&d\in~odd:\quad(0,\Delta_i,\Delta_{i+1},\Delta_j,\Delta_{j+1},\dots)\cup(\Delta_i,\Delta_{i+1},\Delta_j,\Delta_{j+1},\dots,\infty)\notag\\
&d\in~even:\quad(\Delta_i,\Delta_{i+1},\Delta_j,\Delta_{j+1},\dots)
\end{align}
where $\Delta_{i+1}$ is the closet vertex to $\Delta_i$. If we assume continuous specturm,
\eq
(\Delta_{i},\Delta_{i+1})\;\rightarrow\; (\Delta_{i},\dot{\Delta}_{i})\,,
\eqe
where $\dot{\Delta}\equiv \frac{d\vec{\mathbf{G}}_{\Delta}}{d\Delta}$.

\begin{figure}
\begin{center}
\includegraphics[scale=0.35]{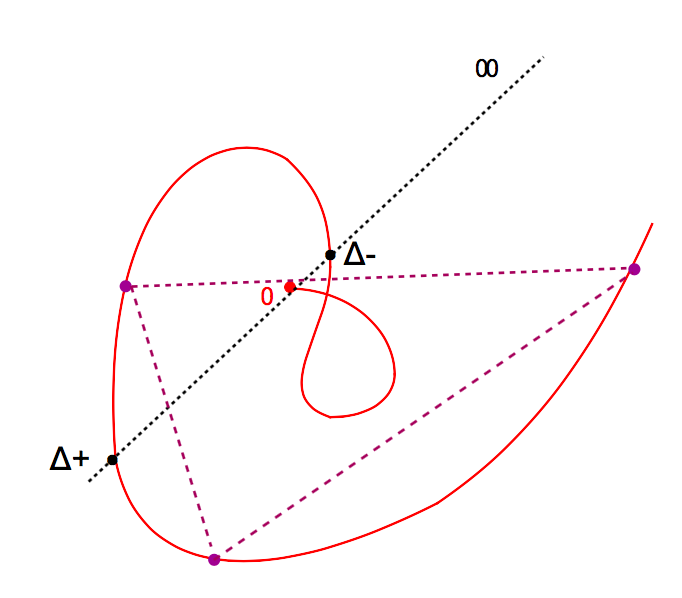}
\caption{Setting $N=2$, the geometry is in $\mathbb{P}^5$. Projecting through $(\mathbf{X},\mathbf{G}_0)$ we have a $2$-dimensional space where the block vectors are displayed on the red curve, which is populated by the CFT operators. A consistent CFT requires that there are at least one triplet of operators in the spectrum such that the corresponding triangle encircles the origin. From the graph one sees that $\Delta_+$, defined by the largest intersection of the line between $\Delta=\infty$ and $\Delta=0$ and the curve, serve as the gap at $N=2$. Since if all operators are above $\Delta_+$, it is impossible to form a triangle encompassing the origin. }
\label{figN=2}
\end{center}
\end{figure}

Having known the facets structures, we come back to pervious problem. Recall that the crossing plane $\mathbf{X}$ will intersect with $N{+}1$-dimensional face of cyclic polytope, which in general takes the form,
\begin{equation}\label{Set}
\{\Delta_{i_1},\Delta_{i_1+1},\Delta_{i_2},\Delta_{i_3},\dots,\Delta_{i_N{+}1}\}\,.
\end{equation}
Using the form of the intersecting point $\vec{\mathbf{A}}$ in eq.(\ref{Intersect1}), we require the point to lie inside the polytope translate to, $\vec{\mathbf{W}_I}\cdot\vec{\mathbf{A}}>0$. Now since in $\mathbb{P}^{2N+1}$, the facet of cyclic polytope is just $(0,i,i+1,j,j+1,\dots)$ and $(i,i+1,j,j+1,\dots,\infty)$, we find that the condition on the set of operators in eq.(\ref{Set}) that formed the intersection face is:
\begin{eqnarray}\label{sign}
\langle\Delta_{i_1+1},\Delta_{i_2},\dots,\Delta_{i_{N{+}1}},\mathbf{X}\rangle,~(-)^{N{+}1}\langle\Delta_{i_2},\Delta_{i_3}\dots,\Delta_{i_{N{+}1}},\Delta_{i_1},\mathbf{X}\rangle,\nonumber\\
(-)^{N{+}1}\langle\Delta_{i_{N{+}1}},\Delta_{i_1},\Delta_{i_1{+}1},\cdots,\Delta_{i_{N{-}1}},\mathbf{X}\rangle,\quad same\;sign
\end{eqnarray}
Since the constraint is given in terms of determinants involving the crossing plane, the relevant geometry is in the space perpendicular to $\mathbf{X}$, which is $N{+}1$-dimensional. Thus the above constraint translates to at fixed $N$, we seek $N{+}2$ operators such that when projected through $\mathbf{X}$, it forms a simplex that incloses the origin. 

Since the identity operator is always present, it is useful to further project this geometry through $\vec{\mathbf{G}}_0$, resulting in an $N$-dimensional space. Thus by projecting through $(\mathbf{X},\mathbf{G}_0)$, we require the CFT spectrum to contain $N{+}1$ operators to form an $N$-dimensional simplex that encloses the origin. This condition corresponds to finding a set of $N{+}1$ operators $(\Delta_{i_1},\Delta_{i_1+1},\Delta_{i_2},\Delta_{i_3},\dots,\Delta_{i_N})$ such that:
  
\eqa\label{sign0}
\displaystyle\langle\mathbf{X},0,\Delta_{i_1+1},\Delta_{i_2},\dots,\Delta_{i_N}\rangle,~(-)^{N}\langle\mathbf{X},0,\Delta_{i_2},\Delta_{i_3},\dots,\Delta_{i_N},\Delta_{i_1}\rangle\nonumber\\
\dots(-)^{N}\langle\mathbf{X},0,\Delta_{i_1},\Delta_{i_1+1},\dots,\Delta_{i_{N-1}}\rangle~~same~sign
\eqae
This will be the primary constraint considered for 1D CFT from now on. Consider for example $N=2$, where projecting through $(\mathbf{X},\mathbf{G}_0)$, results in a two dimensional space. The constraint is then to have three operators in the spectrum that encloses the origin. This is demonstrated in fig.\ref{figN=2}

\subsection{Optimal functionals in the projective geometry}\label{sec:Optimal}
In this subsection we will first give a brief introduction to linear functionals, which gives bound to the lightest operator. In particular, we will focus on the zero patterns of the optimal functional, whose gap is the lowest. In the end we will rephrase the functional in terms our projective geometry.

\begin{figure}
\centering
\includegraphics[width=0.6\textwidth]{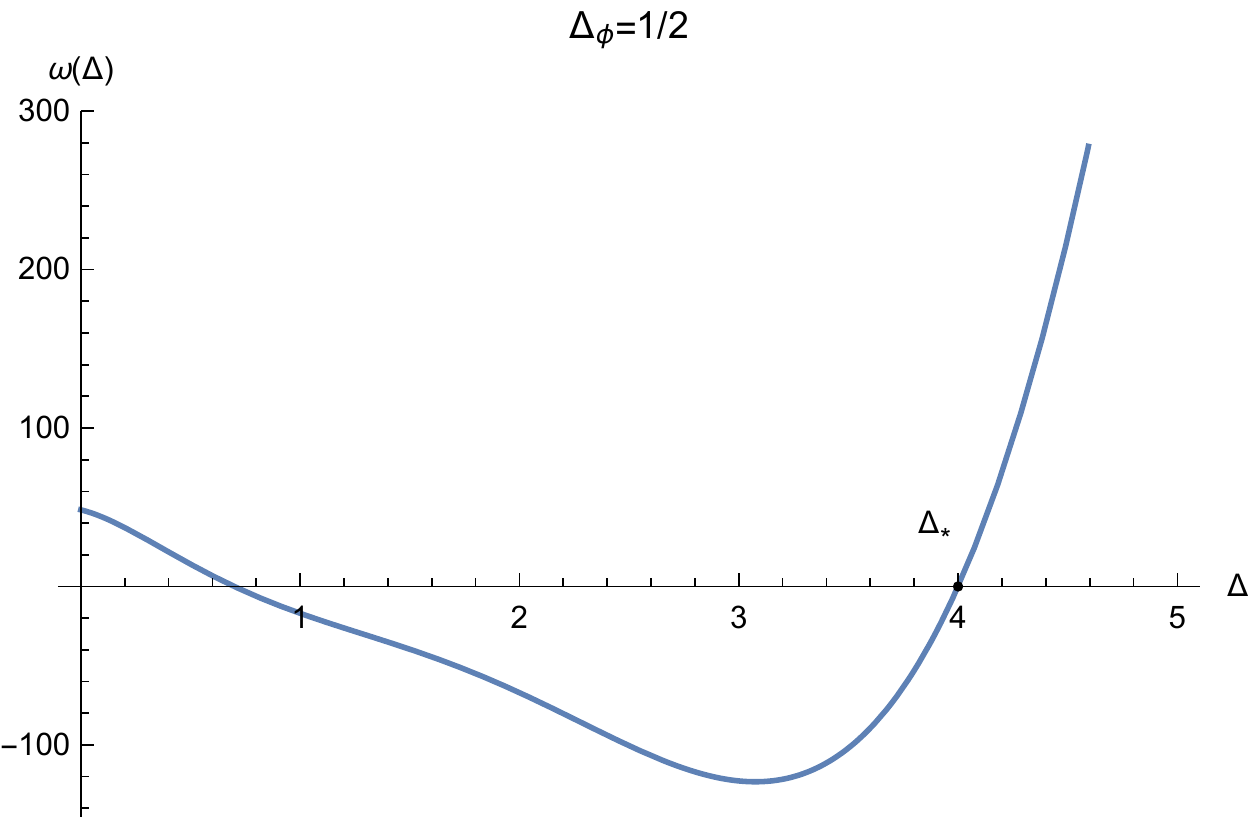}
\caption{\label{fig_fun1}Example of linear functional at $\Delta_\phi=1/2$, satisfying $sign[\omega(F_0^{\Delta_\phi})]=sign[\omega(F_\infty^{\Delta_\phi})]$. The functional is $\omega(\Delta)=(\frac{\mathrm{d}}{\mathrm{d}z}+a_3\frac{\mathrm{d}^3}{\mathrm{d}z^3}+a_5\frac{\mathrm{d}^5}{\mathrm{d}z^5})F_\Delta^{\Delta_\phi}(z)\Big|_{z=1/2}$, where $a_3\approx-1.49,a_5\approx0.01$ And the largest root is $\Delta_*\approx4$}
\end{figure}

First the unitary (\ref{schannel}) and crossing symmetry (\ref{crossing}) conditions can be recast into the following sum rule form,
\begin{equation}\label{sum}
\sum_\Delta c_\Delta^2\Big[z^{-2\Delta_\phi}G_\Delta(z)-(1-z)^{-2\Delta_\phi}G_\Delta(1-z)\Big]=0\Rightarrow\sum_\Delta c_\Delta^2F_\Delta^{\Delta_\phi}(z)=0
\end{equation}
where $F_\Delta^{\Delta_\phi}(z)$ is defined as $F_\Delta^{\Delta}=z^{-2\Delta_\phi}G_\Delta(z)-(z\to1-z)$. Now let's introduce a linear functional $\omega$ which acts on $F_\Delta^{\Delta_\phi}$, $\omega:F_\Delta^{\Delta_\phi}(z)\to \omega[\Delta]$, and satisfying the condition $sign[\omega(F_0^{\Delta_\phi})]=sign[\omega(F_\infty^{\Delta_\phi})]$. In fig.\ref{fig_fun1} shows a example of linear functional with satisfying the sign condition. Now apply 
 this linear functional $\omega$ to the sum rule (\ref{sum})
\begin{equation}\label{FunctionalConstraint}
\omega\Big[\sum_\Delta c_\Delta^2F_\Delta^{\Delta_\phi}(z)\Big]=\omega\Big[F_0^{\Delta_\phi}\Big]+\sum_\Delta c_\Delta^2\omega\Big[F_\Delta^{\Delta_\phi}\Big]=0
\end{equation}
Consider the largest single root of $\omega[\Delta]$, i.e. $\omega[\Delta^*]=0$ and $\omega[\Delta]>0$ for all $\Delta>\Delta^*$, then eq.(\ref{FunctionalConstraint}) cannot be satisfied if all operators are larger than $\Delta^*$, which provides a gap. Thus to have the lowest gap, we seek the functional whose largest single root is the smallest, this is the optimal functional for the problem.

One typical kind of functional used by numerical bootstrap is taking derivatives at $z=1/2$,  
\begin{equation}
\omega(F_\Delta^{\Delta_\phi})=\left[a_1\frac{\mathrm{d}}{\mathrm{d}z}+\frac{a_3}{3!}\frac{\mathrm{d}^3}{\mathrm{d}z^3}+\dots+\frac{a_{2N+1}}{(2N+1)!}\frac{\mathrm{d}^{2N+1}}{\mathrm{d}z^{2N+1}}\right]F_\Delta^{\Delta_\phi}(z)\Big|_{z=1/2}=\vec{\mathbf{\alpha}}\cdot\vec{\mathbf{F}}_\Delta^{\Delta_\phi}
\end{equation}
where $\vec{\mathbf{\alpha}}$ is the coefficient vector $(a_1,a_3,\dots,a_{2N+1})$ and the vector $\vec{\mathbf{F}}_\Delta^{\Delta_\phi}$ is comprised of the Taylor series coefficient of $F_\Delta^{\Delta_\phi}(z)$ at $z=1/2$, which is non-vanishing only for odd number of derivatives due to crossing symmetry. Thus a given functional is characterized by the coefficient vectors $\vec{\mathbf{\alpha}}$.

\begin{figure}
\centering
\includegraphics[width=0.6\textwidth]{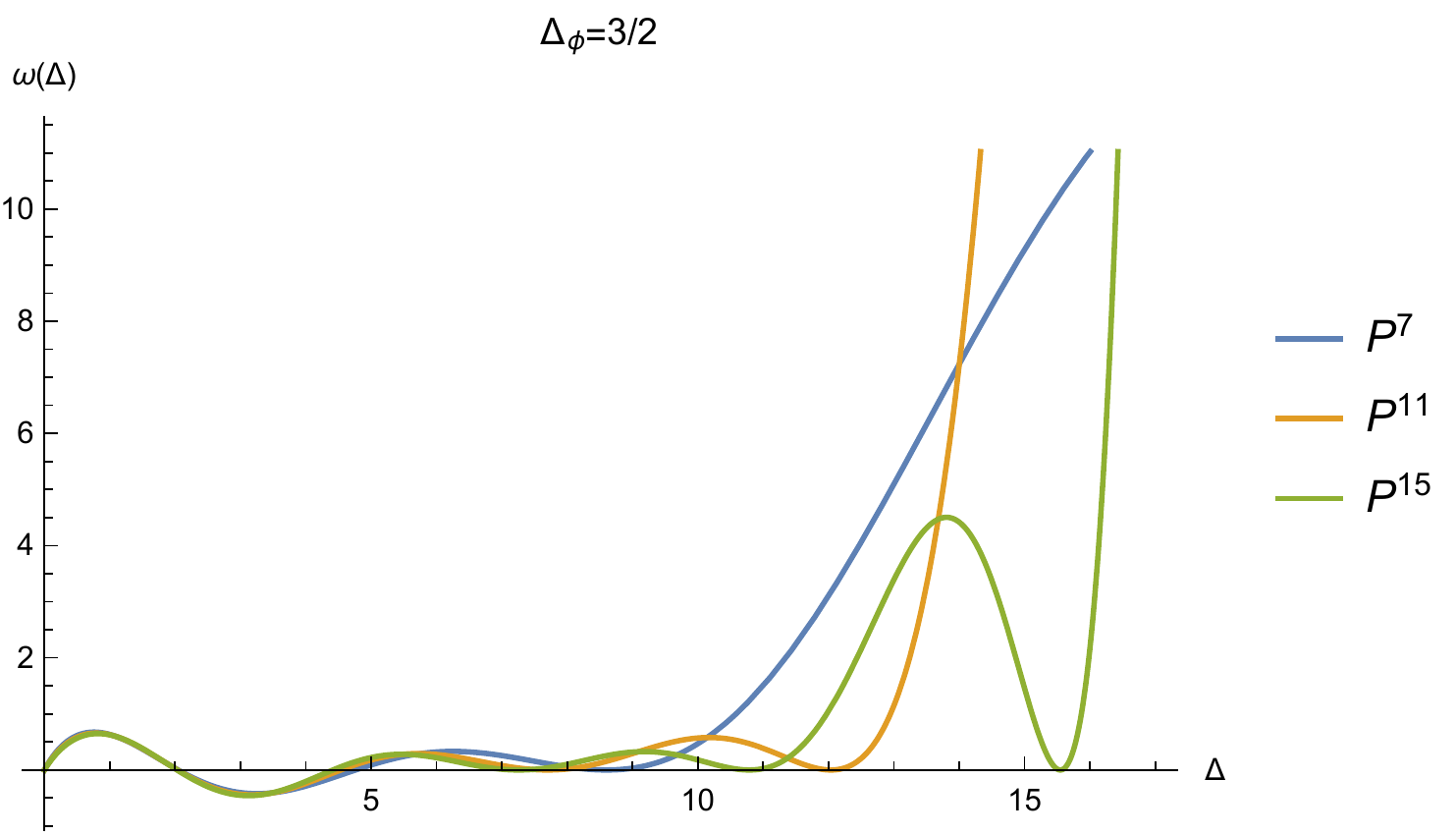}
\caption{\label{fig_optimal1} Optimal functional using 7,11,15 number of derviatives}
\end{figure}

The zero structure of this optimal functional at finite derivative order $n$ reveals interesting simple patterns.\footnote{The simple zero sturcture is also realized in other spinless bootstrap\cite{Afkhami-Jeddi:2019zci}} On the positive real axes, generic functionals will only have three single zeros. One is lying at $\Delta=0$ and the largest one will define the scalar gap. There are also double zeros whose pattern fall in two classes: At $\mathbb{P}^{2N+1}$, when $N=2k+1$ is odd, it will have $k$ double zeros. When $N=2k$ is even, it will have $k-1$ double zeros (Examples are in Fig.\ref{fig_optimal1}). As $N\to\infty$, in the continuum limit, the authors in \cite{Mazac:2016qev, Mazac:2018mdx} construct the extremal functional for 1D CFT as an integral functional, which gives the exact gap $\Delta_{gap}=2\Delta_\phi+1$ for free fermion theory. The functional has an infinite number of double zeros, corresponding to the physical spectrum $\Delta_n=2\Delta_\phi+2n+1$.

Now let us relate the general derivative linear functional to the projective geometry above.  Since $\vec{\mathbf{F}}_\Delta^{\Delta_\phi}$ is given by the difference between the block and its crossing symmetry image, it can be easily identified as projecting the block vectors through the crossing plane, i.e. the part of $\vec{\mathbf{G}}_\Delta$ orthogonal to the crossing plane $\mathbf{X}$. For the functional, a given vector $\vec{\mathbf{\alpha}}$ defines a direction in the complement of $\mathbf{X}$ on which $\vec{\mathbf{F}}_\Delta^{\Delta_\phi}$ is projected on to. At fixed $N$ the complement of $\mathbf{X}$ is $N{+}1$ dimensional, and we can identify  $\vec{\mathbf{\alpha}}$ as the dual vector of an $N$  plane spanned by vectors $\{\mathbf{v}_1,\mathbf{v}_2,\dots,\mathbf{v}_N\}$. With these two observations, we can rewrite the numerical functional $\alpha\cdot\mathbf{F}_\Delta^{\Delta_\phi}$ as
\begin{equation}
\omega(F_\Delta^{\Delta_\phi})=\alpha\cdot\mathbf{F}_\Delta^{\Delta_\phi}=\langle\mathbf{X},\mathbf{W},\Delta\rangle=\langle\mathbf{X},\mathbf{v}_1,\mathbf{v}_2,\dots,\mathbf{v}_N,\Delta\rangle\,.
\end{equation}
Thus in our projective geometry, a functional corresponds to choosing $N$ block vectors to serve as $\{\mathbf{v}_1,\mathbf{v}_2,\dots,\mathbf{v}_N\}$ in the above. We will give a explicit proof of this equation in the appendix.\ref{prof}. Once again, since the identity block is always present, it is natural to choose $\mathbf{v}_1=\vec{\mathbf{G}}_0$, and consider the functional $\langle\mathbf{X},0,\Delta_{i_1},\Delta_{i_2},\dots,\Delta_{i_{N-1}},\Delta\rangle$  Note that $\omega(F_0^{\Delta_\phi})=0$ by construction, so there is no restriction to the sign of functional at $\Delta=\infty$, the largest root of this functional will be a valid scalar gap.

\begin{figure}
\centering
\includegraphics[width=0.6\textwidth]{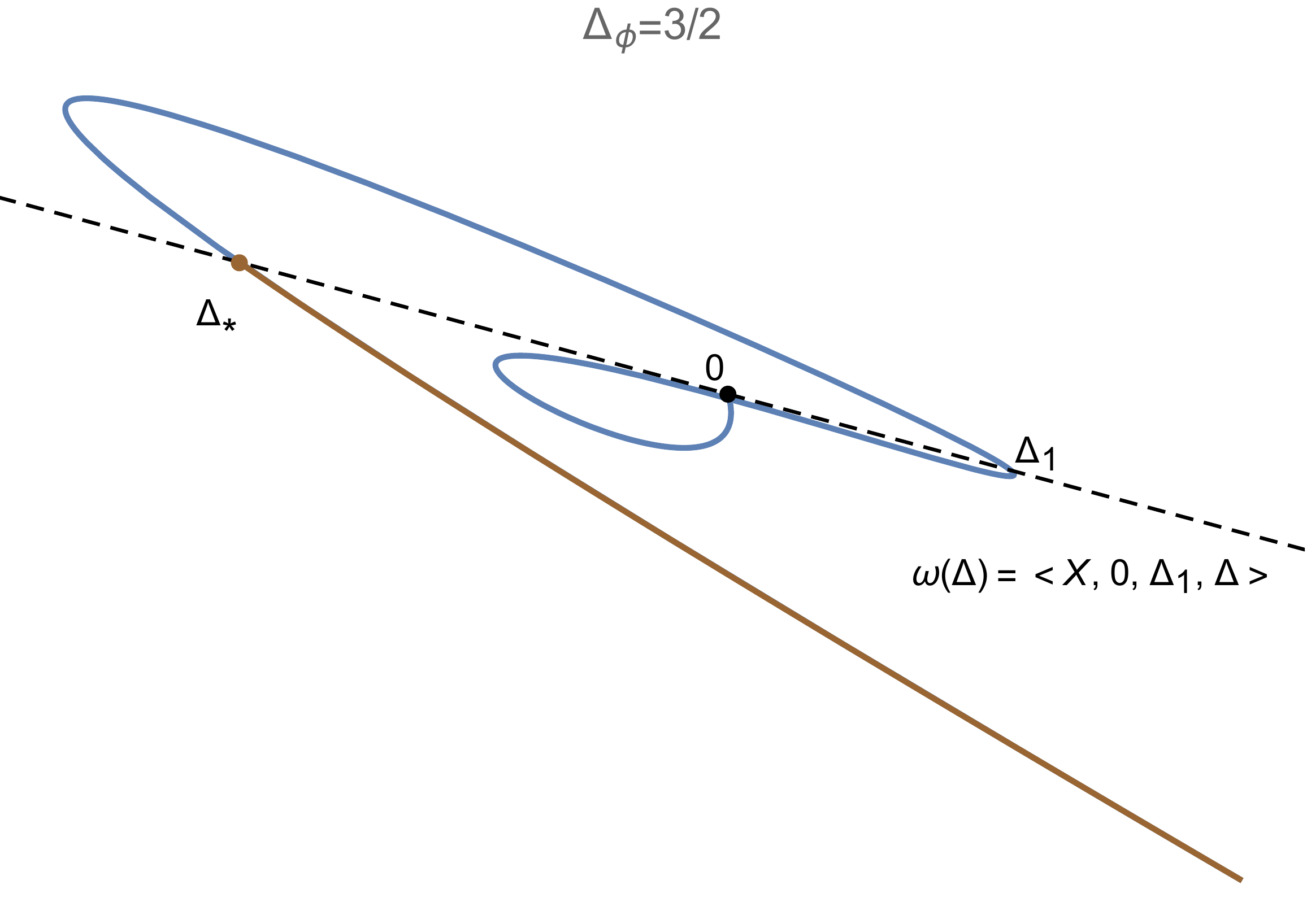}
\caption{\label{fig_p5} 2d geometry of  $\mathbb{P}^5$ in the space orthogonal to $(\mathbf{X},\mathbf{G}_0)$ when $\Delta_\phi=3/2$. The black dot is the origin, the black dashed line intersects the curve at $\langle\mathbf{X},0,\Delta_1,\Delta\rangle=0$, where $\Delta_1=3.3$. The brown dot represent the largest intersection of the line $\Delta=\Delta_*\approx6.90$. The brown region is when $\Delta>\Delta_*$}
\end{figure}

It is instructive to consider why the largest single root of $\langle\mathbf{X},0,\Delta_{i_1},\Delta_{i_2},\dots,\Delta_{i_{N-1}},\Delta\rangle$ correspond to a gap from a pure geometric standpoint. Recall from (\ref{sign0}), projecting through the crossing plane and identity block $(\mathbf{X},\mathbf{G}_0)$,  there must exist $N+1$ operators $\{\Delta_{j_1},\Delta_{j_1+1},\Delta_{j_2},\dots,\Delta_{j_N}\}$ forming $N$-simplex to enclose origin. Now consider the solution space of $\langle\mathbf{X},0,\Delta_{i_1},\Delta_{i_2},\dots,\Delta_{i_{N-1}},\Delta\rangle=0$, it defines a co-dimensional one plane spanned by these vertex $\{\Delta_{i_1},\Delta_{i_2},\cdots,\Delta_{i_{N-1}}\}$ in the same space. So if all the operators in the spectrum have their scaling dimension above the largest single root $\Delta>\Delta_*$, then they all lie on the same side of the plane in discussion which contains the origin (Example in $\mathbb{P}^5$ in Fig.\ref{fig_p5}), and this it is impossible to form a simplex containing origin. This tells us that there must be some point lower than $\Delta_*$ situated at the other side of the plane. In the following we will use the fact that the convex hull of the block vectors form a cyclic polytope to determine the optimal functional.

\subsection{The Optimal functional at $\mathbb{P}^{2N+1}$} 
We will now derive the optimal functional utilizing the fact that the convex hull of block vectors form a cyclic polytope. Recall that from the projective geometry point of view, a functional can be mapped to 
\eq
\omega(F_\Delta^{\Delta_\phi})=\langle \mathbf{X},0,\Delta_{i_1},\Delta_{i_2},\dots,\Delta_{i_{N-1}},\Delta\rangle\,.
\eqe
For the optimal functional, we expect would like to choose the $\Delta_i$s in the above such that the last single root is the smallest. In other words, we would like to find an $\omega(F_\Delta^{\Delta_\phi})$ such that it is positive for a majority of $\Delta$. This naturally leads to the conclusion that $\{0,\Delta_{i_1},\Delta_{i_2},\dots,\Delta_{i_{N-1}}\}$ must be associated with one of the higher co-dimensional boundaries of the unitary polytope. As we will see, with the input of the cyclic polytope nature of $\mathbf{U}_N$, this allows us to uniquely determine the optimal functional!    

\subsubsection{Odd $N=2k+1$}\label{odd}
Let's begin with $N=2k+1$, as we will see later on, the case for $N\in even$ follows trivially. Since the boundaries of the cyclic polytope are known, we begin with the fact that in $2N{+}1$ dimensions, the block vector $\vec{\mathbf{G}}_\Delta$ will always satisfy,
\begin{equation}\label{X0}
\langle 0,i,i{+}1,j,j{+}1,\cdots,\Delta\rangle>0
\end{equation}
 So the natural ansatz of optimal functional for $N=2k+1$ is 
 \begin{equation}\label{optimal1}
\omega(F_\Delta^{\Delta_\phi})=\langle\mathbf{X},0,\Delta_{i_1},\Delta_{i_1+1},\dots,\Delta_{i_k},\Delta_{i_k+1},\Delta\rangle=\langle\mathbf{X},0,\Delta_{i_1},\dot{\Delta}_{i_1},\dots,\Delta_{i_k},\dot{\Delta}_{i_k},\Delta\rangle
\end{equation}
This can be viewed as projecting eq.(\ref{X0}) through $\mathbf{X}$, for which the properties associated with cyclic polytope may no longer hold true. For example, the determinant may no longer be positive. However for the region $\Delta\gg \Delta_\phi$, since the components for the block vectors are comparatively large with respect to the vectors on the crossing plane, we expect that the determinant will be positive satisfying our criteria. (Schematic explaination in appendix.\ref{positive}) Note that since by $(i,i{+}1)$ we are referring to $(\vec{\mathbf{G}}_{\Delta_i}, \dot{\vec{\mathbf{G}}}_{\Delta_i})$, by construction this functional gives double zeros at $k$ positions. Thus the presence of double zeros can be linked to the property of cyclic polytope! 

\begin{figure}
\centering
\begin{minipage}[t]{0.48\textwidth}
\centering
\includegraphics[width=6cm]{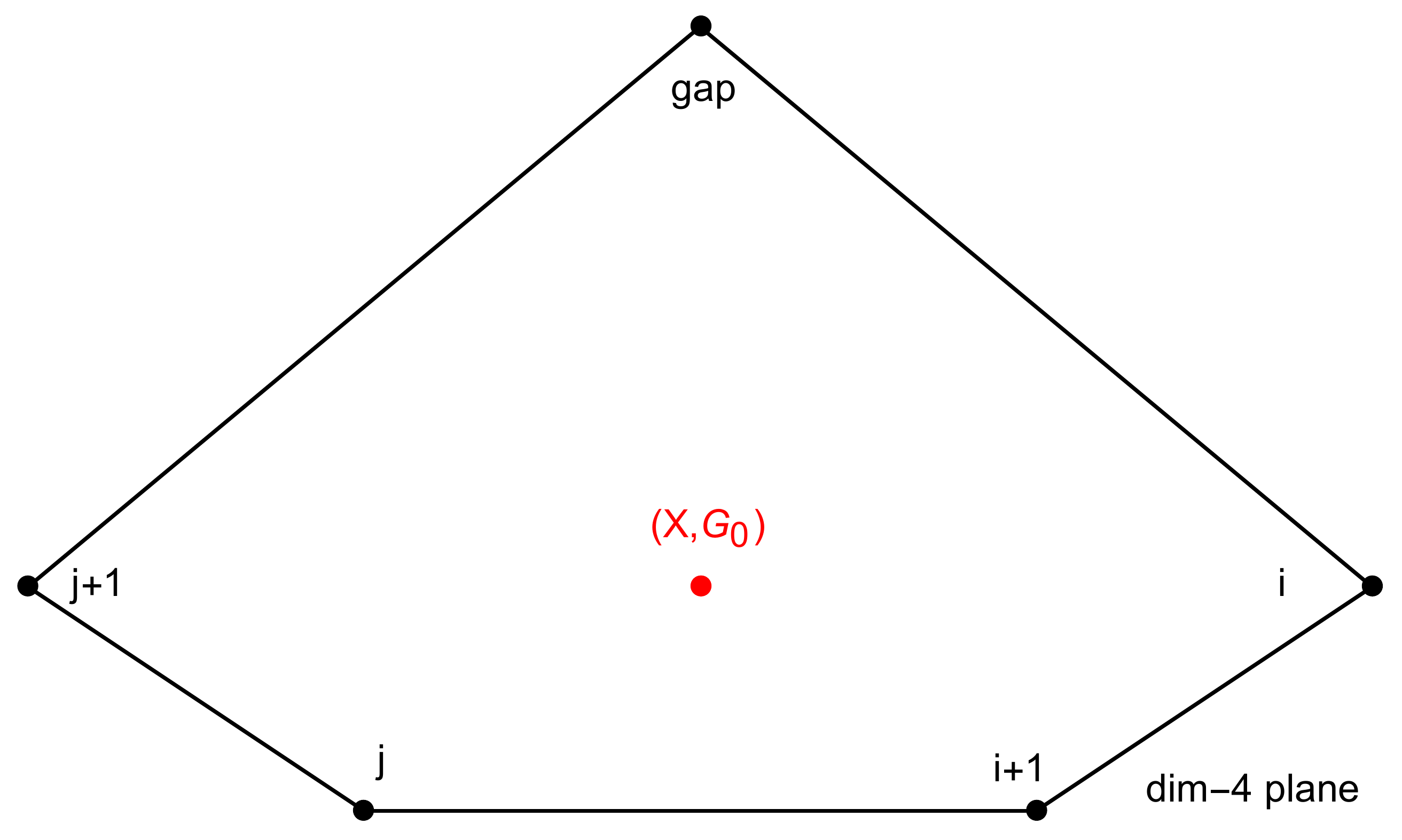}
\end{minipage}
\begin{minipage}[t]{0.48\textwidth}
\centering
\includegraphics[width=5cm]{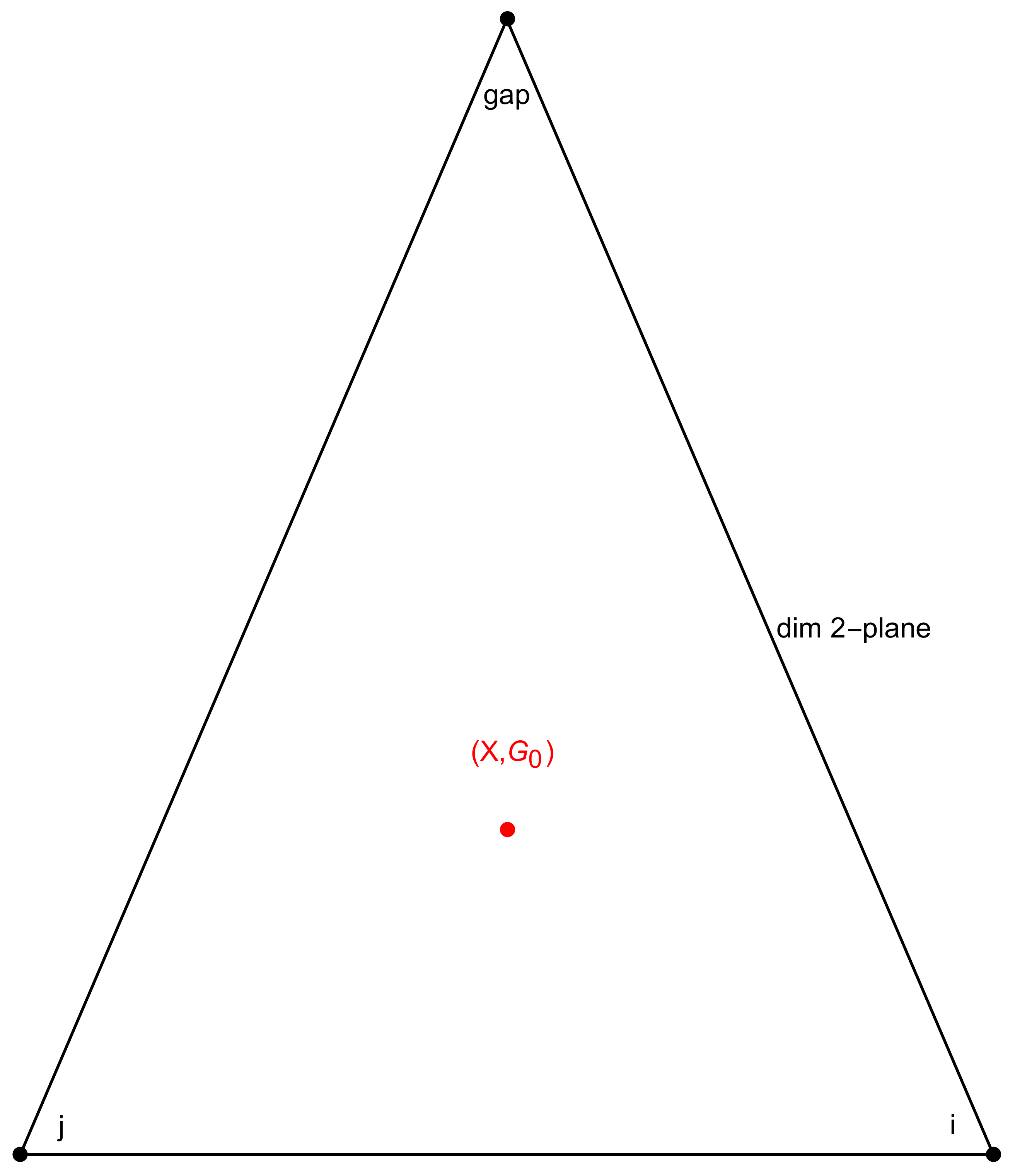}
\end{minipage}
\caption{Here we show two type of simplex in the four-dimensional space orthogonal to $(\mathbf{X},\mathbf{G}_0)$  that contains the origin, denoted as the red dot. On the left hand side, the origin is lying in the 4-dimensional simplex spanned by $\{\mathbf{G}_{\Delta_{gap}},\mathbf{G}_{\Delta_{i}},\mathbf{G}_{\Delta_{i+1}},\mathbf{G}_{\Delta_{j}},\mathbf{G}_{\Delta_{j+1}}\}$. On the right hand side, the two pairs of vertices become degenerate and the simplex shrinks to a triangle spanned by $\{\mathbf{G}_{\Delta_{gap}},\mathbf{G}_{\Delta_{i}},\mathbf{G}_{\Delta_{j}}\}$. }
\label{fig_kplane}
\end{figure}

The problem of the preferred functional now boils down to the selection of $k$-points. Let us step back and recall for a spectrum to be consistent, eq.(\ref{sign0}) tells us that one should have $2k+2$ operators $\{\Delta_1,\Delta_2,\cdots,\Delta_{2k+2}\}$ forming a $(2k+1)$-simplex which containing the origin $\mathbf{0}$. Now for the optimal case, where one has the true gap for the given derivative order, we expect that the spectrum which includes the gap to be degenerate such that any modification of the operator on the gap will render the origin out side of the simplex. This tells us that 1. the gap must form one of the vertices of this simplex, and 2. the origin must be on its boundary. This precisely translate to 
\eq
\langle\mathbf{X},0,\Delta_{i_1},\Delta_{i_1+1},\dots,\Delta_{i_k},\Delta_{i_k+1},\Delta_{Gap}\rangle=0\,.
\eqe
Now since we have the adjacent block vectors being in the continuous limit, each pair of vertices collapses into one point $(\vec{\mathbf{G}}_{\Delta_{i_k}},\vec{\mathbf{G}}_{\Delta_{i_k}+1})\to(\vec{\mathbf{G}}_{\Delta_{i_k}},\vec{\mathbf{G}}_{\Delta_{i_k}}+\delta\Delta\dot{\vec{\mathbf{G}}}_{\Delta_{i_k}})$. Thus for this $2k$-plane to inclose the origin, we must have that the $k$-vectors along with the gap form a $k$-plane that incloses the origin ! (see Fig.\ref{fig_kplane} for illustration) Thus in conclusion, the optimal functional at  $N=2k+1$ is given by 
\eq
\langle\mathbf{X},0,\Delta_{i_1},\Delta_{i_1+1},\dots,\Delta_{i_k},\Delta_{i_k+1},\Delta\rangle
\eqe
where its last single root $\Delta^*$, when combined with $\{\Delta_{i_1},\Delta_{i_2},\cdots,\Delta_{i_k}\}$ forms a $k$-plane containing zero.

\begin{figure}
\centering
\includegraphics[width=0.4\textwidth]{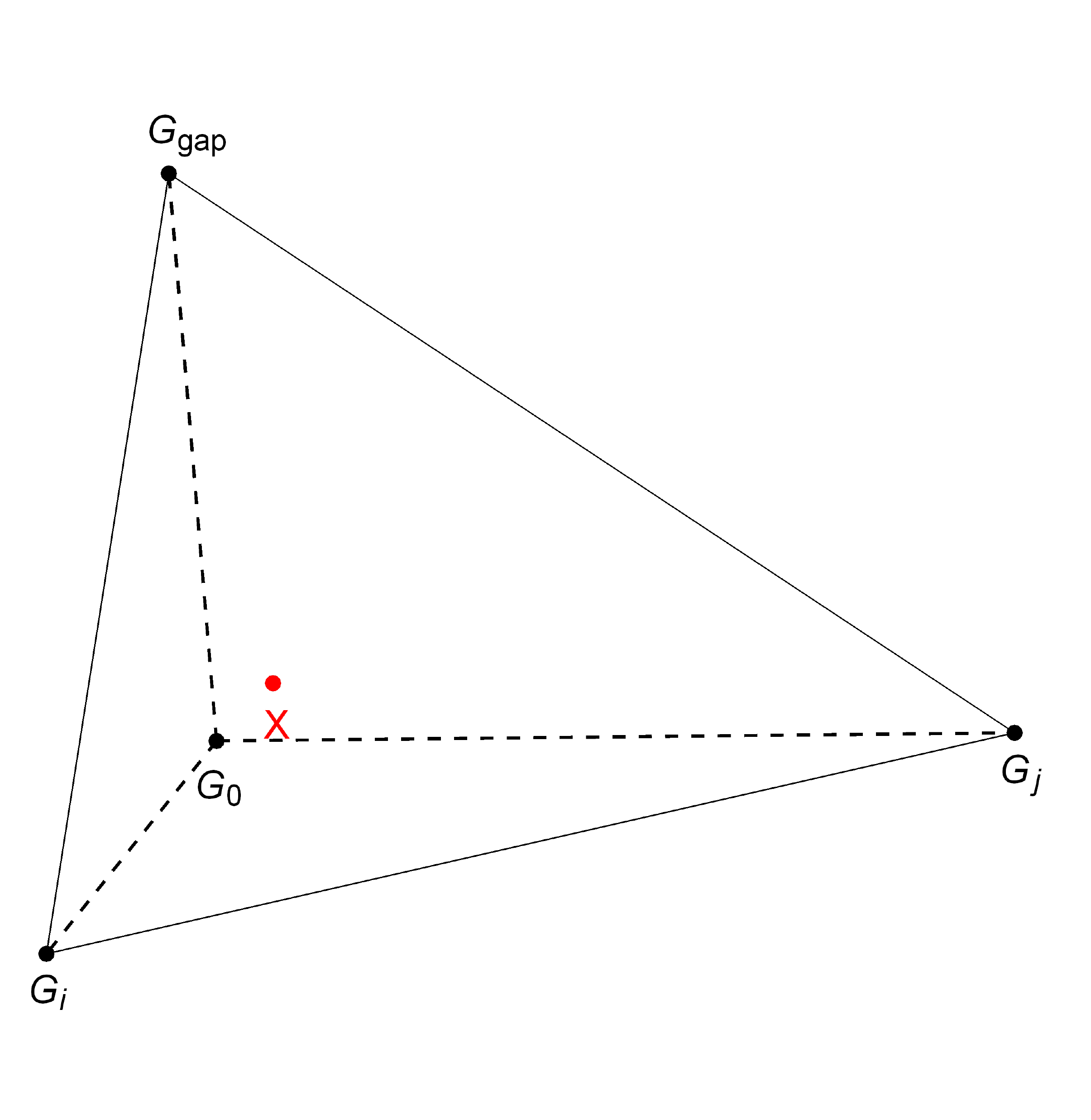}
\caption{\label{fig_select} Here we present the 3-dimensional polytope made by vertices $\{\mathbf{G}_0,\mathbf{G}_{\Delta_i},\mathbf{G}_{\Delta_j},\mathbf{G}_{\Delta_{gap}}\}$ in the space orthorognal to $\mathbf{X}$. Origin is the red dot, and it's inside the 3-dimensional polytope}
\end{figure}

Now the name of the game is to find a set of $k{+}1$-vectors spanning a $k$-plane enclosing the origin. In practice, the co-plane condition implies,
\begin{equation}\label{subplane}
\langle\mathbf{X},0,\Delta_{gap},\Delta_{i_1},\Delta_{i_2},\dots,\Delta_{i_k},\mathbf{v}_1,\mathbf{v}_2,\dots\mathbf{v}_k\rangle=0,\quad for~abritary~\mathbf{v}_1,\mathbf{v}_2,\dots,\mathbf{v}_k\,,
\end{equation}
while the requirement that the origin is ``inside"  the $k$-dimensional simplex  (inside the 3d simlpex of Fig.\ref{fig_select}) eq.(\ref{sign}) implies the following consistent sign pattern:
\begin{align}\label{select}
&~\langle\mathbf{X},\Delta_{gap},\Delta_{i_1},\Delta_{i_2},\dots,\Delta_{i_k},\mathbf{v}_1,\dots,\mathbf{v}_{k+1}\rangle,\quad-\langle\mathbf{X},0,\Delta_{i_1},\Delta_{i_2},\dots,\Delta_{i_k},\mathbf{v}_1,\dots,\mathbf{v}_{k+1}\rangle,\notag\\
&\dots(-1)^{k+1},\langle\mathbf{X},0,\Delta_{i_1},\Delta_{i_2},\dots,\Delta_{i_k},\mathbf{v}_1,\dots,\mathbf{v}_{k+1}\rangle~same~sign~for~arbitrary~\mathbf{v}_1,\mathbf{v}_2,\dots,\mathbf{v}_{k+1}
\end{align}
This is summarized in the following two steps: 
\begin{itemize}
\item Step1: Take $k{+}1$ linear independent set of $\mathbf{v}_i$ s and use eq.(\ref{subplane}) to solve for all the possible solutions of $\{\Delta_{gap},\Delta_{i_1},\dots,\Delta_{i_k}\}$.
\item Step2: From the above solutions, select only those that satisfies the condition (\ref{select}). So that the vertices will form a simplex enclosing origin $\mathbf{0}$.
\item Step3: The operator with lowest scaling dimension in the solution is $\Delta_{gap}$, while the remaining $k$ operators serve as $\{\Delta_{i_1},\dots\Delta_{i_k}\}$ in the optimal functional (\ref{optimal1}). 
\end{itemize}
We have explicitly verified up to $\mathbb{P}^{23}$ that this uniquely determines the functional and matches with the optimal functional from the numeric bootstrap ! Let us consider the example in $\mathbb{P}^{7}$ as an illustration of the procedure.

\begin{figure}
\centering
\includegraphics[width=0.6\textwidth]{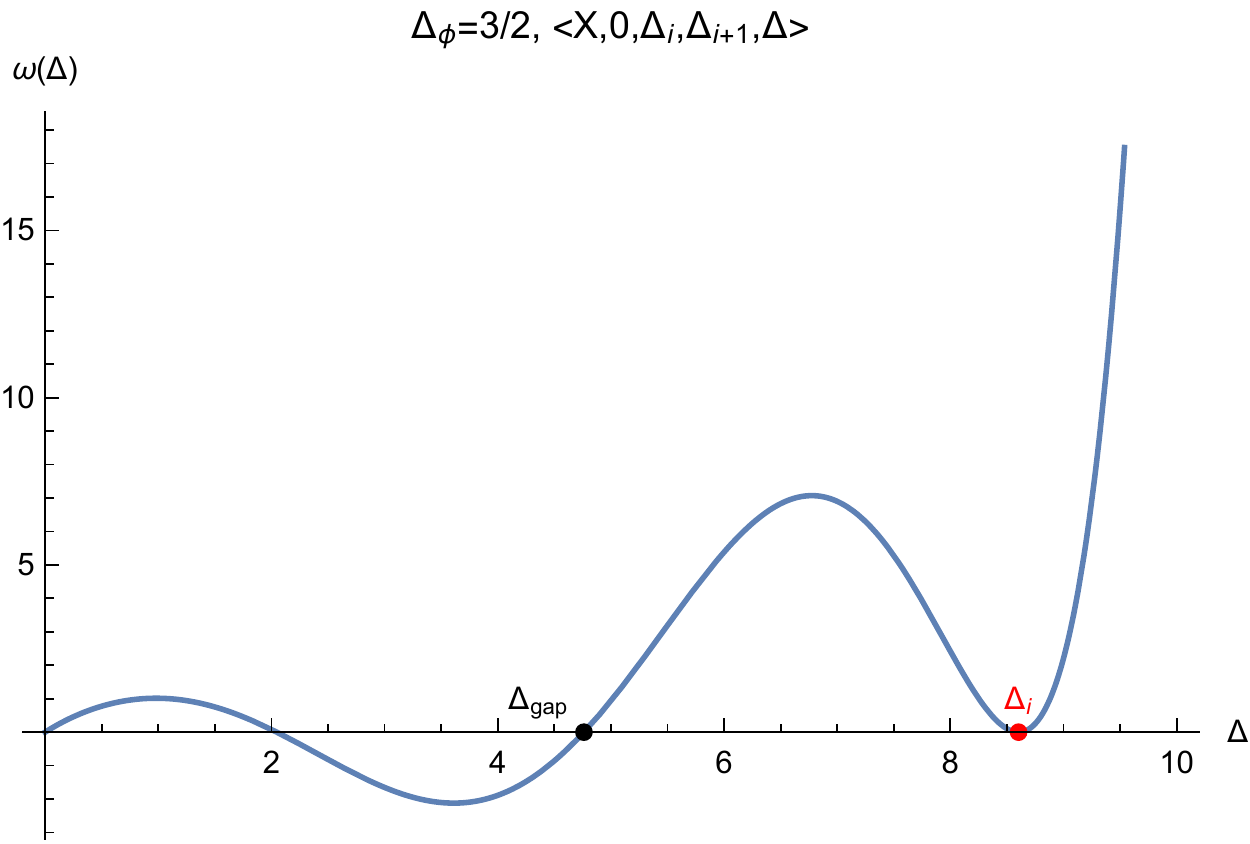}
\caption{\label{fig_p7fun} The plot of functional $\omega(F_\Delta^{\Delta_\phi})=\langle\mathbf{X},0,\Delta_i,\Delta_{i+1},\Delta\rangle$ when $\Delta_\phi=3/2$. The black dot corresponds to $\Delta_{gap}\approx4.762$, while the red dot corresponds to $\Delta_i\approx8.603$}
\end{figure}

{\large\textbf{Example in $\mathbb{P}^7$ with $\Delta_\phi=3/2$}:} In $\mathbb{P}^7$, crossing plane $\mathbf{X}$ is 3-dimensional, so the space orthogonal to $(\mathbf{X},\mathbf{G}_0)$ is also 3-dimensinal. The first thing is to find all the special 1-dimensional plane which intersecting origin $\mathbf{0}$ and two other opearators $(\Delta_a,\Delta_b)$ on it. So we can just set $k=1$ in (\ref{subplane}) to get the condition,
\begin{equation}
For~abitrary~\mathbf{v},~\langle\mathbf{X},0,\Delta_a,\Delta_b,\mathbf{v}\rangle=0
\end{equation}
So we can just chose two random, linear independent vectors $\mathbf{v}_1,\mathbf{v}_2$,
\begin{equation}\label{subplane7}
\langle\mathbf{X},0,\Delta_a,\Delta_b,\mathbf{v}_1\rangle=0,\quad\langle\mathbf{X},0,\Delta_a,\Delta_b,\mathbf{v}_2\rangle=0
\end{equation}
Solving these two non-linear equation we get three sets of solutions, they are
\begin{equation}
(\Delta_a,\Delta_b)\approx\{(4.762,8.603),(2.047,8.567),(2.044,4.680)\}
\end{equation}
Next we use condition (\ref{select}) to find the right 1-simplex, this condition in $\mathbb{P}^7$ simplifies to
\begin{equation}\label{select7}
For~abitrary~\mathbf{v}_1,\mathbf{v}_2,~\langle\mathbf{X},\Delta_a,\Delta_b,\mathbf{v}_1,\mathbf{v}_2\rangle,~-\langle\mathbf{X},0,\Delta_b,\mathbf{v}_1,\mathbf{v}_2\rangle,~\langle\mathbf{X},0,\Delta_a,\mathbf{v}_1,\mathbf{v}_2\rangle~same~sign
\end{equation}
We found that only one of the solution $(\Delta_a,\Delta_b)\approx(4.762,8.603)$ will satisfy this constraint. In this solution, the smallest root will correpsonds the scalar gap $\Delta_{gap}\approx4.762$ in $\mathbb{P}^7$, And the remaining one will correspond to $\Delta_i\approx8.603$ in the optimal functional,
\begin{equation}
\omega(F_\Delta^{\Delta_\phi})=\langle\mathbf{X},0,\Delta_i,\Delta_{i+1},\Delta\rangle=\langle\mathbf{X},0,\Delta_i,\dot{\Delta}_i,\Delta\rangle
\end{equation} 
The plot of this funcitional is in Fig.\ref{fig_p7fun}. This largest root of this functional exactly equals to $\Delta_{gap}$. One can also use linear programming to optimize the functional which has the following form, and it will exactly correspond to the functional $\langle\mathbf{X},0,\Delta_i,\Delta_{i+1},\Delta\rangle$
\begin{equation}
\tilde{\omega}(F_\Delta^{\Delta_\phi})=\left(\frac{\mathrm{d}}{\mathrm{d}z}+\frac{a_3}{3!}\frac{\mathrm{d}^3}{\mathrm{d}z^3}+\frac{a_5}{5!}\frac{\mathrm{d}^5}{\mathrm{d}z^5}+\frac{a_7}{7!}\frac{\mathrm{d}^7}{\mathrm{d}z^7}\right)F_\Delta^{\Delta_\phi}(z)\Big|_{z=1/2}
\end{equation}
where $\{a_3,a_5,a_7\}\approx\{-0.187,0.0230,-0.00141\}$ \footnote{Precise value of zeros in the optimal functional from $\mathbb{P}^7$ to $\mathbb{P}^{23}$ in $1D$ CFT are in the attachment ``functional.nb". 
Also examples from $2D$ CFT, spinless modular bootstrap are also included.} 

\subsubsection{Even $N=2k$}
In $\mathbb{P}^{4k+1}$, we observed that the roots structure (i.e. the positions of single zeros and double zeros) of optimal functional is the same as $\mathbb{P}^{4k-1}$. So the natrual guess for the functional is,
\begin{equation}
\omega(F_\Delta^{\Delta_\phi})=\langle\mathbf{X},0,\Delta_{i_1},\Delta_{i_1+1},\dots,\Delta_{i_{k-1}},\Delta_{i_{k-1}+1},\infty,\Delta\rangle
\end{equation}
where $\{\mathbf{G}_{\Delta_{i_1}},\mathbf{G}_{\Delta_{i_2}},\dots,\mathbf{G}_{\Delta_{i_{k-1}}}\}$ are the same vertexs in optimal functional of $\omega(F_\Delta^{\Delta_\phi})$ in $\mathbb{P}^{4k-1}$. 
\begin{align}
\omega(F_\Delta^{\Delta_\phi})&=\mathrm{Det}[\mathbf{X},\mathbf{G}_0,\mathbf{G}_{\Delta_{i_1}},\dots,\mathbf{G}_\infty,\mathbf{G}_\Delta]
=\mathrm{Det}[\mathbf{X}_1,\mathbf{X}_2,\dots,\mathbf{X}_{2k+1},\mathbf{G}_0,\mathbf{G}_{\Delta_{i_1}},\dots,\mathbf{G}_\infty,\mathbf{G}_\Delta]\notag\\
&=\mathrm{Det}
\begin{pmatrix}
1&0&\cdots&0&&&&0\\
4\Delta_\phi&0&\cdots&0&&&&\vdots\\
0&1&\cdots&\vdots&&&&\vdots\\
\frac{16}{3}(\Delta_\phi-4\Delta_\phi^3)&4\Delta_\phi&\cdots&\vdots&&&&\vdots\\
0&0&\cdots&\vdots&\mathbf{G}_0&\mathbf{G}_{\Delta_{i_1}}&\cdots&\vdots&\mathbf{G}_\Delta\\
\frac{64}{15}\Delta_\phi(32\Delta_\phi^4-20\Delta_\phi^2+3)&\frac{16}{3}(\Delta_\phi-4\Delta_\phi^3)&\cdots&\vdots&&&&\vdots\\
\vdots&\vdots&\ddots&\vdots&&&&\vdots\\
\cdots&\cdots&\cdots&1&&&&0\\
\cdots&\cdots&\cdots&4\Delta_\phi&&&&1
\end{pmatrix}
_{(4k)\times(4k)}
\end{align}
where $\{\mathbf{X}_1,\mathbf{X}_2,\dots,\mathbf{X}_{2k+2}\}$ are $2k+2$ linear independent vectors that span the crossing plane $\mathbf{X}$. Observing that $\mathbf{G}_\infty=\{0,0,\dots,1\}$ and the last component of crossing plane is $\mathbf{X}_{2k+1}=\{0,0,\dots,1,4\Delta_\phi\}$, we get this determinant equals to,
\begin{align}
\omega(F_\Delta^{\Delta_\phi})&=-4\Delta_\phi\mathrm{Det}[\mathbf{X}_1,\mathbf{X}_2,\dots,\mathbf{X}_{2k-1},\mathbf{G}_0,\mathbf{G}_{\Delta_{i_1}},\dots,\mathbf{G}_{\Delta_{i_{k-1}+1}},\mathbf{G}_\Delta]\,.
\end{align}
In other words, the determinant is reduce from that of an $4k\times4k$ matrix to an  $4k{-}2\times4k{-}2$, i.e. it will be proportional to $\langle\mathbf{X},0,\Delta_{i_1},\Delta_{i_1+1},\dots,\Delta_{i_{k-1}},\Delta_{i_{k-1}+1},\Delta\rangle$, with is exactly the optimal functional in $\mathbb{P}^{4k-3}$. This has also been verified to match with linear programming up to  $\mathbb{P}^{25}$.
\subsubsection{Implication of subplane condition}\label{explainsub}
Note that when the functional becomes optimal, condition (\ref{subplane}) and (\ref{select}) implies a vector equation for $\vec{\mathbf{F}}_\Delta^{\Delta_\phi}$ which will allow us to approximate the OPE coefficients. First we tranform the condition (\ref{subplane}) to the equation related to $\vec{\mathbf{F}}_\Delta^{\Delta_\phi}$, using (\ref{appA}) from appendix.\ref{prof}
\eqa
\langle\mathbf{X},0,\Delta_{gap},\Delta_{i_1},\Delta_{i_2},\dots,\Delta_{i_k},\mathbf{v}_1,\mathbf{v}_2,\dots\mathbf{v}_k\rangle
\propto\langle\mathbf{F}_0^{\Delta_\phi},\mathbf{F}_{\Delta_{gap}}^{\Delta_\phi},\mathbf{F}_{\Delta_{i_1}}^{\Delta_\phi},\dots,\mathbf{F}_{\Delta_{gap}}^{\Delta_\phi},\tilde{\mathbf{v}}_1,\dots\tilde{\mathbf{v}}_k\rangle\,,\nonumber\\
\eqae
where $\tilde{\mathbf{v}}_i$ is just some GL rotation acting on $\mathbf{v}_i$. So if the above vanishes for arbitrary $\tilde{\mathbf{v}}_1,\dots\tilde{\mathbf{v}}_k$, it implies that the vectors $\{\mathbf{F}_{\Delta_{gap}}^{\Delta_\phi},\dots,\mathbf{F}_{\Delta_{i_k}}^{\Delta_\phi}\}$ are linear dependent.
\begin{equation}\label{sumrule}
\vec{\mathbf{F}}_0^{\Delta_\phi}+\alpha_{gap}\vec{\mathbf{F}}_{\Delta_{gap}}^{\Delta_\phi}+\sum_{l=1}^k\alpha_l\vec{\mathbf{F}}_{\Delta_{i_l}}^{\Delta_\phi}=\vec{0}
\end{equation}
Furthermore, the simplex selection condition (\ref{select}) implies that all the coefficient above will be positive, 
\begin{equation}\label{ope}
\alpha_{gap},\alpha_1,\dots,\alpha_k>0
\end{equation}
In other words, the two conditions give a solution to the OPE coefficient for the truncated crossing equation! 
\begin{align}\label{spectOPE}
\text{Spectrum}:~\Delta_{gap},\Delta_{i_1},\dots,\Delta_{i_k}\notag\\
\text{OPE coefficient}:~\alpha_{gap},\alpha_1,\dots,\alpha_k
\end{align}
Note that since we are at finite $N$, the corresponding gap is not exact and is higher than the true gap, there are no physical theories living on the gap for finite $N$. Thus the solution is an approximation for the OPE coefficients. 

The fact that optimal functionals yield approximations for the OPE coefficients was discussed in \cite{ElShowk:2012hu}. There, the authors introduced the \emph{Extremal Functional Method} (EFM) which extracts the spectrum and OPE coeffcient of CFT that lives in the boundary of extremal functional. Basically, it contains the following three steps:
\begin{itemize}
\item Find the optimal linear functional $\omega(F_\Delta^{\Delta_\phi})$, which is equivelent to finding $\alpha$ in $\alpha\cdot\mathbf{F}_\Delta^{\Delta_\phi}$
\item Compute the vectors $\mathbf{F}_\Delta^{\Delta_\phi}$ which are zeros of $\omega(F_\Delta^{\Delta_\phi})$
\item Solve for the linear combination of $\mathbf{F}_\Delta^{\Delta_\phi}$' s which gives the identity vector. The coefficients are the square of the OPE coefficients
\end{itemize}
In 1D CFT, we can use the (\ref{subplane}) and (\ref{select}) to get the spectrum $\{\Delta_{gap},\Delta_{i_1},\dots,\Delta_{i_k}\}$. But to compute the OPE coefficient, it requires minimization procedure introduced in section 3.2 of \cite{ElShowk:2012hu} since the number of OPE coefficient is smaller than the number of constraints equations. Let's use $N=odd$ as an illustration. In general, $\mathbf{F}_\Delta^{\Delta_\phi}$ in $\mathbb{P}^{4k+3}$ is a $2k+2$ dimensional vector, but the number of double zeros appearing in the optimal functional is $k$, so the sum rule will be
\begin{equation}\label{efmeq1}
\mathbf{F}_0^{\Delta_\phi}+c_{\Delta_{gap}}^2\mathbf{F}_{\Delta_{gap}}^{\Delta_\phi}+\sum_{i=1}^kc_{\Delta_i}^2\mathbf{F}_{\Delta_i}^{\Delta_\phi}=\mathbf{0}
\end{equation} 
It has $2k+2$ linear equation but only has $k+1$ numbers of variables $(c_{\Delta_{gap}}^2,c_{\Delta_1}^2,\dots,c_{\Delta_k}^2)$, so in general it's unsolvable! But from (\ref{sumrule}) and (\ref{ope}) we know that there only exist a unique positive solution of $\{c_{\Delta_{gap}}^2,c_{\Delta_1}^2,\dots,c_{\Delta_k}^2\}$ to equation (\ref{efmeq1}). Thus the co-plane condition for the optimal functional guarantees that one has a solution for the OPE coefficients (simplex selection condition guarantees its positivity), without the need for minimization procedure. \footnote{When applying the EFM method to extract the approximated CFT data of 2D ising model, the authors of \cite{ElShowk:2012hu} argue that, using different component of vectors equations one may get different OPE coefficient. This implies maybe the optimal functional in this case do not satisfy this co-plane condition. } In the table \ref{ope23} we give the result of approximated spectrum and OPE coefficient in $\mathbb{P}^{23}$.
\begin{table}
\centering
\begin{tabular}{|l|l|l|l|}
\hline
\multicolumn{2}{|c|}{Exact} & \multicolumn{2}{|c|}{Approximate}\\
\hline
Scaling dimension & OPE & Scaling dimension & OPE\\
\hline
4 & 3 & 4.283 & 3.577\\
\hline
6 & 1.667 & 6.829 & 1.379\\
\hline
8 & 0.441 & 9.736 & 0.161\\
\hline 
10 & 0.0814 & 13.136 & $7.167\times10^{-3}$\\
\hline
12 & 0.0123 & 17.266 & $1.052\times10^{-4}$\\
\hline
14 & 0.00157 & 22.720 & $2.784\times10^{-7}$\\
\hline
\end{tabular}
\caption{\label{ope23} Exact and approximate CFT data using optimal functional in $\mathbb{P}^{23}$ at $\Delta_\phi=3/2$}
\end{table}

\subsection{Optimal functional at $N\rightarrow \infty$} 
\begin{figure}
\centering
\includegraphics[width=0.6\textwidth]{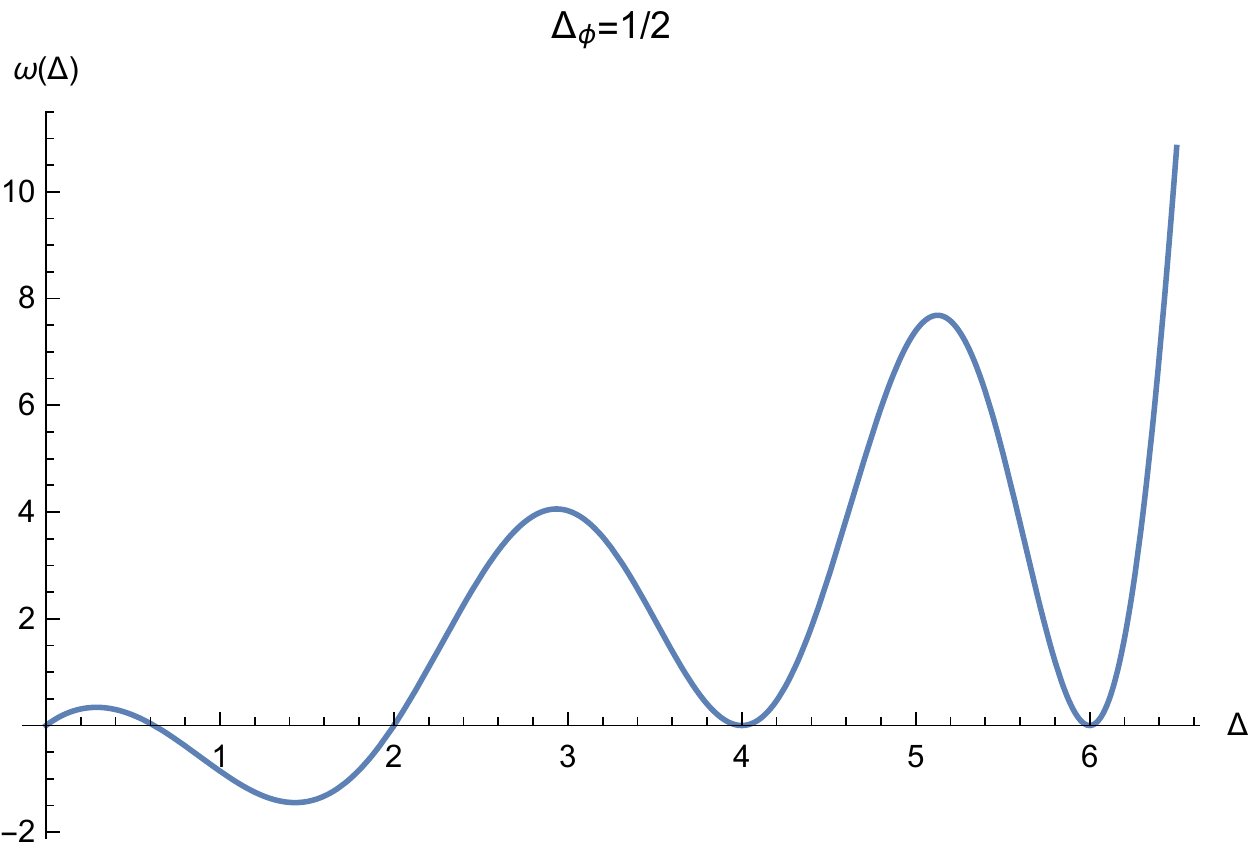}
\caption{\label{fig_extrefun} Here we present the extremal functional when $\Delta_\phi=1/2$ using eq(4.34) in \cite{Mazac:2016qev}}
\end{figure}

In this subsection, we will consider the case $N\to\infty$. In finite $\Delta_\phi$, the block vectors are very complicated, so calculation of $N\to\infty$ is impossible. But in the limit $(\Delta_\phi,\Delta_i)\to\infty$, we can compute both block vectors and the functional at leading order of $\Delta_\phi$. So to simplify the analysis we will consider the limit $(\Delta_\phi,\Delta_i)\to\infty$. The optimal functional for 1D CFT in the continuous limit was discussed in  \cite{Mazac:2016qev,Mazac:2018mdx}. There, the authors have construct the a integral functional which gives exactly the lowest scalar gap $\Delta_{gap}=2\Delta_\phi+1$. The property of this extremal functional is (see Fig.\ref{fig_extrefun} for example)
\begin{align}\label{extrezero}
&\omega(F^{\Delta_\phi}_0)=0\notag\\
&\omega(F^{\Delta_\phi}_{2\Delta_\phi+1})=0,~for~\Delta\geq2\Delta_\phi+1,~\omega(F^{\Delta_\phi}_{\Delta})\geq0\notag\\
&\omega(F^{\Delta_\phi}_{2\Delta_\phi+2n+1})=0,~\frac{\mathrm{d}}{\mathrm{d}\Delta}\omega(F^{\Delta_\phi}_{2\Delta_\phi+2n+1})=0
\end{align}
And when $\Delta_\phi\to\infty$, the functional will have a single zero at exactly $\Delta=\sqrt{2}\Delta_\phi$.

In order to compare our definition of optimal functional with that \cite{Mazac:2016qev,Mazac:2018mdx}, we have to evaluate the determinant of infinite dimensional matrix $\omega(\Delta)=\langle\mathbf{X},0,\Delta_{i_1},\Delta_{i_1+1},\dots,\Delta_{i_k},\Delta_{i_k+1},\Delta\rangle$. It turns out that at leading order in $\Delta_\phi$, we can get a simple approximate expression. Consider $F_\Delta^{\Delta_\phi(n)}(1/2)$ at large $\Delta_\phi, \Delta$ limit \footnote{Similar expression of $F_\Delta^{\Delta_\phi(n)}(1/2)$ has also been derived in appendix D of \cite{Paulos:2016fap}}
\begin{equation}
\frac{1}{{2^{2\Delta_\phi}G_\Delta(1/2)}}\frac{\mathrm{d}^n}{\mathrm{d}z^n}F_{\Delta}^{\Delta_\phi}(z)\Big|_{z=1/2}=\Delta_\phi^n\left[2^{3n/2+1}(m-\sqrt{2})^n+\mathcal{O}\left(\frac{1}{\Delta_\phi}\right)\right]
\end{equation}
where $m=\Delta/\Delta_\phi$. We see that after divided by a definite positive factor $2^{2\Delta_\phi}G_\Delta(1/2)$, the leading contribution of $F_\Delta^{\Delta_\phi(n)}(1/2)\sim(m-\sqrt{2})^n$. So the vectors $\vec{\mathbf{F}}_\Delta^{\Delta_\phi}$ is approximately points on a moment curve!
\begin{equation}
\vec{\mathbf{F}}_\Delta^{\Delta_\phi}=
\begin{pmatrix}
4\sqrt{2}\Delta_\phi(m-\sqrt{2})\\
\frac{16\sqrt{2}}{3}\Delta_\phi^3(m-\sqrt{2})^3+\mathcal{O}(\Delta_\phi^2)\\
\frac{32\sqrt{2}}{15}\Delta_\phi^5(m-\sqrt{2})^5+\mathcal{O}(\Delta_\phi^4)\\
\vdots
\end{pmatrix}
\end{equation}
In such case, our optimal functional becomes:
\begin{align}
&\langle\mathbf{X},0,\Delta_{i_1},\Delta_{i_1+1},\dots,\Delta_{i_k},\Delta_{i_k+1},\Delta\rangle\propto\langle\vec{\mathbf{F}}_0^{\Delta_\phi},\vec{\mathbf{F}}_{\Delta_{i_1}}^{\Delta_\phi},\vec{\mathbf{F}}_{\Delta_{i_1}+1}^{\Delta_\phi},\dots,\vec{\mathbf{F}}_\Delta^{\Delta_\phi}\rangle\notag\\
&\rightarrow\Bigg(m(m-\sqrt{2})(m-2\sqrt{2})\prod_{l=1}^k\Big[(m-m_{i_l})^2(m-2\sqrt{2}+m_{i_l})^2\Big]+\mathcal{O}\left(\frac{1}{\Delta_\phi}\right)\Bigg)
\end{align}
The roots structure of this functional is,
\begin{align}
&single~roots:~m=0,m=\sqrt{2},m=2\sqrt{2}\notag\\
&double~roots:~m=m_{i_l}~m=2\sqrt{2}-m_{i_l}
\end{align}
The single root at $m=\Delta/\Delta_\phi=\sqrt{2}$ is exactly also present in \cite{Mazac:2016qev}. Now the value for $m_{i_l}$ is determined from the co-plane condition (\ref{subplane}). For this, we consider the form of $\langle\mathbf{X},0,\Delta_1,\dots,\Delta_N\rangle$ at leading order in $\Delta_\phi$,
\begin{equation}
\langle\mathbf{X},0,\Delta_1,\dots,\Delta_N\rangle\propto\Bigg(\prod_{i=1}^N\Big[m_i(m_i-\sqrt{2})(m_i-2\sqrt{2})\Big]\prod_{i<j}\Big[(m_i-m_j)(m_i-2\sqrt{2}+m_j)\Big]+\mathcal{O}\left(\frac{1}{\Delta_\phi}\right)\Bigg)
\end{equation}
The co-plane condition is now written as the vanishing of
\begin{align}\label{largefun}
\langle\mathbf{X},0,\Delta_{gap},\Delta_{i_1},\dots,\Delta_{i_k},\tilde{\Delta}_1,\dots,\tilde{\Delta}_k\rangle\sim\prod_{l=1}^k\Big[m_{i_l}(m_{i_l}-\sqrt{2})(m_{i_l}-2\sqrt{2})\Big]+\mathcal{O}\left(\frac{1}{\Delta_\phi}\right)
\end{align}
for arbitrary $\{\tilde{\Delta}_1,\dots,\tilde{\Delta}_k\}$. This leads to 
\begin{equation}
m_{gap},m_{i_k}=2\sqrt{2}+\mathcal{O}(\Delta_\phi^{-1})
\end{equation}
But this deviates from the exact gap and spectrum,
\begin{equation}
m_{gap}=2+\frac{1}{\Delta_\phi},\quad m_n=2+\frac{2n+1}{\Delta_\phi}~(n\in\mathbb{Z})\,.
\end{equation}
Thus we see that at leading order in $\Delta_\phi$ the functional eq.(\ref{largefun}) will give the correct first single zero $(m=\Delta/\Delta_\phi=\sqrt{2})$, but largest single zero and all the double zeros will be shifted. This shift is an artifact of the large $\Delta_\phi$ approximation. Indeed one can compare the roots of the optimal functional in the large $\Delta_\phi$ approximation at finite $N$. We present the result for $\mathbb{P}^3$ and $\mathbb{P}^7$ at $\Delta_\phi=200.1$ in Table.\ref{largeroot}. We see that at different dimension $N$, the large $\Delta,\Delta_\phi$ approximation does not shift location of the first root, while the second one will get shifted as $N$ gets large. We see both cases first root is closed to $\sqrt{2}\Delta_\phi\approx282.984$, and the second root of approximated functional is always close to $2\sqrt{2}\Delta_\phi\approx565.968$
\begin{table}
\centering
\begin{tabular}{|c|c|c|c|c|}
\hline
&\multicolumn{2}{|c|}{Exact} &\multicolumn{2}{|c|}{Approximate}\\
\hline
&First root& Second root ($\Delta_{gap}$)&First root& Second root ($\Delta_{gap}$)\\
\hline
$\mathbb{P}^3~\&~\mathbb{P}^5$ & 282.9534 & 566.9987 & 282.8633 & 567.0891\\
\hline
$\mathbb{P}^7$ & 282.9526 & 554.1857 & 282.9521 & 566.3068 \\
\hline
\end{tabular}
\caption{\label{largeroot} The location of the first root and second root $(\Delta_{gap}$) in the exact and approximate ($\Delta,\Delta_\phi\to\infty$) optimal functional at $\Delta_\phi=200.1$}
\end{table}

\section{Extensions to 2D diagonal limit and modular bootstrap}\label{2DCFT}
There is reasonable expectation that the geometry behind the optimal functionals in 1D CFT can be found in the diagonal limit of 2D CFT and the modular bootstrap problem. For one, as discussed previously, the convex hull of scalar blocks in the diagonal limit was shown to be a cyclic polytope in~\cite{Sen:2019lec}. Similarly, as we will review shortly, the same holds for modular bootstrap for the rectangular torus~\cite{ShuHeng}. Remarkably, not only will we see that this is indeed the case, it's validity extends to the case where spins are included.  
\subsection{2D CFT in the diagonal limit $(z=\bar{z})$} 
In 2D CFT, the local opearator now will both labeled by scaling dimension $\Delta$ and spin $l$. Also, we will have two independent cross ratio $u$, $v$,
\begin{equation}
u=\frac{x_{12}^2x_{34}^2}{x_{13}^2x_{24}^2},\quad v=\frac{x_{14}^2x_{23}^2}{x_{13}^2x_{24}^2}
\end{equation}
So the 4-pt indentical scalar function $\langle\phi(x_1)\phi(x_2)\phi(x_3)\phi(x_4)\rangle$ has this form
\begin{equation}
\langle\phi(x_1)\phi(x_2)\phi(x_3)\phi(x_4)\rangle=\frac{\mathcal{G}(z,\bar{z})}{|x_{12}|^{2\Delta_\phi}|x_{34}|^{2\Delta_\phi}}
\end{equation}
where $u=z\bar{z},v=(1-z)(1-\bar{z})$. If we constraint it on $z=\bar{z}$ line, the (\ref{schannel}) and (\ref{crossing}) conditions now are,
\begin{align}\label{2dcon}
&Unitarity:~\mathcal{G}(z,z)=\sum_{O\in\phi\times\phi}c_{\Delta_O,l}^2G_{\Delta_O,l}(z)\notag\\
&Crossing:~\mathcal{G}(z,z)=\left(\frac{z}{1-z}\right)^{2\Delta_\phi}\mathcal{G}(1-z,1-z)
\end{align}
where $G_{\Delta,l}(z,z)$ is the $d=2$ conformal blocks on $z=\bar{z}$ line,
\begin{equation}
G_{\Delta,l}(z,z)=2k_{\Delta+l}(z)k_{\Delta-l}(z),\quad k_\beta(z)=z^{\beta/2}{}_2F_1(\beta/2,\beta/2,\beta,z)\notag\\
\end{equation}
Expand (\ref{2dcon}) at $z=1/2$ we also get the convex hull condition,
\begin{equation}
\vec{\mathcal{G}}=\sum_\Delta c_{\Delta,0}^2\vec{\mathbf{G}}_{\Delta,0}
+\sum_\Delta c_{\Delta,2}^2\vec{\mathbf{G}}_{\Delta,2}+\dots
\end{equation}
Notice that, when setting $z=\bar{z}$, the crossing relation in (\ref{2dcon}) is the same as 1D case, so the crossing plane $\mathbf{X}$ in unchanged.  We start with the most simplest case, assuming that there are only scalar primaries exchanged in OPE $\phi\times\phi$. So the problem is simplified to just having scalar vertexs in the convex hull, $\vec{\mathcal{G}}=\sum_\Delta c_{\Delta,0}^2\vec{\mathbf{G}}_{\Delta,0}$. The positivity of $\{\vec{\mathbf{G}}_{\Delta_1,0},\vec{\mathbf{G}}_{\Delta_2,0},\dots,\vec{\mathbf{G}}_{\Delta_N,0}\}$ has been justified in Fig.4(b) of \cite{Sen:2019lec}. Next we briefly summerize the result,
\begin{itemize}
\item For large external scaling dimension case $\Delta_\phi>0.1$, we found that the functional in $\mathbb{P}^{2N+1}$ is similar to 1D case, namely,
\begin{align}\label{2dfun}
&\omega(\Delta)=\langle\mathbf{X},0,\Delta_{i_1},\Delta_{i_1+1},\dots,\Delta_{i_k},\Delta_{i_k+1},\Delta\rangle,~when~N~is~odd,~N=2k+1\notag\\
&\omega(\Delta)=\langle\mathbf{X},0,\Delta_{i_1},\Delta_{i_1+1},\dots,\Delta_{i_k},\Delta_{i_k+1},\infty,\Delta\rangle,~when~N~is~even,~N=2k+2
\end{align}
Using condition (\ref{subplane}) and (\ref{select}) we can determine the remaining vertexs $\{\vec{\mathbf{G}}_{\Delta_{i_1}},\vec{\mathbf{G}}_{\Delta_{i_2}},\dots,\vec{\mathbf{G}}_{\Delta_{i_k}}\}$
\item For small external scaling dimension case $\Delta_\phi<0.1$, the conjecture no longer holds true. The change in optimal functional indicates the presence of a kink, and our analysis implies that the position of the kink can be stated as a geometric configuration as well~\cite{ToBePublish}
\end{itemize}

\begin{figure}
\centering
\includegraphics[width=0.6\textwidth]{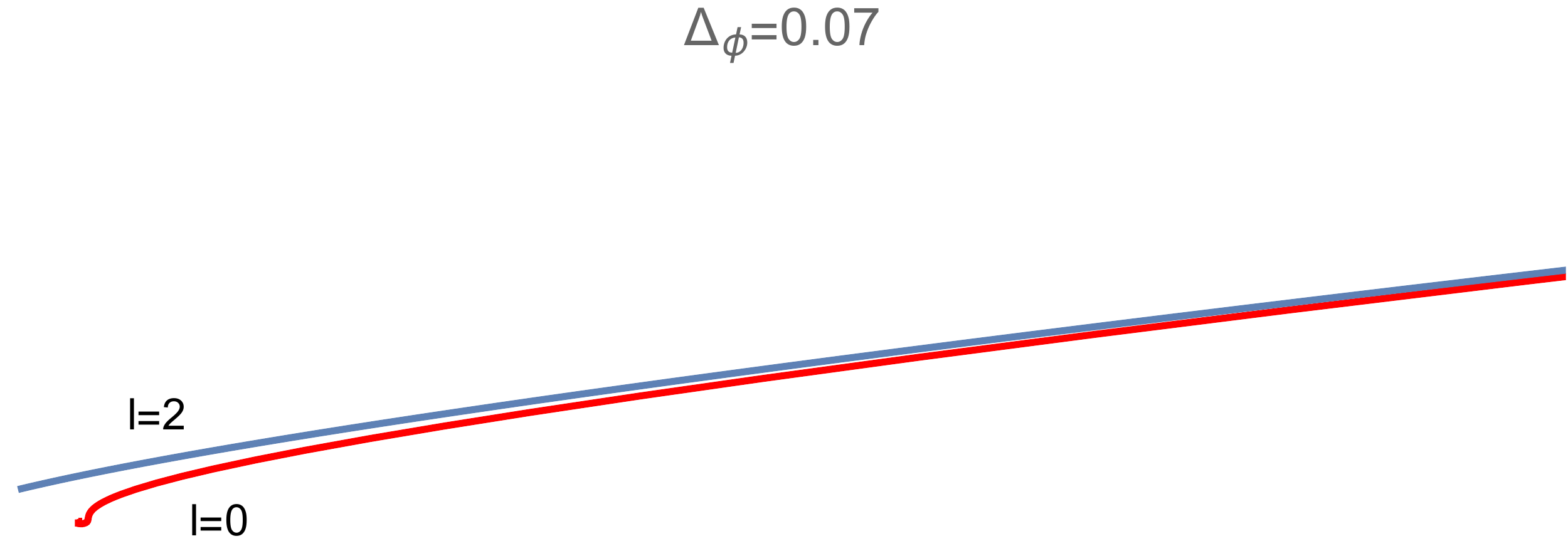}
\caption{\label{fig_p5spin} Here we present the 2d geometry in the space orthorognoal to $(\mathbf{X},\vec{\mathbf{G}}_{0,0})$. Scalar section (red curve) is starting from $\Delta=0$ while spin-2 (blue curve) is from $\Delta=2$ (unitary bound). From the curve we see that the convex hull of spin-2 blocks is not inside the convex hull of scalar blocks. }
\end{figure}

Next we consider the real 2D CFT with all the higher spins. When including different spin sectors, the convex hull of unitary polytope $\mathbf{U}_N$ now is the Minkowski sum of polytopes of different spin sector. Here we mention two important geometrical features of the individual spin sector:
\begin{itemize}
\item
For every individual spin sector, the positivity has been discussed in \cite{Sen:2019lec}. From Fig.4(b) we see that if requiring cyclicity of polytope in $\mathbb{P}^7$ in 2d, the sufficient condition is $g_1,g_2,\dots,g_7>0$. We can read from this plot that when $\Delta>0.13$, $\mathbb{P}^7$ polytope is a \emph{cyclic} polytope.
\item Using (2.3) in \cite{Sen:2019lec},
\begin{equation}
G_{d,\Delta,l}^{approx}(z,z)=G_{d,\Delta,0}^{approx}(z,z)\frac{\Gamma(d+l+2)\Gamma(d/2-1)}{\Gamma(d/2+l-1)\Gamma(d-2)}[1+\mathcal{O}(\Delta^{-1})]
\end{equation}
we see that when $\Delta\gg\Delta_\phi$, the higher spin blocks just differ by constant factor with the scalar! Geometrically, this tell us that the convex hull of higher spin sectors have the same convex hull as scalar one when $\Delta\gg\Delta_\phi$.
\end{itemize}
Despite of these nice features, the Minkowski sum of \emph{cyclic polytope} is in general not a \emph{cyclic polytope}. Of course if all the higher spin sector of polytope is inside the convex hull of scalar sector, then the unitary polytope $\mathbf{U}_N$ is still \emph{cyclic polytope}. But already in $\mathbb{P}^5$ we see that this is not true (Fig.\ref{fig_p5spin}).

But even if we don't know the boundary of the unitary polytope $\mathbf{U}_N$, we can still construct a linear functional to get a valid scalar gap. And to find the conditions for suitable functional, we write down the sum rule in s-channel,
\begin{equation}
F_{0,0}^{\Delta_\phi}(z)+\sum_{\Delta,l}c_{\Delta,l}^2F_{\Delta,l}^{\Delta_\phi}(z)=0
\end{equation}
where $F_{\Delta,l}^{\Delta_\phi}(z)=z^{-2\Delta_\phi}G_{\Delta,l}(z,z)-(z\to1-z)$. Unitarity will imply that the scaling dimension of primary opearators exachaged in OPE must satisfy,
\begin{equation}
\Delta_{min}(l)=l,\quad \mathrm{if}~l=2,4,6,\dots,\quad\mathrm{and}\quad\Delta_{min}(0)=0
\end{equation}
So if the functional $\omega(\Delta,l)$ satisfies the following condition,
\begin{equation}
\mathrm{Sign}[\omega(0,0)]=\mathrm{Sign}[\omega(0,\infty)]=\mathrm{Sign}[\omega(\Delta,l)]~for~all~\Delta\ge l
\end{equation}
Then the largest single zero $\Delta^*$ of linear functional acting of scalar $\omega(\Delta,0)$ will provide a valid scalar gap. We will consider this kind of fucntional $\omega(\Delta,l)=\vec{\alpha}\cdot\vec{\mathbf{F}}_{\Delta,l}^{\Delta_\phi}$, and the equivalence between functional $\omega(\Delta,l)=\langle\mathbf{X},0,\vec{\mathbf{G}}_{\Delta_1,l_1},\vec{\mathbf{G}}_{\Delta_2,l_2},\dots,\vec{\mathbf{G}}_{\Delta,l}\rangle$ and $\vec{\alpha}\cdot\vec{\mathbf{F}}^{\Delta_\phi}_{\Delta,l}$ still holds in this case using the proof in appendix.\ref{prof}. The result of optimal functional in $\mathbb{P}^3$ to $\mathbb{P}^9$ are summerized in the table.\ref{2dspinfun}
\begin{table}
\centering
\begin{tabular}{|l|l|l|l|}
\hline
& Small $\Delta_\phi$ region & Large $\Delta_\phi$ region & Transition point\\
\hline
$\mathbb{P}^3$& $\langle\mathbf{X},\vec{\mathbf{G}}_{0,0},\vec{\mathbf{G}}_{\Delta,l}\rangle$ &$\langle\mathbf{X},\vec{\mathbf{G}}_{0,0},\vec{\mathbf{G}}_{\Delta,l}\rangle$ & No transition \\
\hline
$\mathbb{P}^5$& $\langle\mathbf{X},\vec{\mathbf{G}}_{0,0},\vec{\mathbf{G}}_{2,2},\vec{\mathbf{G}}_{\Delta,l}\rangle$ & $\langle\mathbf{X},\vec{\mathbf{G}}_{0,0},\infty,\vec{\mathbf{G}}_{\Delta,l}\rangle$ & 0.088 \\
\hline
$\mathbb{P}^7$& $\langle\mathbf{X},\vec{\mathbf{G}}_{0,0},\vec{\mathbf{G}}_{2,2},\infty,\vec{\mathbf{G}}_{\Delta,l}\rangle$ & $\langle\mathbf{X},\vec{\mathbf{G}}_{0,0},\vec{\mathbf{G}}_{i,4},\vec{\mathbf{G}}_{i+1,4},\vec{\mathbf{G}}_{\Delta,l}\rangle$ & 0.089\\
\hline
$\mathbb{P}^9$& $\langle\mathbf{X},\vec{\mathbf{G}}_{0,0},\vec{\mathbf{G}}_{2,2},\vec{\mathbf{G}}_{i,6},\vec{\mathbf{G}}_{i+1,6},\vec{\mathbf{G}}_{\Delta,l}\rangle$ & $\langle\mathbf{X},\vec{\mathbf{G}}_{0,0},\vec{\mathbf{G}}_{i,4},\vec{\mathbf{G}}_{i+1,4},\infty,\vec{\mathbf{G}}_{\Delta,l}\rangle$ & 0.09\\
\hline
\end{tabular}
\caption{\label{2dspinfun} List of optimal functional $\omega(\Delta,l)$ from $\mathbb{P}^3$ to $\mathbb{P}^9$ in 2d CFT when $z=\bar{z}$}
\end{table}
In large $\Delta_\phi$ region, we see the functional is very similar to the case in 1D CFT, excepct the double zero position is not at the scalar section (Like $\mathbb{P}^7$, the double zero of optimal functional is at $l=4$ sector). And furthermore, combined with the co-plane condition (\ref{subplane}) and simplex selection condition (\ref{select}), the remaining solution is not unique! This means that are more than one degenerate 1-simplexs. So in this case, we need a addition condition to select out the one corresponding to optimal. We will take $\Delta_\phi=1/8,~\mathbb{P}^7$ as a concrete example.

\begin{figure}
\centering
\begin{minipage}[t]{0.48\textwidth}
\centering
\includegraphics[width=6cm]{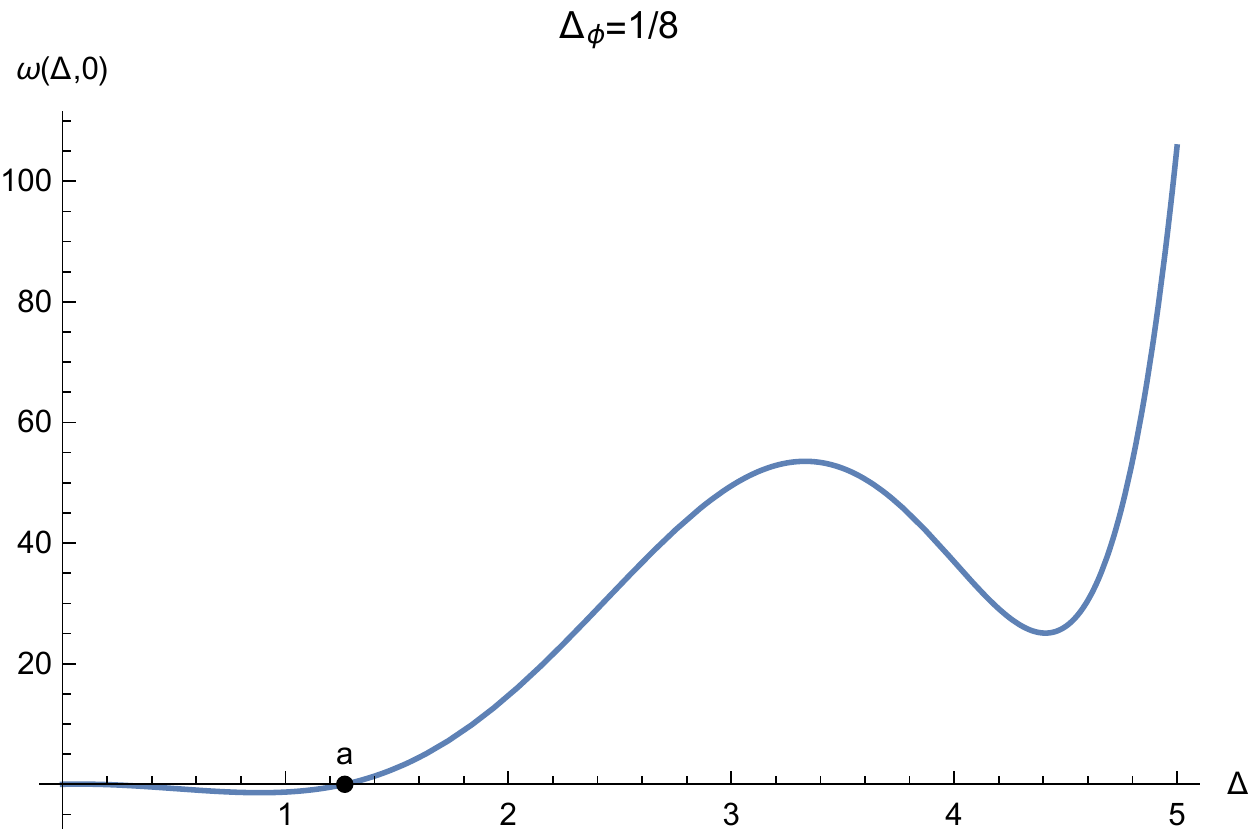}
\end{minipage}
\begin{minipage}[t]{0.48\textwidth}
\centering
\includegraphics[width=6cm]{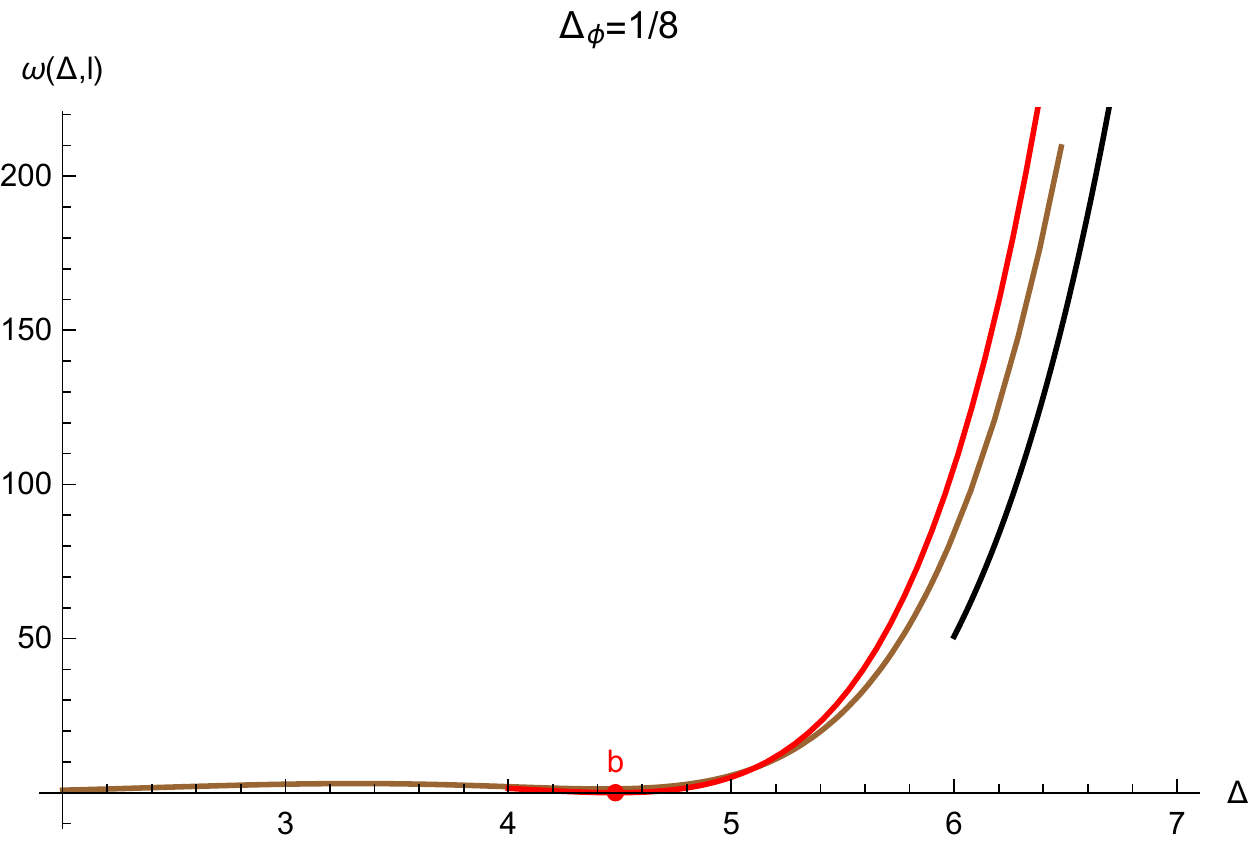}
\end{minipage}
\caption{Here we present 2d optimal functional when $\Delta_\phi=1/8$ in $\mathbb{P}^7$, which is $\omega(\Delta,l)=\langle\mathbf{X},\vec{\mathbf{G}}_{0,0},\vec{\mathbf{G}}_{b,4},\vec{\mathbf{G}}_{b+1,4},\vec{\mathbf{G}}_{\Delta,l}\rangle$. LHS: Plot of functional at when $l=0$, where $a\approx1.266$. RHS: Plot of functional when $\Delta>l$ at $l=2$(brown), $l=4$(red), $l=6$(black).  $b\approx4.481$ is the double zero of $\omega(\Delta,4)$.} 
\label{fig_spinfun}
\end{figure}

In $\mathbb{P}^7$ including spin, the condition (\ref{subplane}) and (\ref{select}) get slightly modified,
\begin{align}\label{2dcoplane}
&co-plane:~For~abitrary~\vec{\mathbf{v}},~\langle\mathbf{X},\vec{\mathbf{G}}_{0,0},\vec{\mathbf{G}}_{a,0},\vec{\mathbf{G}}_{b,l},\vec{\mathbf{v}}\rangle=0\notag\\
&simplex~selection:~For~abitrary~\vec{\mathbf{v}}_1,\vec{\mathbf{v}}_2,\langle\mathbf{X},\vec{\mathbf{G}}_{a,0},\vec{\mathbf{G}}_{b,l},\vec{\mathbf{v}}_1,\vec{\mathbf{v}}_2\rangle,-\langle\mathbf{X},\vec{\mathbf{G}}_{0,0},\vec{\mathbf{G}}_{b,l},\vec{\mathbf{v}}_1,\vec{\mathbf{v}}_2\rangle,\langle\mathbf{X},\vec{\mathbf{G}}_{0,0},\vec{\mathbf{G}}_{a,0},\vec{\mathbf{v}}_1,\vec{\mathbf{v}}_2\rangle\notag\\
&\qquad\qquad\qquad same~sign
\end{align}
For each spin $l$, we found that there are one degenerate 1-simplex, the solutions are in Table.\ref{table_sol}. $(\vec{\mathbf{G}}_{\Delta_a,0},\vec{\mathbf{G}}_{\Delta_b,l})$ are the vertexs of these 1-simplexs. Two remarks of this table which will be important later,
\begin{table}
\centering
\begin{tabular}{|c|c|}
\hline
$\Delta_a$ & $(\Delta_b,l)$\\
\hline
1.2646 & (4.2862,0)\\
\hline
1.2650 & (4.3172,2)\\
\hline
1.2662 & (4.4806,4)\\
\hline
1.2711 & (5.3463,6)\\
\hline
1.2762 & (7.1251,8)\\
\hline
1.2787 & (9.0744,10)\\
\hline
\vdots & \vdots\\
\hline
\end{tabular}
\caption{\label{table_sol} List of all the possible solutions satisfying (\ref{2dcoplane})}
\end{table}
\begin{itemize}
\item As spin $l$ increases, $\Delta_a$ is also increasing
\item When $l\ge6$, $\Delta_b$ is below the unitarity bound i.e $(\Delta_b<l)$
\end{itemize}
Accroding to the discussion in seciton.\ref{explainsub}, finding such a 1-simplex is equivalent to finding a solution with positive $c_{a,0},c_{b,l}>0$ to the vector equation in $\mathbb{P}^7$,
\begin{equation}
\vec{\mathbf{F}}_{\Delta_0,0}+c_{a,0}\vec{\mathbf{F}}_{\Delta_a,0}+c_{b,l}\vec{\mathbf{F}}_{\Delta_b,l}=\vec{0}
\end{equation}
Because of unitarity, the solutions with $l\ge6$ is not a valid solution to this vector equation since its scaling dimension is below than the unitaritiy bound. So we're left with three solutions with spin $l=0,2,4$. And if the simplex corresponds the optimal functional, its the smallest vertex is exactly the scalar gap $\vec{\mathbf{G}}_{\Delta_{gap},0}$. It turns out the only consistent choice is $\Delta_a=1.2662,(\Delta_b,l)=(4.4806,4)$, which have the largest scaling dimension of smallest vertexs in the simplex.\footnote{If choose the $\Delta_a=1.2646,(\Delta_b,l)=(4.2862,0)$, then the scalar gap is assumed to be $\Delta_{gap}=\Delta_a$. But the scaling dimension of the smallest vertex  of simplex in $l=2$ has bigger than $\Delta_{gap}$, so it's contradictory to the assumption. } We also show the plot of optimal functional in Fig\ref{fig_spinfun}.

\subsection{Spinless modular bootstrap} 
Let us begin with the torus partition function in 2D CFT with central charge $c$, given by
\eq
Z(q,\bar{q})\equiv {\rm Tr} q^{L_0-\frac{c}{24}}\bar{q}^{\bar{L}_0-\frac{c}{24}}\,,
\eqe
where $q=e^{i 2\pi\tau}$, and $\tau$ the torus modulus. Modular invariance on the torus tells us that:
\eq\label{modular}
Z(\tau,\bar{\tau})=Z(-1/\tau,-1/\bar{\tau})\,.
\eqe
The torus partition function can be expanded on Virasoro characters, which sums up the contribution to the partition function from the descendants of each Virasoro primary. With $c>1$, it is given by 
\eq
\chi_0(q)=q^{-\frac{(c-1)}{24}}\frac{1-q}{\eta(q)},\quad \chi_h(q)=q^{h-\frac{(c-1)}{24}}\frac{1}{\eta(q)}\quad \forall h>0\,,
\eqe
where $\eta(q)$ is the Dedekind eta function, satisfying $\eta(-\frac{1}{\tau})=\sqrt{-i\tau}\eta(\tau)$. The partition function then admits the following expansion in terms of these characters
\eq
Z_{q,\bar{q}}=\sum_{h,\bar{h}}n_{h,\bar{h}}\chi_h(q)\chi_{\bar{h}}(q)
\eqe
where $n_{h,\bar{h}}$ is the degeneracy number, i.e. it is a positive integer with $n_{0,\bar{0}}=1$.

Let us consider the case where $\tau=i\beta$, i.e. a rectangular torus. In such case $q=\bar{q}$, and the dependence on $h,\bar{h}$ comes in the combination $\Delta=h+\bar{h}$, i.e. it only depends on the scaling dimension and not spin. In this case the partition function becomes 
\eq\label{character}
Z(q)=q^{-\frac{(c{-}1)}{12}}\frac{(1{-}q)^2}{\eta^2(q)}{+}\frac{q^{-\frac{(c{-}1)}{12}}}{\eta^2(q)}\sum_{\Delta>0}n_{\Delta}q^{\Delta}\,.
\eqe
It's conveninent to define the reduce partition function $\hat{Z}(\beta)$ using $\beta=-i\tau$ variable,
\begin{equation}
\hat{Z}(\beta)=|\eta(i\beta)|^2|i\beta|^{1/2}Z(\beta)
\end{equation}
Combining modular invariance (\ref{modular}) and proberty of Dedekind eta function $\eta\left(-\frac{1}{\tau}\right)=\sqrt{-i\tau}\eta(\tau)$, we get
\begin{equation}
\hat{Z}(\beta)=\hat{Z}\left(\frac{1}{\beta}\right)
\end{equation}
And accroding to (\ref{character}), the reduce partition function $\hat{Z}(\beta)$ can be expanded into
\begin{equation}\label{reducep}
\hat{Z}(\beta)=G_0(\beta)+\sum_{\Delta>0}n_\Delta G_\Delta(\beta)
\end{equation}
where the blocks are
\begin{equation}
G_\Delta(\beta)=\beta^{1/2}\mathrm{exp}\left[-2\pi\beta(\Delta-\frac{c-1}{12})\right],\quad G_0(\Delta)=\beta^{1/2}e^{2\pi\beta\frac{c-1}{12}}(1-e^{-2\pi\beta})^2
\end{equation}
We define linear opearator $\mathcal{F}_k$,
\begin{equation}
\mathcal{F}_k=\frac{1}{2k!}\left[\frac{1}{2}(1+\beta)^2\partial_\beta\right]^k\frac{1+\beta}{2\sqrt{\beta}}\Big|_{\beta=1}
\end{equation}
Then we act $\mathcal{F}_k$ to both side of (\ref{reducep}), packaging different equation labeled by $k$, we finally get a vector equation,
\begin{equation}
\vec{Z}=\vec{\mathbf{G}}_0+\sum_\Delta n_\Delta\vec{\mathbf{G}}_\Delta\Rightarrow
\begin{pmatrix}
Z_0\\
Z_1\\
\vdots\\
Z_k
\end{pmatrix}
=
\begin{pmatrix}
\mathcal{F}_0(0)\\
\mathcal{F}_1(0)\\
\vdots\\
\mathcal{F}_k(0)
\end{pmatrix}
+\sum_\Delta n_\Delta
\begin{pmatrix}
\mathcal{F}_0(\Delta)\\
\mathcal{F}_1(\Delta)\\
\vdots\\
\mathcal{F}_k(\Delta)
\end{pmatrix}
\end{equation}
Using modular invariance $\hat{Z}(\beta)=\hat{Z}(1/\beta)$, the even component $Z_{2k}$ is automatically vanished, while we have no constraint on the odd one $Z_{2k+1}$. So the crossing plane $\mathbf{X}$ is simply,
\begin{equation}
\mathbf{X}=
\begin{pmatrix}
1&0&0&\cdots&0\\
0&0&0&\cdots&0\\
0&1&0&\cdots&0\\
0&0&0&\cdots&0\\
0&0&1&\cdots&0\\
\vdots&\vdots&\vdots&\vdots&\vdots\\
0&0&0&\cdots&1
\end{pmatrix}
\end{equation}
And the vector components now are \cite{Afkhami-Jeddi:2019zci},
\begin{align}
&\mathcal{F}_k(0)=e^{1/6\pi(c-25)}\Big[L_k(4\pi x_0)-2e^{-2\pi}L_k(4\pi(x_0+1))+e^{-4\pi}L_k(4\pi(x_0+2))\Big]\notag\\
&\mathcal{F}_k(\Delta)=e^{1/6\pi(c-12\Delta-1)}L_k(4\pi x)
\end{align} 
where $x=\Delta-\frac{c-1}{12},x_0=-\frac{c-1}{12}$, $L_k(x)$ is Laguerre polynomials. So again, we can transform the modular bootstrap problem into the fixed plane $\mathbf{X}$ intersecting with a convex polytope! Next we'll show that this polytope is exactly a \emph{cyclic polytope}.

Because $L_k(4\pi x)$ is degree $k$ polynomial in $\Delta$, after diving by a overall positive factor $e^{1/6\pi(c-12\Delta-1)}$, block vectors $\vec{\mathbf{G}}_\Delta$ are related by moment curve just by a constant matrix $M(c)$ which depends on central charge. For example in $\mathbb{P}^3$,
\eqa
&&M(c)\cdot\vec{\mathbf{G}}_\Delta=\nonumber\\
&&\begin{pmatrix}
 1 & 0 & 0 & 0 \\
 \frac{\pi  (c{-}1){+}3}{12 \pi } & {-}\frac{1}{4 \pi } & 0 & 0 \\
 \frac{\pi  (c{-}1) (\pi  (c{-}1){+}6){+}18}{144 \pi ^2} & \frac{\pi{-}\pi 
   c{-}6}{24 \pi ^2} & \frac{1}{8 \pi ^2} & 0 \\
 \frac{\pi  (c{-}1) (\pi  (c{-}1) (\pi  (c{-}1){+}9){+}54){+}162}{1728 \pi
   ^3} & {-}\frac{\pi  (c{-}1) (\pi  (c{-}1){+}12){+}54}{192 \pi ^3} &
   \frac{\pi  (c{-}1){+}9}{32 \pi ^3} & {-}\frac{3}{32 \pi ^3} \\
\end{pmatrix}
\begin{pmatrix}
L_0(4\pi x)\\
L_1(4\pi x)\\
L_2(4\pi x)\\
L_3(4\pi x)
\end{pmatrix}
=
\begin{pmatrix}
1\\
\Delta\\
\Delta^2\\
\Delta^3
\end{pmatrix}\nonumber\\
\eqae
For moment curve, the ordered determinant is definite positive, which implies the definite positivity of block vectors $\vec{\mathbf{G}}_\Delta$
\begin{equation}
\langle\vec{\mathbf{G}}_{\Delta_1},\vec{\mathbf{G}}_{\Delta_1},\dots,\vec{\mathbf{G}}_{\Delta_n}\rangle=\mathrm{det}[M(c)]\prod_{i<j}(\Delta_j-\Delta_i)>0,\quad for~\Delta_1<\Delta_2<\dots<\Delta_n
\end{equation}
 Next we move to the optimal functional at $\mathbb{P}^{2n+1}$. When central charge $c>1$, we found that the functional is exactly the same as 1D and 2D CFT when $\Delta_\phi$ becomes large (\ref{2dfun}). Then after imposing condition (\ref{subplane}) and (\ref{select}), we are left with finite number of solutions, which is exactly the case in 2D CFT that the degenerate simplex is not unique.  And to select out the optimal one, we just choose the simplex with the largest scaling dimension of the smallest vertexs. Still take a example in $\mathbb{P}^7$, when setting $c=12$, we found 6 sets of solution to equation (\ref{subplane7}),
\begin{equation}
(\Delta_a,\Delta_b)\approx\{(0.933,1.470),(0.954,1.115),(0.999,3.74),(0.100,2.015),(1.044,1.509),(2.231,3.804)\}
\end{equation}
Then using simplex selection condition (\ref{select7}) we are left with three 1-simplexs,
\begin{equation}
(\Delta_a,\Delta_b)\approx\{(0.954,1.115),(0.999,3.74),(2.231,3.804)\}
\end{equation}
So the optimal one will be $(\Delta_{gap},\Delta_{i})\approx(2.231,3.804)$. 
\section{The ``Theory Space"}\label{sec:TheorySpace}
The bootstrap equations not only impose bounds on the gap of the first operator, it imposes global constraints on the entire spectrum. Recall that from the view point of a cyclic polytope intersecting the crossing plane, at each $N$ the constraint of the crossing equation is encapsulated by the statement that there are \textit{at least} $N{+}1$ operators that forms an $N$-dimensional simplex that encloses the origin. Let us refer to this simplex as $\tilde{\mathbf{U}}_N$, as it is the subspace of $\mathbf{U}_N$ that is relevant for the intersection. Importantly, the vertices that form $\tilde{\mathbf{U}}_N$ need not be the first $N{+}1$ lowest lying operators, and hence the constraint is global. We will analyze the global constraint from two directions:  (1) for fixed $N$, the bounds on the entire spectrum depending on the dimensions of the low lying operators, and  (2) assuming that $\mathbf{U}_N$ is constructed from  the lowest dimension operators, the modification to their bounds as we increase $N$.

These bounds can be incapsulated by the recursive construction discussed in~\cite{Arkani-Hamed:2018ign}. There, it was observed that the vertex $\infty$, being the only universal vertex in the subspace project through $(\mathbf{X},0)$, played an outsize role in determining the validity of the spectrum. For fixed $N$, consider all sets of operators $\{\Delta_1,\Delta_2,\cdots, \Delta_{N}\}$ such that when combined with $\infty$, the corresponding simplex incloses the origin. Denote this set as $\mathbf{S}_N$, and the claim is that it can be built recursively in $N$. To see this, for each element in $\mathbf{S}_N$, we denote the space of all possible $\Delta_{N{+}1}$, for which when combined with the element in $\mathbf{S}_N$ the resulting simplex incloses zero, as $\mathbf{T}_N$. Then the union $\mathbf{S}_N\cup \mathbf{T}_N$, gives all possible sets of operators $\{\Delta_1,\Delta_2,\cdots, \Delta_{N}, \Delta_{N{+}1}\}$, such that the origin is inclosed, i.e. $\mathbf{S}_N\cup \mathbf{T}_N=\tilde{\mathbf{U}}_N$. Now let's move to $N{+}1$, and consider $\mathbf{S}_{N+1}$. The space of all possible $\{\Delta_1,\cdots, \Delta_{N{+}1}\}$, such that:
\eqa\label{eq:n+1}
&&\langle \mathbf{X},0,\Delta_1,\Delta_2,\cdots, \Delta_{N}\rangle>0\nonumber\\
&&\langle \mathbf{X},0,\Delta_1,\Delta_2,\cdots, \Delta_{N{-}1},\infty\rangle<0,\quad\langle \mathbf{X},0,\Delta_2,\cdots, \Delta_{N{+}1},\infty\rangle>0,\cdots,\quad{\rm e.t.c.}\,. 
\eqae  
Now, the conditions in the second line are the ones that involve $\infty$, and as discussed previously this projects the geometry down to $\mathbb{P}^{2N{-}1}$. Furthermore, these constraint are precisely the complete constraint for the set of $N$ operators inclosing the origin in ${N{-}1}$-dimensions, i.e. they correspond to $\tilde{\mathbf{U}}_N$ ! Thus the space for $\mathbf{S}_{N{+}1}$ is given by $\tilde{\mathbf{U}}_N=\mathbf{S}_n\cup \mathbf{T}_n$ \textit{plus} the extra constraint in the first line of eq.(\ref{eq:n+1}). In other words, not all elements in $\mathbf{S}_N$ will survive when the $N{+}1$-dimensional geometry is considered.

On the other hand, given a set of $N$ ordered operators $\{\Delta_i\}$, we can devise a functional to bound $\Delta_{N{+}1}$, if it satisfies 
\eq
sign\left[\frac{\omega\left(F^{\Delta_\phi}_{\Delta_1}\right)}{\omega\left(F_{0}\right)}\right]=sign\left[\frac{\omega\left(F^{\Delta_\phi}_{\Delta_2}\right)}{\omega\left(F_{0}\right)}\right]=\cdots=sign\left[\frac{\omega\left(F^{\Delta_\phi}_{\Delta_N}\right)}{\omega\left(F_{0}\right)}\right]= sign\left[\frac{\omega\left(F^{\Delta_\phi}_{\infty}\right)}{\omega\left(F_{0}\right)}\right]\,,
\eqe
where once again, the bound for $\Delta_{N{+}1}$ is given by the largest single root of the functional $\omega\left(F^{\Delta_\phi}_{\Delta}\right)$. In the following, we will demonstrate that the optimal functionals derived numerically can be analytically reproduced by considering the projection of $\mathbf{S}_{N}$ onto the space of lowest dimension operators. For the leading $N$ operators, by definition $\mathbf{S}_{N}$ defines the region in which no other operators (except $\infty$) are needed for consistency. Outside of this region, one can find various distinct subregions where only a subset of operators combine with $\Delta_{N{+}1}$ to form $\mathbf{S}_{N}$, and the optimal functional exactly corresponds to the simplex constraint on $\Delta_{N{+}1}$! For the boundaries of allowed  $(\Delta_1,\Delta_2,\cdots,\Delta_{N{-}1})$, one simply project $\mathbf{S}_{N}$ along $\Delta_N$ onto the subspace $(\Delta_1,\Delta_2,\cdots,\Delta_{N{-}1})$. The bound for $\Delta_N$ can similarly be read off directly.

\subsection{Theory space in $\mathbb{P}^5$} 
From $N=1$, the condition is simply,
\begin{equation}
\langle\mathbf{X},0,\Delta\rangle,\quad-\langle\mathbf{X},0,\infty\rangle~~\mathrm{same~sign}
\end{equation}
Solve it we get $\Delta_-<\Delta<\Delta^+$, where $\Delta_-$ and $\Delta_+$ is the first and second root of function $\langle\mathbf{X},0,\Delta\rangle$.  So it means that there must be one operator between $\Delta_-<\Delta<\Delta_+$ such that with $\infty$ we have a line that encloses the origin. In other words $\mathbf{S}_1$ consists of all operators in the range $\Delta_-<\Delta<\Delta_+$. 

\begin{figure}
\centering
\includegraphics[scale=0.6]{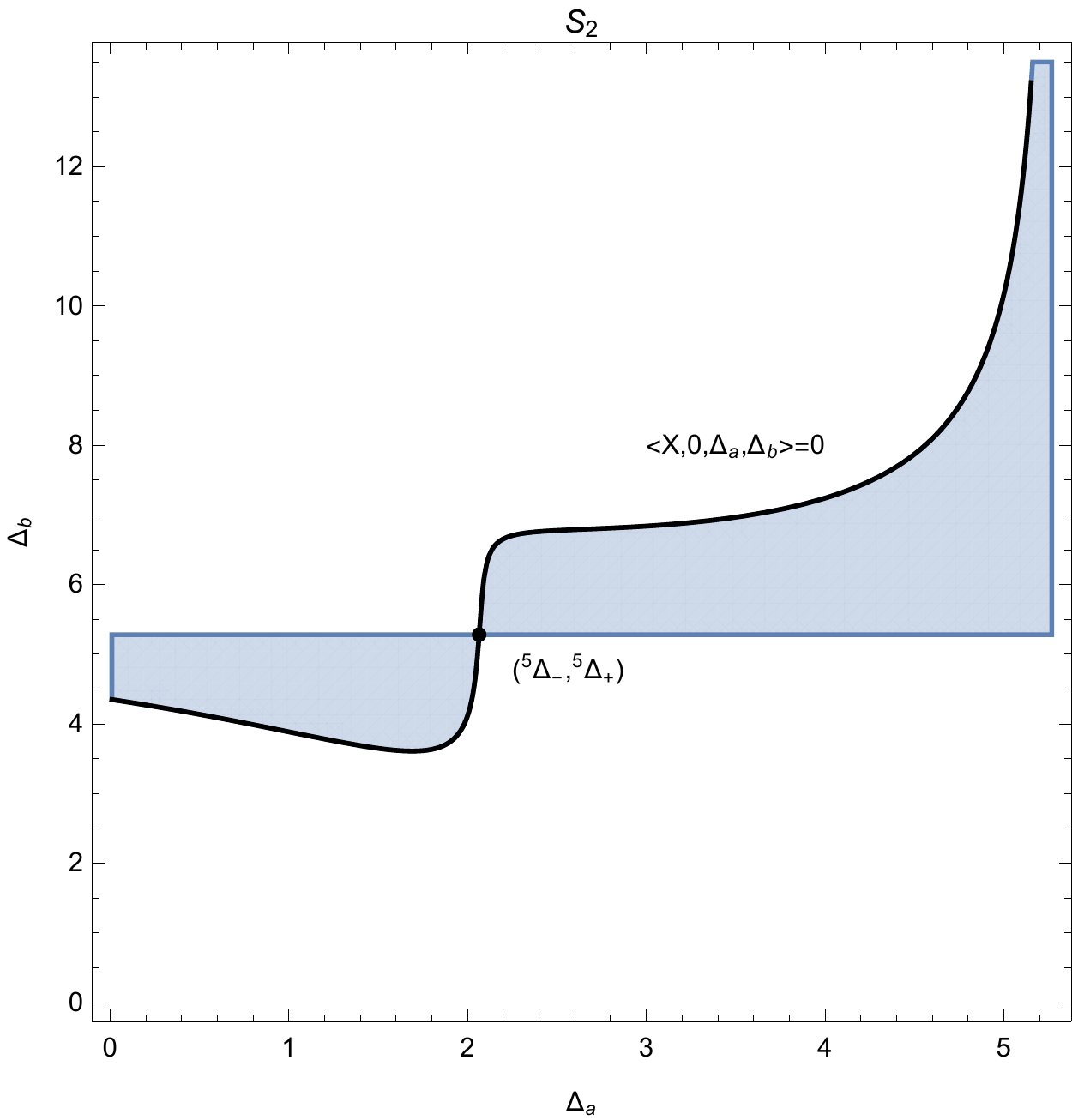}
\caption{\label{S2Fig}. The region $\mathbf{S}_2$ defined by eq.(\ref{S2Sign}). The black curve is given by $\langle\mathbf{X},0,\Delta_a,\Delta_b\rangle=0$. And when $\Delta_a<{}^5\Delta_-$, this curve is the second largest root of $\langle\mathbf{X},0,\Delta_a,\Delta\rangle$. When $\Delta_a<{}^5\Delta_-$, this curve is given by the largest root of $\langle\mathbf{X},0,\Delta_a,\Delta\rangle$}
\end{figure}

Now lets consider the theory space for $\mathbb{P}^5$, i.e. for $N=2$, and $\Delta_\phi=3/2$. Here we focus on $\mathbf{S}_2$, which corresponds to space of $\Delta_a<\Delta_b$ such that,
\eq\label{S2Sign}
\langle \mathbf{X},0,\Delta_a,\Delta_b\rangle,\quad \langle \mathbf{X},0,\Delta_b,\infty\rangle,\quad \langle \mathbf{X},0,\infty,\Delta_a\rangle\quad {\rm same \,sign} 
\eqe
By the fact that $\langle \mathbf{X},0,\Delta_b,\infty\rangle$ and $\langle \mathbf{X},0,\Delta_a,\infty\rangle$ having opposite signs means that this is allowed only if $(\Delta_a,\Delta_b)$ is lying on opposite sign regions separated by the  roots of $\langle \mathbf{X},0,\infty,\Delta\rangle=0$ denoted as $\Delta_+$ and $\Delta_-$. Thus we see that $\mathbf{S}_2$ consists of two region of interest:
\eq
 \;\;(\Delta_a< \Delta_-, \;\;\Delta_-< \Delta_b< \Delta_+),\quad  \;\;(\Delta_-< \Delta_a< \Delta_+, \;\; \Delta_+< \Delta_b )
\eqe
Note that if we only consider the last two sign constraint in eq.(\ref{S2Sign}), then we have reproduced $\mathbf{S}_1$. The first constraint further requires $\langle \mathbf{X},0,\Delta_a,\Delta_b\rangle$ to have a preferred sign, thus giving an extra boundary when it is set to zero. This sets the region for $\mathbf{S}_2$ shown in fig.\ref{S2Fig}.

\begin{figure}
\centering
\includegraphics[width=1.0\textwidth]{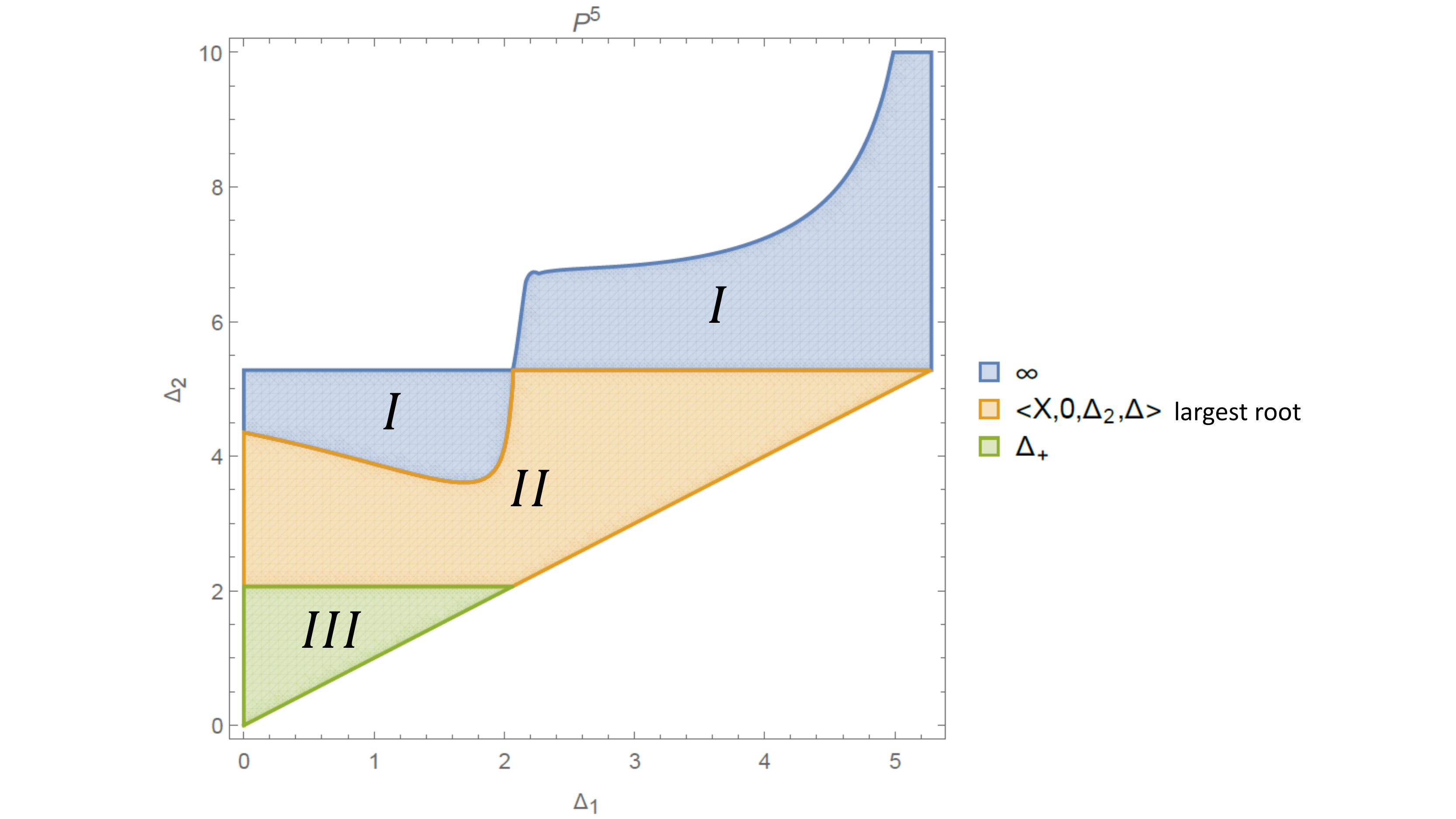}
\caption{\label{figP52d} Different allowed region for $\{\Delta_1, \Delta_2\}$ along with the associated bounds on $\Delta_3$ , with $\Delta_\phi=3/2$. }
\end{figure}

\begin{figure}
\centering
\begin{minipage}[t]{0.48\textwidth}
\centering
\includegraphics[width=6cm]{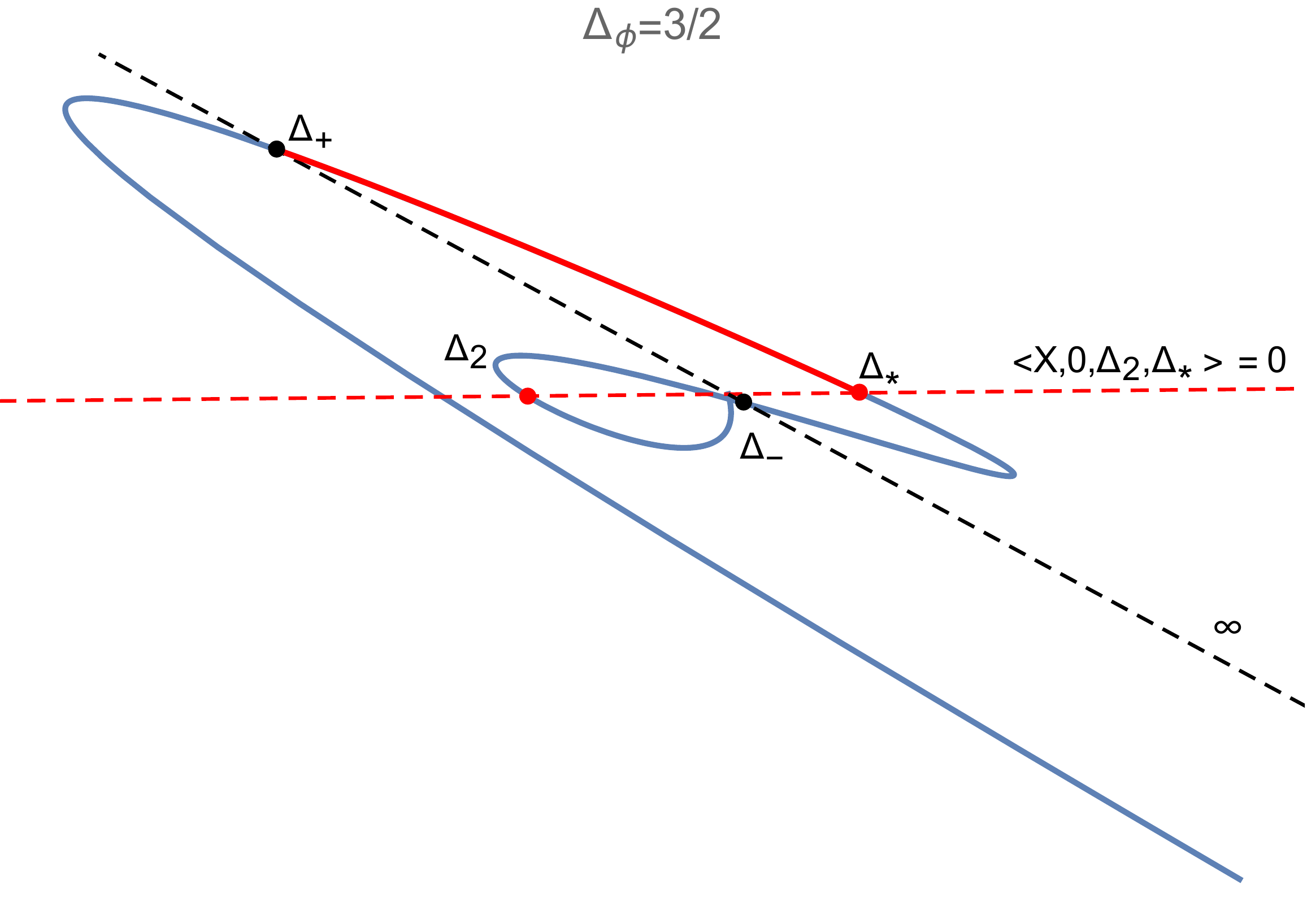}
\end{minipage}
\begin{minipage}[t]{0.48\textwidth}
\centering
\includegraphics[width=6cm]{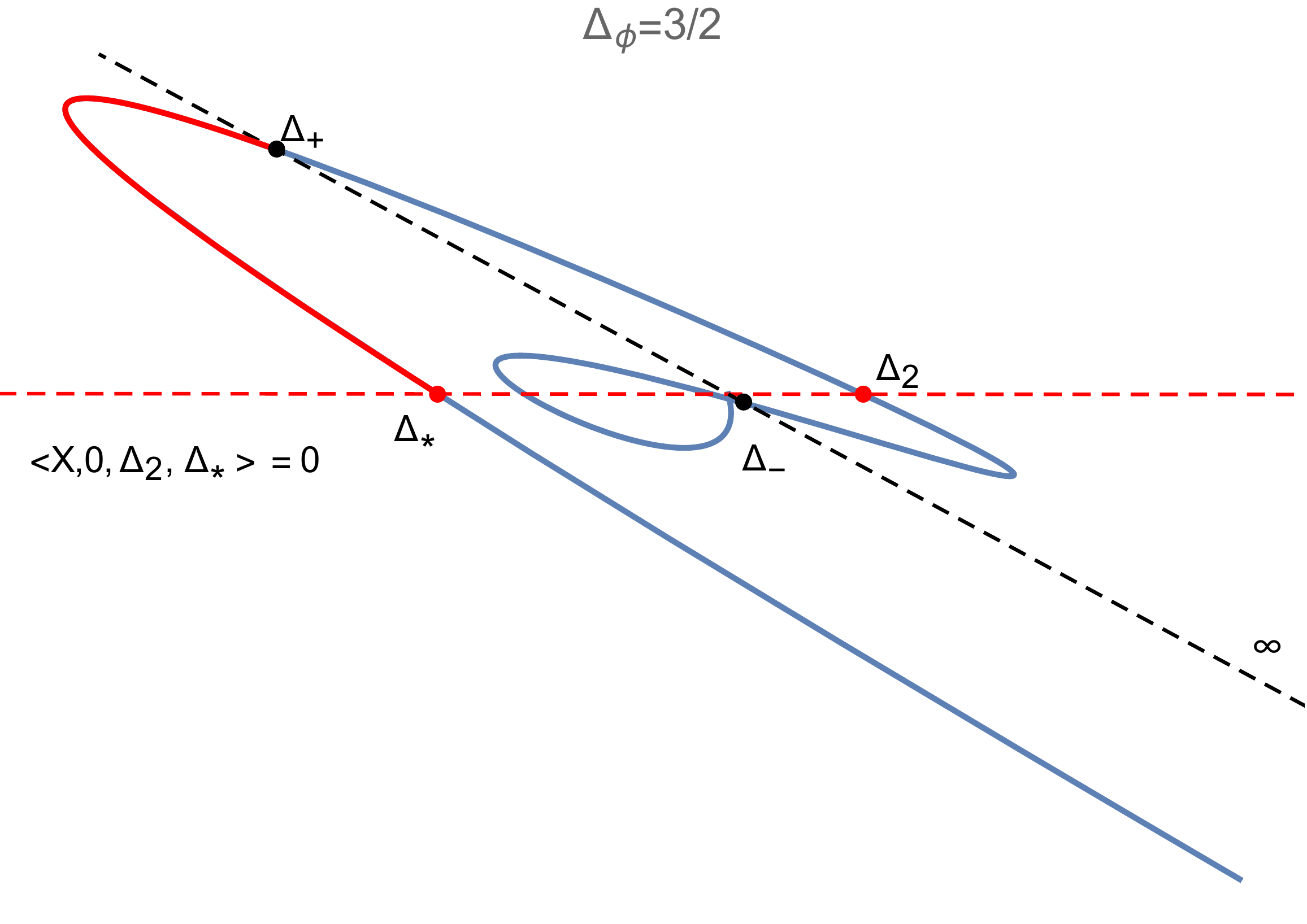}
\end{minipage}
\caption{\label{fig_s2p5} Here we present the region of $II$ (RHS) and $III$ (LHS) in (\ref{region23}) using the 2d geometry of $\mathbb{P}^5$ after projecting through $(\mathbf{X},\vec{\mathbf{G}_0})$. The red dashed line represents the intersection point of $\langle\mathbf{X},0,\Delta_2,\Delta\rangle=0$, and $\Delta_*$ is the largest intersection point. The allowed region in both figure is labeled by the red curve} 
\end{figure}

\begin{figure}
\centering
\includegraphics[width=1.2\textwidth]{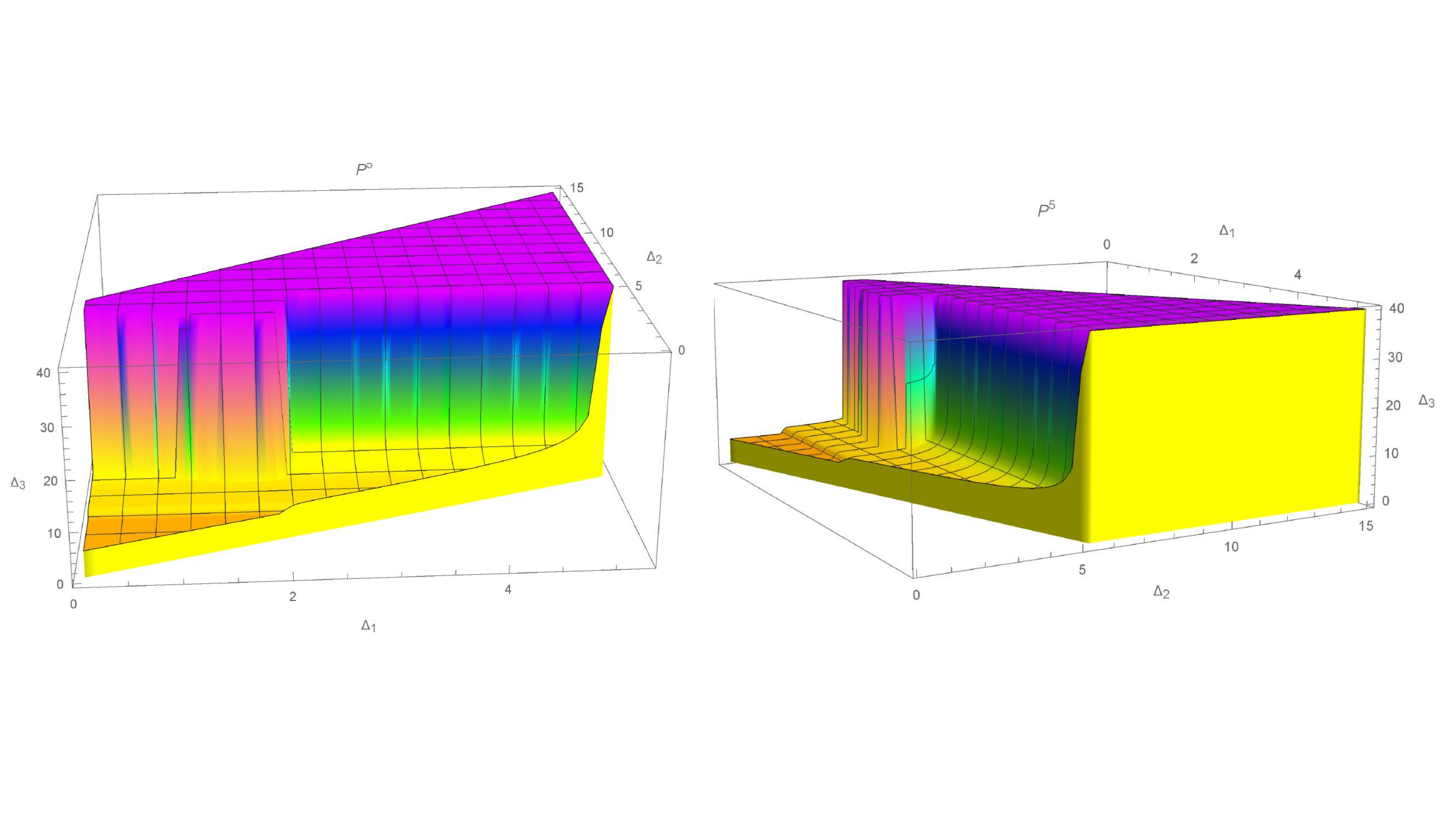}
\caption{\label{figP53dfs} In $\mathbb{P}^5$, $\Delta_\phi=3/2$, $ \{\Delta_1,\Delta_2,\Delta_3\}$ is allowed below the surface.The purple (dark) region is to be understood to extend to infinity}
\end{figure}

Now let us see how the knowledge of $\mathbf{S}_2$ impose itself on the spectrum. Let's take the view point from the first two operators $(\Delta_1,\Delta_2)$. The allowed region for the two operators are simply given by the area projected from $\mathbf{S}_2$ along the direction of $\Delta_2$ as shown in fig.\ref{figP52d}. Note that the allowed region is separated into different subregions reflecting the distinct bounds on the next operators $\Delta_3$. If we identify $(\Delta_1,\Delta_2)$ as $(\Delta_a,\Delta_b)$, then it is simply the projection of $\mathbf{S}_2$ on to the $(\Delta_1,\Delta_2)$ plane, and since when combined with $\infty$ we already have a simplex that encloses the origin, there is no constraint on the remaining spectrum in this region. This is denoted as  $I_1$ and $I_2$ in fig.\ref{figP52d}. Outside of this region, we can have $\Delta_2$ and $\Delta_3$ forming $\mathbf{S}_2$. First, for $\Delta_-<\Delta_2<\Delta_+$, we must have 
\eq\label{region23}
II:\quad (\Delta_-< \Delta_2< \Delta_+, \;\; \Delta_+< \Delta_3, \quad \langle \mathbf{X},0,\Delta_2,\Delta_3\rangle>0 )
\eqe
This sets the boundary for $\Delta_3$ at $\langle \mathbf{X},0,\Delta_2,\Delta_3\rangle=0$. Finally as $\Delta_2<\Delta_-$, we must have 
\eq
III:\quad (\Delta_2< \Delta_-, \;\; \Delta_-< \Delta_3<\Delta_+, \quad \langle \mathbf{X},0,\Delta_2,\Delta_3\rangle<0 )
\eqe
In this case we can see that the bound for $\Delta_3$ is simply $\Delta_+$. (We present the allowed region of $II$ and $III$ in fig.\ref{fig_s2p5}). The plot for allowed regions for $\Delta_1,\Delta_2,\Delta_3$ is presented in fig.\ref{figP53dfs} for all regions. In principle, it may be $\Delta_1$ and $\Delta_3$ forming $\mathbf{S}_2$, you may substitute $\Delta_b$ for $\Delta_a$ in eq.(\ref{region23}). But in this situation, it generates a lower bound on $\Delta_3$.

\subsection{Theory space in $\mathbb{P}^7$} 
\begin{figure}
\begin{center}
\includegraphics[scale=0.3]{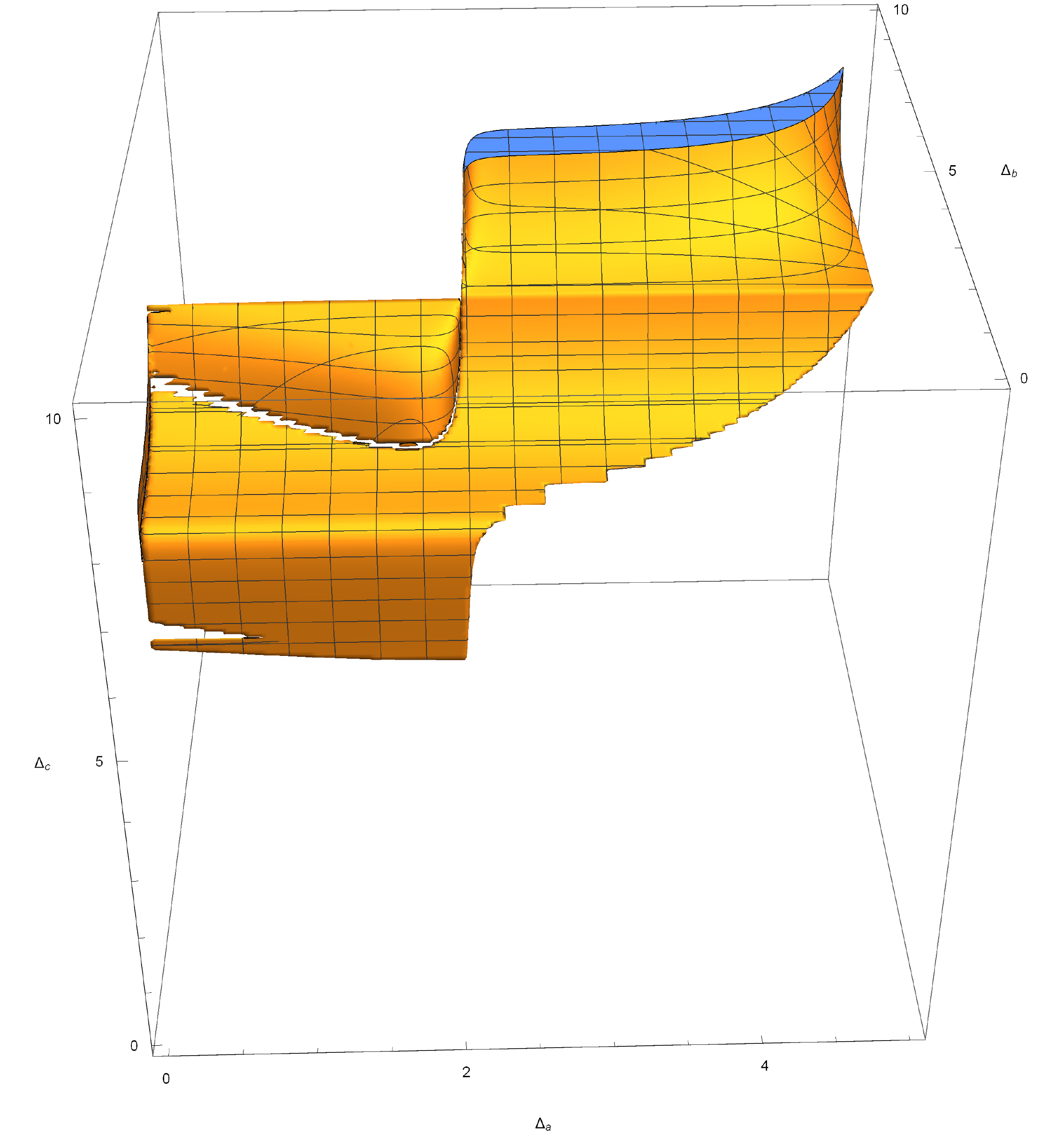}\includegraphics[scale=0.3]{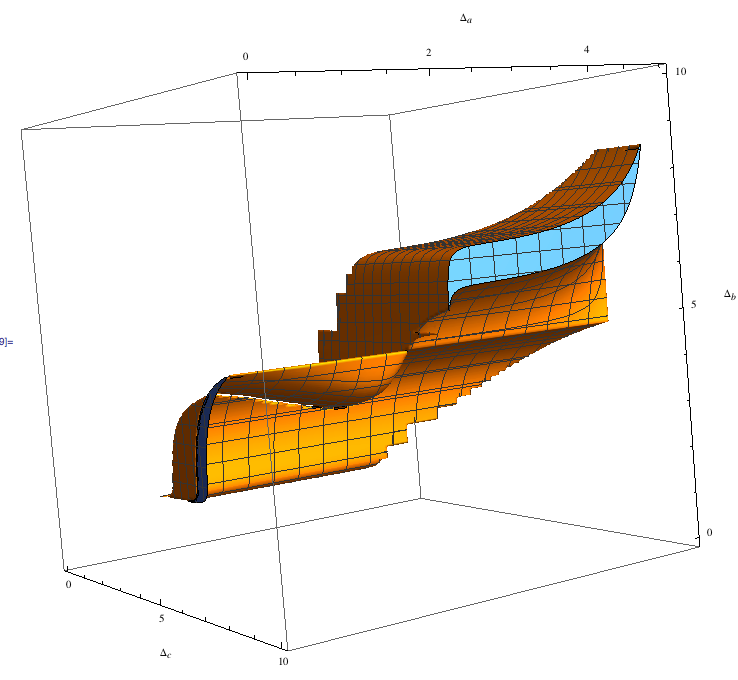}
\caption{Different views of  $\mathbf{S}_3$ defined through eq.(\ref{P7condition}).}
\label{S3Plot}
\end{center}
\end{figure}

\begin{figure}
\begin{center}
\includegraphics[scale=0.25]{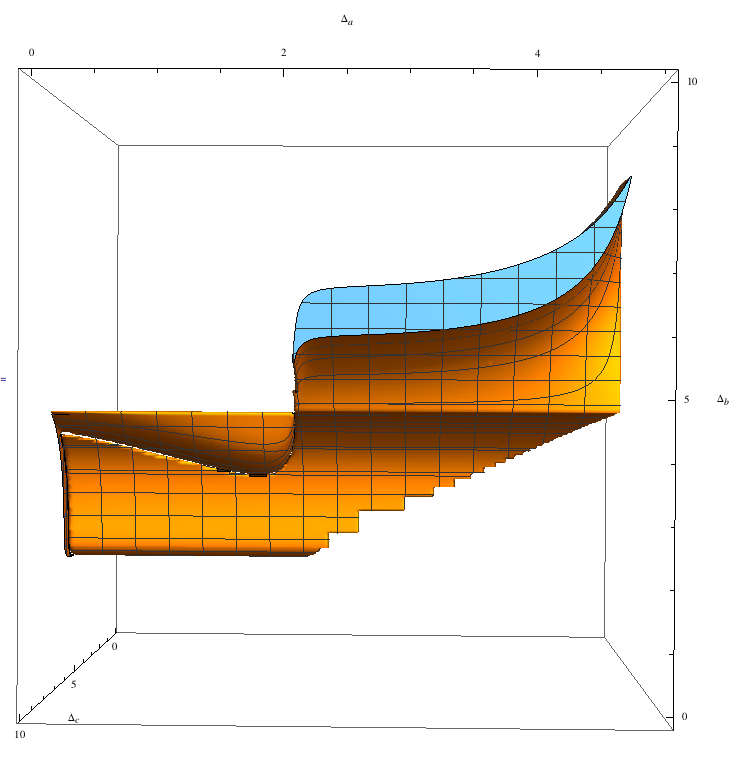}\includegraphics[scale=0.35]{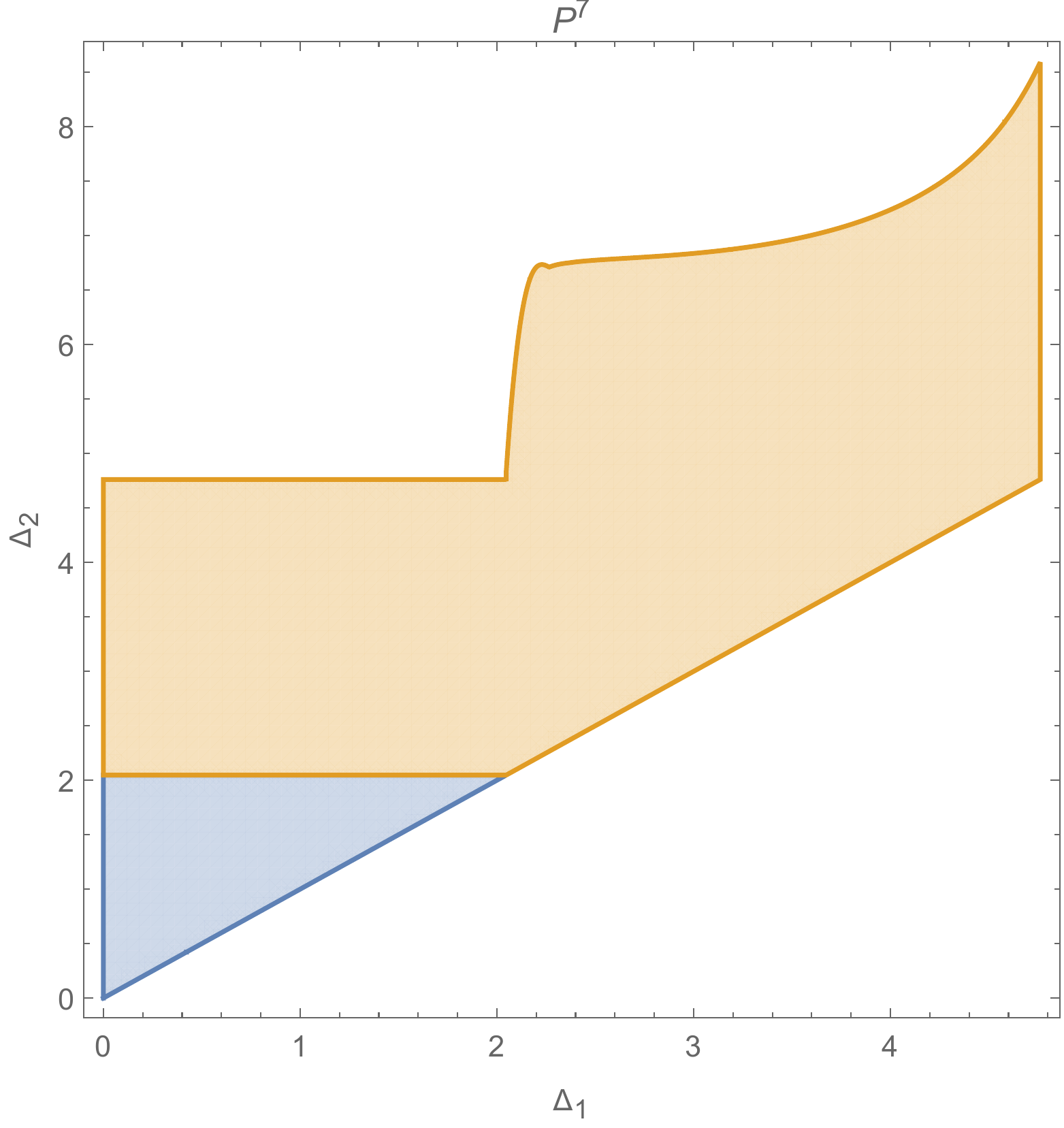}
\caption{On the left, we view it from the plane of $\Delta_1, \Delta_2$ where the surface is interpreted as the boundary of $\Delta_3$ for a given set of $(\Delta_1, \Delta_2)$. The right is a graphical representation of this projection with different subregions singled out to reflect the bounds on $\Delta_3$. Note that the bottom left corner of the left figure is empty. This implies that $(\Delta_1, \Delta_2)$ does not participate in forming $\mathbf{S}_2$, and $\Delta_3$ takes up the role of $\Delta_a$, which is why it is bounded by $\Delta^7_+$  as indicated on the right. }
\label{S3Plot2}
\end{center}
\end{figure}

Now, let's consider $\mathbb{P}^7$. For $N=3$, we construct $\mathbf{S}_3$ given by the uniform sign for:
\eqa\label{P7condition}
\textcircled{1}\;\langle \mathbf{X},0,\Delta_a,\Delta_b,\Delta_c\rangle,\quad \textcircled{2}\;-\langle \mathbf{X},0,\Delta_a,\Delta_b,\infty\rangle,\quad \textcircled{3}\;\langle \mathbf{X},0,\Delta_a,\Delta_c,\infty\rangle,\nonumber\\
\textcircled{4}\; -\langle \mathbf{X},0,\Delta_b,\Delta_c,\infty\rangle, 
\eqae 
with $\Delta_a<\Delta_b<\Delta_c$. \footnote{The full 3D plot of $\mathbf{S}_3$ has also been included as a mathematica notebook ``s3plot.nb" as an attachment}The resulting region is displayed in fig.\ref{S3Plot}. Again, we analyze the constraint on $(\Delta_1, \Delta_2, \Delta_3)$ from $\mathbf{S}_3$. By identifying the first three operators with $(\Delta_a,\Delta_b,\Delta_c)$, we can project along $\Delta_3$ onto the $(\Delta_1,\Delta_2)$ plane, as shown in the left figure in fig.\ref{S3Plot2}. Note that there is a section that is empty! This does not imply that $(\Delta_1,\Delta_2)$ is ruled out for this region, but rather that it does not participate in forming $\mathbf{S}_2$. In other words, in this region $\Delta_3$ takes up the role of $\Delta_a$, and hence is bounded by $\Delta^7_+$, the gap in $\mathbb{P}^7$. This is indicated on the RHS of fig.\ref{S3Plot2}. As discussed in the beginning, the last three sign constraint is equivalent to saying $\{\Delta_a,\Delta_b,\Delta_c\}$ are in  $\tilde{\mathbf{U}}_2$ which gives us the exact bound as in $\mathbb{P}^5$. However, the additional first sign constraint implies that some part of  $\tilde{\mathbf{U}}_2$ might be ruled out, indicating the gap, here referred to as $\Delta^7_+$, differs from that of $\mathbb{P}^5$. Let us begin by analyzing how this occurs in detail, and derive the analytic expression for $\Delta^7_+$. This in turn gives an alternative ``derivation" of the optimal functional at $\mathbb{P}^7$. For convenience, we denote the previous $\Delta_+$, $\Delta_-$ in $\mathbb{P}^5$ as $\Delta^5_+$, $\Delta^5_-$.

\begin{figure}
\centering
\includegraphics[width=0.6\textwidth]{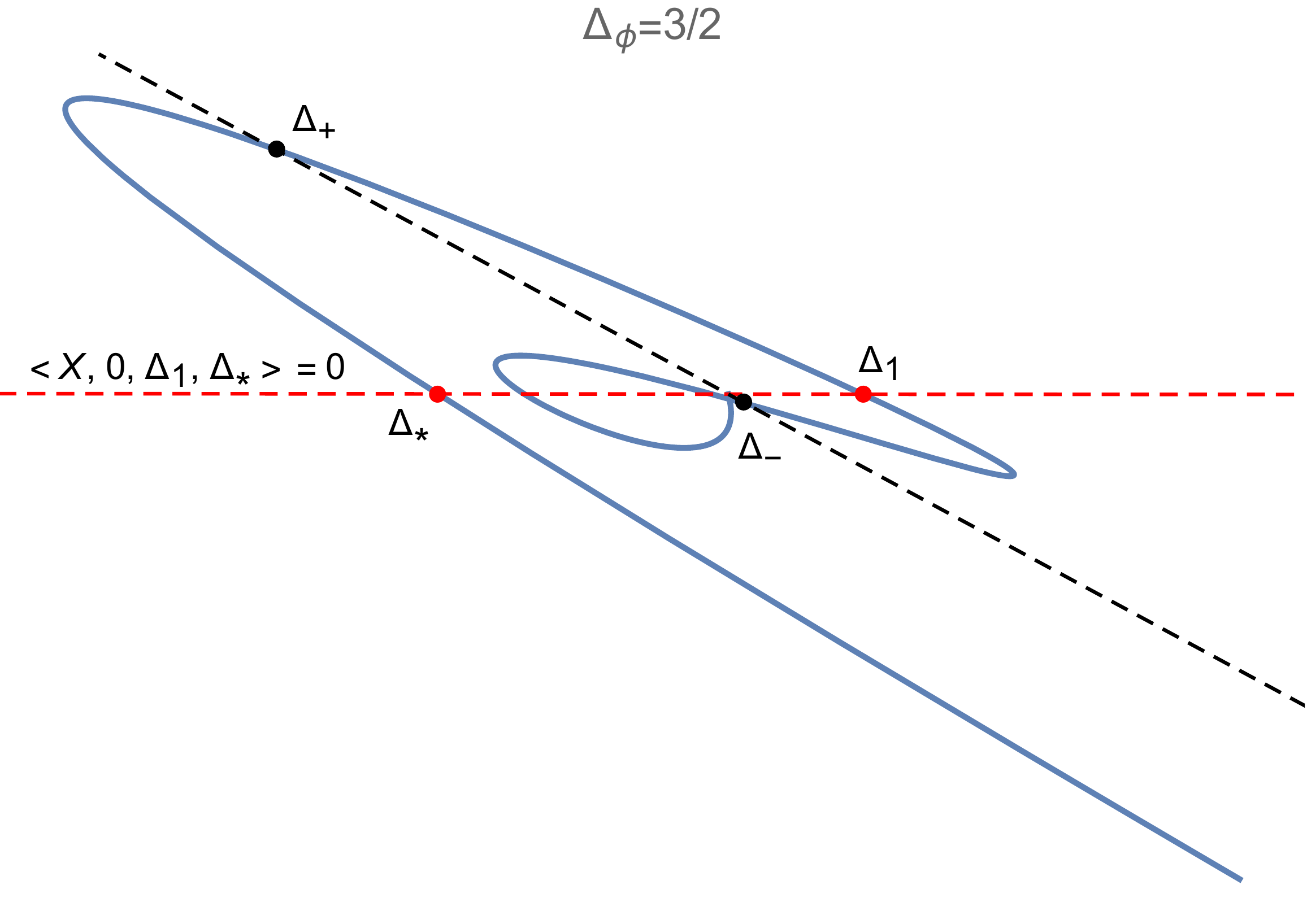}\caption{\label{figX0adelinf} When $\Delta_1$ is between $\Delta^5_+$ and $\Delta^5_-$, $\langle \mathbf{X},0,\Delta_1,\Delta,\infty\rangle$ has only one root($\Delta_*$) above $\Delta_1$. Thus, $\Delta_b$ must be below $\Delta_*$ and $\Delta_c$ must be above $\Delta_*$ to satisfy $\langle \mathbf{X},0,\Delta_1,\Delta,\infty\rangle$ having opposite sign at $\Delta_b$ and $\Delta_c$. }
\end{figure}

First we derive the upper bound on $\Delta_1$.  The can be obtained by setting $\Delta_1=\Delta_a$ in $\mathbf{S}_3$, and see the upper bound of $\Delta_a$. First we use condition \textcircled{2},\textcircled{3},\textcircled{4}, which is essentially the simplex condition in $\mathbb{P}^5$, leading to $\Delta_a\le\Delta_5^+$. Now in anticipation that some part of this region will run into trouble with \textcircled{1}, we consider $\Delta^7_+\le\Delta_a\le\Delta_+^5$, where $\Delta^7_+$ is the threshold for consistency with \textcircled{1}. In this region the function $\langle \mathbf{X},0,\Delta_1,\Delta,\infty\rangle$ has only one root above $\Delta_1$ (See fig.\ref{figX0adelinf} for explaination). As \textcircled{2},\textcircled{3} indicates that $\Delta_b, \Delta_c$ must be on opposite side of this root, we have:
\begin{equation}
\Delta_c\ge\langle \mathbf{X},0,\Delta_a,\Delta,\infty\rangle_{l.r}
\end{equation}
where $\langle\mathbf{X},\dots,\Delta\rangle_{l.r}$ corresponds to the largest root of $\langle\mathbf{X},\dots,\Delta\rangle$. Next, \textcircled{1},\textcircled{2}, indicates $\Delta_c$ must be smaller than $\langle\mathbf{X},0,\Delta_a,\Delta_b,\Delta\rangle_{l.r}$. Combining this two condition we get,
\eq\label{P7critical}
\quad \langle \mathbf{X},0,\Delta_a,\Delta_b,\Delta\rangle_{l.r} \geq \Delta_c \geq \langle \mathbf{X},0,\Delta_a,\Delta,\infty\rangle_{l.r}\,.
\eqe
Thus we conclude that the region $\Delta^7_+\le\Delta_a\le\Delta_+^5$ is ruled out must be due to the contradiction of the above equation, i.e. $\langle \mathbf{X},0,\Delta_a,\Delta_b,\Delta\rangle_{l.r}<\langle \mathbf{X},0,\Delta_a,\Delta,\infty\rangle_{l.r}$. Thus $\Delta^7_+$ is determined by the critical point,
\eq
\quad \langle \mathbf{X},0,\Delta_a,\Delta_b,\Delta\rangle_{l.r} = \langle \mathbf{X},0,\Delta_a,\Delta,\infty\rangle_{l.r}
\eqe
And by setting $\{\mathbf{v}_1,\mathbf{v}_2\} = \{\Delta_b, \infty\}$, we find that this exactly corresponds to solution to co-plane condition condition eq.(\ref{subplane7}). In this way we use $\mathbf{S}^3$ to reproduce the scalar gap in optimal functional.

\begin{figure}
\centering
\includegraphics[width=0.6\textwidth]{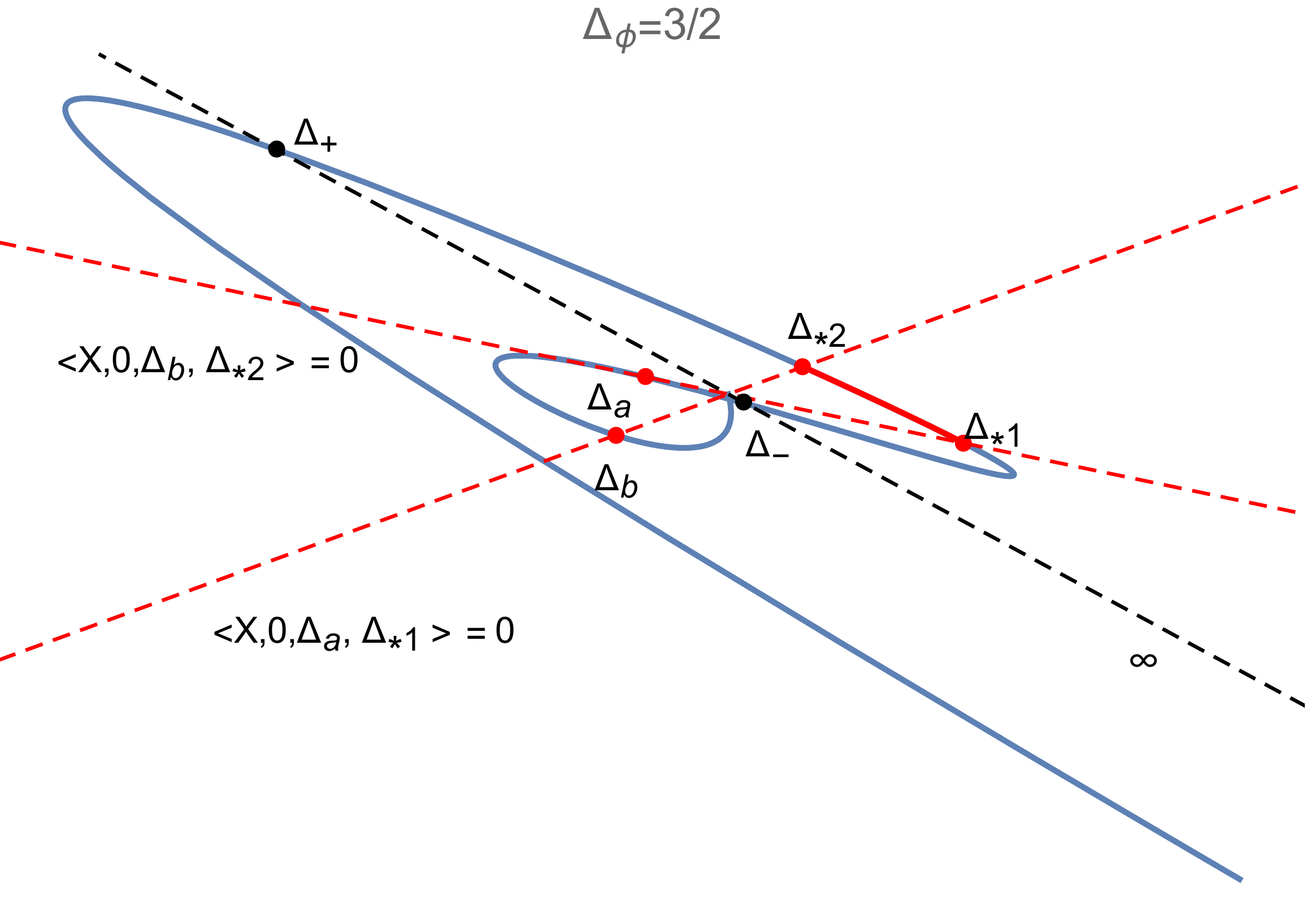}\caption{\label{figX0adelinf2} When $\Delta_1$ is below $\Delta^5_-$, $\Delta_3$ has to sit between $\Delta_{*1}$ and $\Delta_{*2}$ so that \{$\Delta_1, \Delta_2, \Delta_3$\} is in $\mathbb{S}_3$ }
\end{figure}

Before we move to the $\Delta_2$-bound, we define another important quantity $\Delta^7_-$, which is defined by the shadow region in fig.\ref{S3Plot2}, i.e. once $\Delta_a\le\Delta_b\le\Delta^7_-$, it cannot be part of $\mathbf{S}_3$. First, this only occurs when $\Delta_a,\Delta_b\le\Delta_-^5$, so we set $0\le\Delta_a\le\Delta_-^5$, and compute the upper bound for $\Delta_b$. Use $\mathbb{P}^5$ conditions \textcircled{2},\textcircled{3},\textcircled{4}, we see that when $\Delta_a\le\Delta_b\le\Delta_-^5$, the constraint for $\Delta_c$ is (see fig.\ref{figX0adelinf2})
\begin{equation}
\langle\mathbf{X},0,\Delta_a,\infty,\Delta\rangle_{s.l.r}\le\Delta_c\le\langle\mathbf{X},0,\Delta_b,\infty,\Delta\rangle_{s.l.r}
\end{equation}
where $\langle\mathbf{X},\dots,\Delta\rangle_{s.l.r}$ represents the second largest root of $\langle\mathbf{X},\dots,\Delta\rangle$. And also when assuming $\Delta_a\le\Delta_b\le\Delta_-^5$, condition \textcircled{1},\textcircled{2} gives that,
\begin{equation}
\langle\mathbf{X},0,\Delta_a,\Delta_b,\Delta\rangle_{s.l.r}\le\Delta_c\le\langle\mathbf{X},0,\Delta_a,\Delta_b,\Delta\rangle_{l.r}
\end{equation}
 So for given $\Delta_a$, the critical point $\Delta_*$ for $\Delta_b$ happens at,
\eq\label{P7delta-critical}
\quad \langle \mathbf{X},0,\Delta_b,\Delta,\infty\rangle_{s.l.r} = \langle \mathbf{X},0,\Delta_a,\Delta_b,\Delta\rangle_{s.l.r}
\eqe
It turns out that for any $\Delta_a$, the critical point $\Delta_*$ is the same, and solving the critical point we get $\Delta_*\approx$ 2.047. So we get the value of $\Delta_-^7$ is $\Delta_*$, whose value is exactly the same as the optimal functional in $\mathbb{P}^7$

Now lets' move on to find $\Delta_2$-bound, we can first inherit the bound from $\mathbb{P}^5$. Then we can set $\Delta_1=\Delta_a$, $\Delta_2=\Delta_b$ in $\mathbf{S}_3$ to get the bound on $\Delta_2$. The upper bound of $\Delta_b$ happens when for given $\Delta_a,\Delta_b$, there is no solution to $\Delta_c$ in $\mathbf{S}_3$. And from analysis of bound on $\Delta_1$, we know that $\Delta_a$ must below $\Delta_a\le\Delta_+^7$, so we separate the region into three part: $0\le\Delta_a\le\Delta_-^7$ , $\Delta_-^7\le\Delta_a\le\Delta_-^5$ and $\Delta_-^5\le\Delta_a\le\Delta_+^7$. In the following we'll see that in different region, the bound of $\Delta_b$ will correspond to roots of different funtions. 

\begin{figure}
\centering
\includegraphics[width=0.4\textwidth]{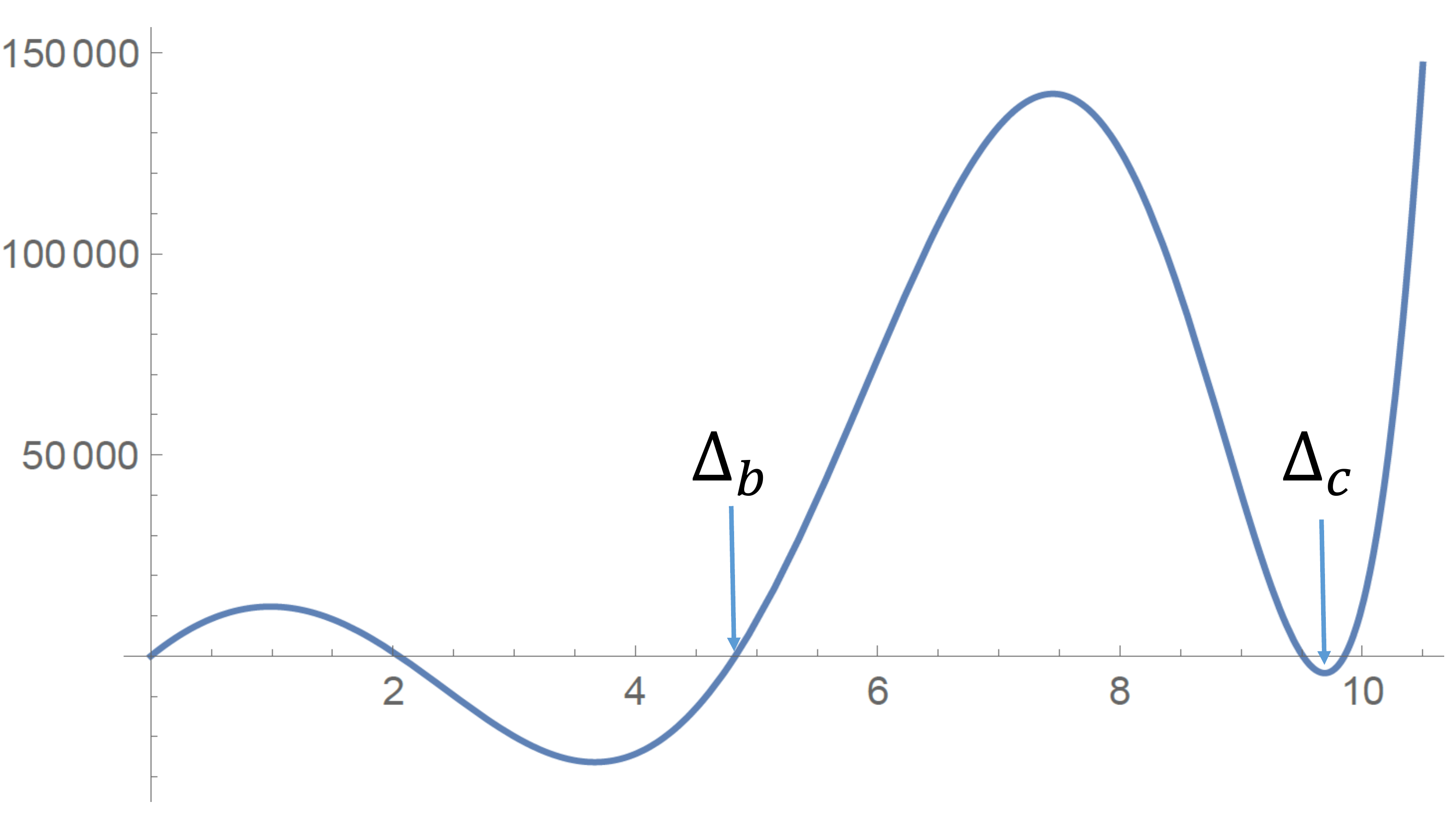}\includegraphics[width=0.4\textwidth]{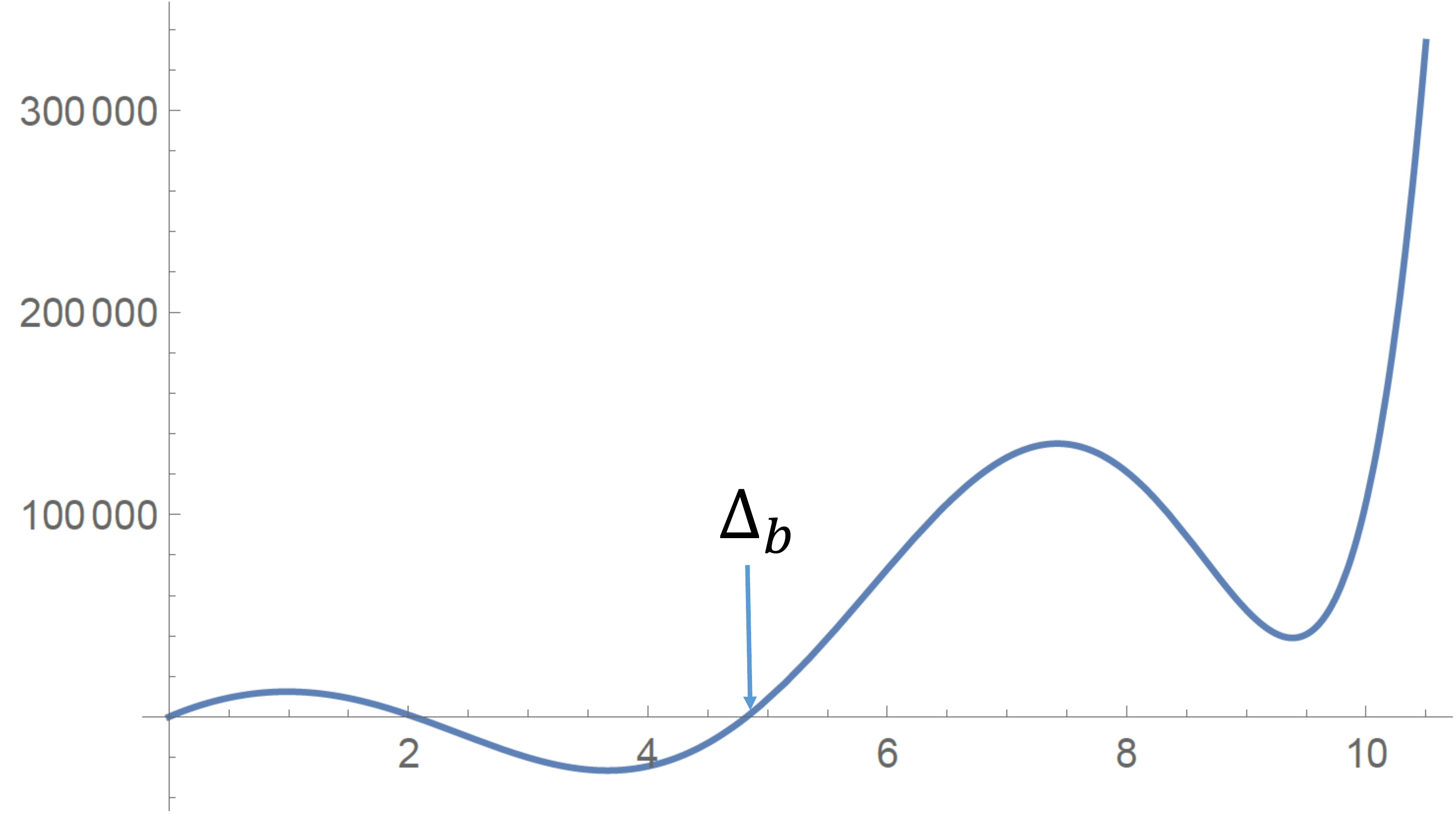}\caption{\label{figcritical2} Plot of function $\langle \mathbf{X},0,\Delta_a,\Delta_b,\Delta\rangle$. LHS: $\Delta_c$ must be in the region between last two zeros in order to satisfy condition eq.(\ref{p7s2con1}). RHS: In this case we can't find any $\Delta_c$ satisfying eq.(\ref{p7s2con1})}
\end{figure}

\begin{figure}
\centering
\includegraphics[width=1.0\textwidth]{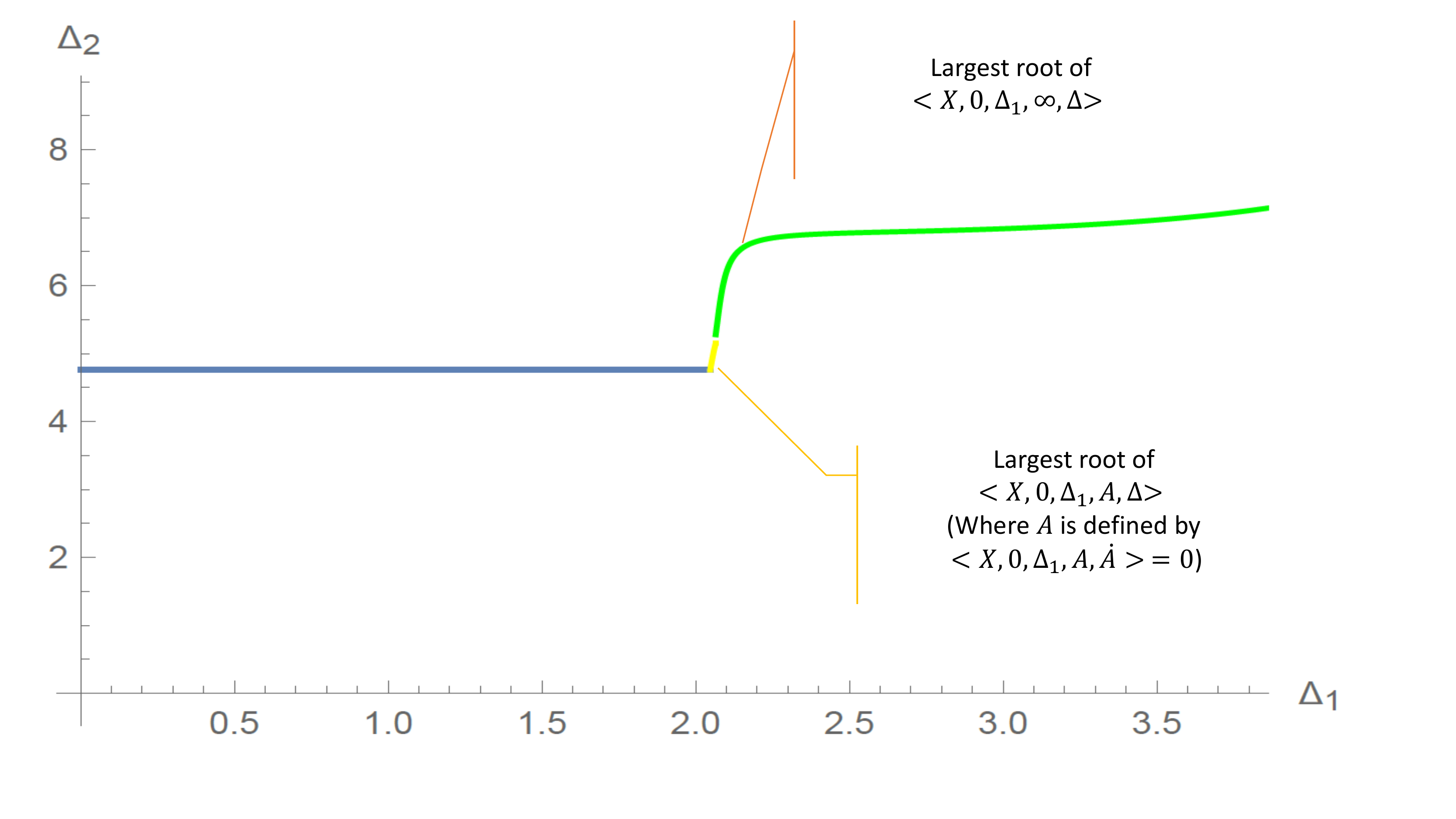}
\caption{\label{figP7b2}Bounding $\Delta_2$ in $\mathbb{P}^7$, $\Delta_\phi=3/2$. When $\Delta_1$ is below $\Delta^7_-$, $\Delta_2$ bound is $\Delta_{gap}$ in $\mathbb{P}^7$. When $\Delta_1$ is between $\Delta^7_-$ and $\Delta^5_-$, the bound is given by $\langle \mathbf{X},0,\Delta_a, A,\infty\rangle_{l.r}$ where $\Delta_i$ satisfied $\langle \mathbf{X},0,\Delta_a, \Delta_i, \dot{\Delta}_i\rangle = 0$. And when $\Delta_a$ is larger than $\Delta^5_-$, the bound is $\langle \mathbf{X},0,\Delta_a,\Delta,\infty\rangle_{l.r}$ }
\end{figure}

\begin{figure}
\centering
\includegraphics[width=0.6\textwidth]{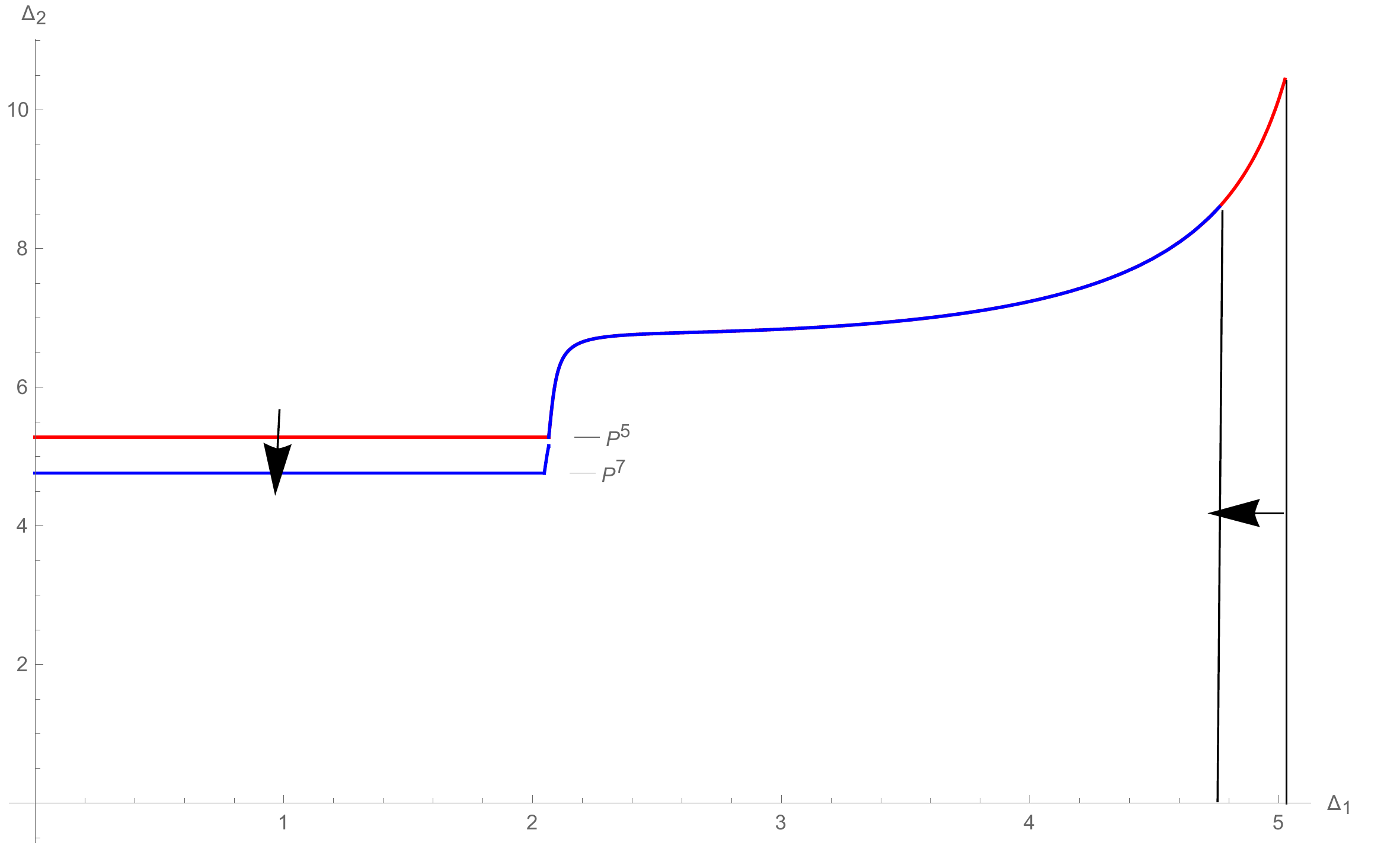}
\caption{\label{figP57lap} Overlapping $\Delta_2$ bounds in $\mathbb{P}^5$ and $\mathbb{P}^7$}
\end{figure}

\begin{itemize}
\item $0\le\Delta_a\le\Delta_-^7$

Consider condition \textcircled{1},\textcircled{2}, we get that,
\eq\label{P7condition+-}
\quad \Delta_c\le\langle \mathbf{X},0,\Delta_a,\Delta_b,\Delta\rangle_{l.r}
\eqe
Next we consider condition \textcircled{2} and \textcircled{3}, $\Delta_b$ and $\Delta_c$ must stay on the both sides of some zero of funciton $\langle\mathbf{X},0,\Delta_a,\infty,\Delta\rangle$. And in order to get the largest value of $\Delta_b$, that zero has to be the largest root, which implies,
\begin{equation}
\Delta_c\ge\langle\mathbf{X},0,\Delta_b,\infty,\Delta\rangle_{l.r}
\end{equation}
Combine this two together we get
\begin{equation}\label{p7s2con1}
\langle \mathbf{X},0,\Delta_b,\infty,\Delta\rangle_{l.r}\le\Delta_c\le\langle\mathbf{X},0,\Delta_a,\Delta_b,\Delta\rangle_{l.r}
\end{equation}
Solving for the critical value is when $\Delta_b$ satisfies $\langle\mathbf{X},0,\Delta_a,\Delta_b,\Delta\rangle = \langle \mathbf{X},0,\Delta_b,\Delta,\infty\rangle$. This value is the same as $\Delta^7_+$. So the bound of $\Delta_2$ in this region is $\Delta_2\le\Delta_+^7$. 

\item $\Delta^7_-\le\Delta_a\le\Delta^5_-$

Next, we focus on $\Delta^7_-\le\Delta_a\le\Delta^5_-$ region.  Notice that the analysis for constraint eq.(\ref{p7s2con1}) is still valid in this region. When $\Delta_a$ stays in this region, we vary $\Delta_b$ to check whether eq.(\ref{p7s2con1}) can be satisfied. And we found that when $\Delta_b$ above the critical value $\Delta^*$, eq.(\ref{p7s2con1}) cannot be satisfied.(Near the critical point, the plot of function $\langle\mathbf{X},0,\Delta_a,\Delta_b,\Delta\rangle$ is in fig.\ref{figcritical2}). From the plot, the critical point happens when there is a double zero in the functional. We can use this condition to determine position of double zero $\Delta_i$,
\begin{equation}
\langle\mathbf{X},0,\Delta_a,\Delta_i,\dot{\Delta}_i\rangle=0
\end{equation}
And the bound of $\Delta_2$ in this case is defined by the $\Delta_2\le\Delta^*=\langle\mathbf{X},0,\Delta_a,\Delta_i,\Delta\rangle_{l,r}$.

\item $\Delta_-^5\le\Delta_a\le\Delta_+^7$

When $\Delta_-^5\le\Delta_a\le\Delta_+^7$, we found as long as $\Delta_b\le\langle\mathbf{X},0,\Delta_b,\infty,\Delta\rangle_{l.r}$, we can always find $\Delta_c$ satisfying condition \textcircled{1}. So the bound of $\Delta_2$ is $\Delta_2\le\langle\mathbf{X},0,\Delta_b,\infty,\Delta\rangle_{l.r}$, which is the same as $\mathbb{P}^5$.
\end{itemize}
The plot of bound on $\Delta_2$ is at Fig.\ref{figP7b2}. And also we compare the bound on $\Delta_2$ both at $\mathbb{P}^7$ and $\mathbb{P}^5$, the result is in Fig.\ref{figP57lap}, in whihc we see that the allowed region for \{$\Delta_1$, $\Delta_2$\} is shrinking. Also notice that, while $\Delta_1\to\Delta_+^5$, the bound on $\Delta_2$ approaches $\infty$ in $\mathbb{P}^5$. But in $\mathbb{P}^7$, the largest value of $\Delta_2$-bound happens when $\Delta_1\to\Delta_+^7$, and the bound approaches a finite number $\Delta_2\to\Delta_i$, which is the first double zero in optimal functional in $\mathbb{P}^7$. Physcially, it means in $\mathbb{P}^7$, the bounds of first two opeartors shrinks to a finite region!

\begin{figure}
\centering
\includegraphics[width=1.0\textwidth]{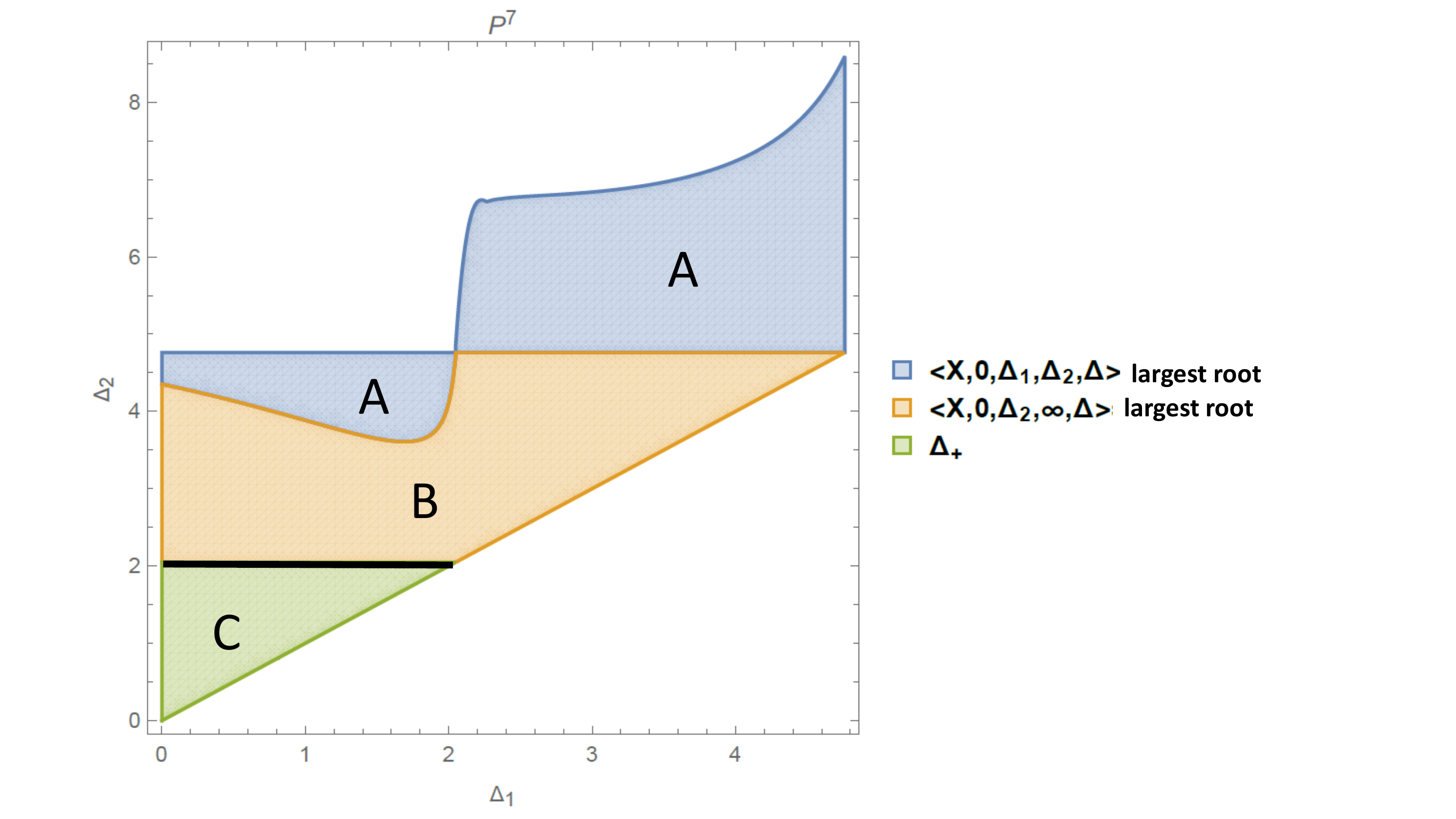}\caption{\label{figP72d} Bounds on $\Delta_3$. Note that there's a thin stripe between A and B. When $\Delta_2$ is between $\Delta_-$ defined in $\mathbb{P}^7$ and $\mathbb{P}^5$, bound is $\langle\mathbf{X},0,\Delta_2,\Delta_B,\Delta\rangle_{l.r}$, where $\Delta_B$ is the root of $\langle\mathbf{X},0,\Delta_2,\Delta_B,\dot{\Delta}_B\rangle$}
\end{figure}

\begin{figure}
\centering
\includegraphics[width=1.2\textwidth]{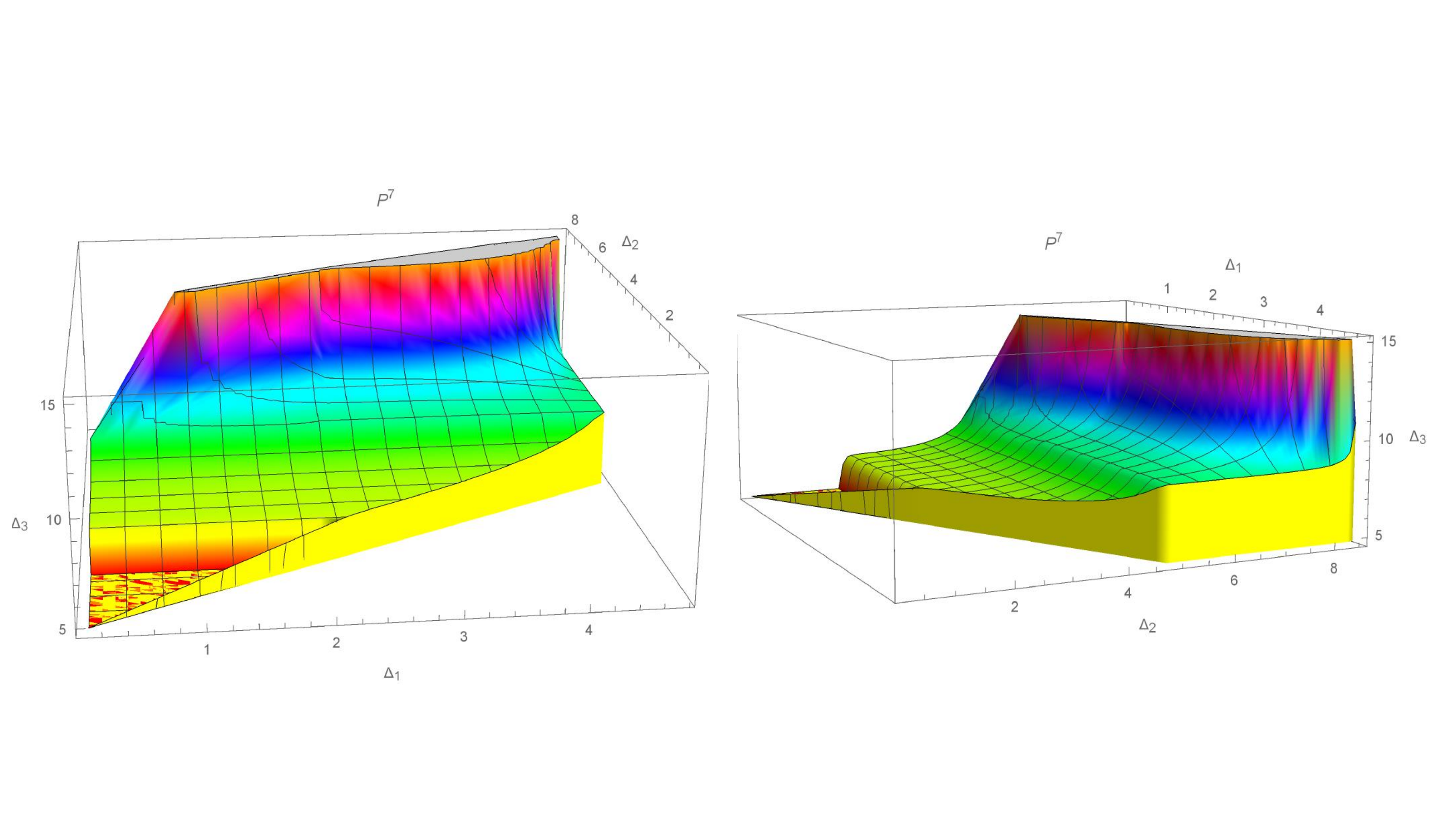}
\caption{\label{figP73dfs}In $\mathbb{P}^7$, $\Delta_\phi=3/2$, $ \{\Delta_1,\Delta_2,\Delta_3\}$ is allowed below the surface.}
\end{figure}

We briefly mention the bound on $\Delta_3$. The result is in Fig.\ref{figP72d} and Fig.\ref{figP73dfs} (3D plot). For region $A$ and $B$, we get the bound for $\Delta_3$ by setting $(\Delta_1,\Delta_2)=(\Delta_a,\Delta_b)$ , in $\mathbf{S}_3$, and looking for the bound on $\Delta_c$.
While in region $C$. fig.\ref{figP72d}, the bound can be obtained by setting $\Delta_2=\Delta_a$, and look for bound of $\Delta_b$ or setting $\Delta_1=\Delta_a$ and look for bounds on $\Delta_a$.

\subsection{Application: uniqueness of 1D fermionic free field theory}
\begin{figure}
\centering
\includegraphics[width=0.6\textwidth]{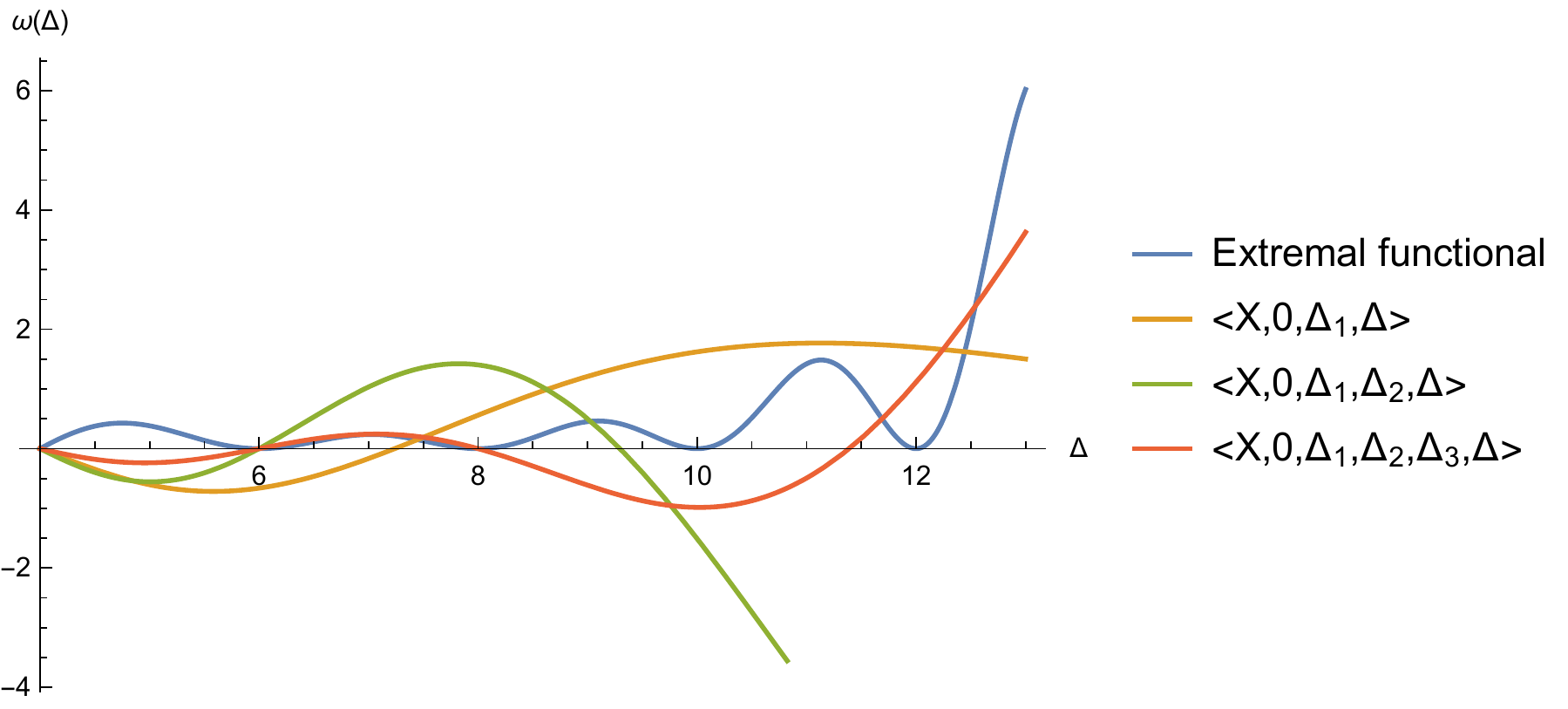}\caption{\label{fig_fer} Plot of extremal functional in \cite{Mazac:2018mdx} and $\omega_1(\Delta),\omega_2(\Delta),\omega_3(\Delta)$ in (\ref{testfun}) when $\Delta_\phi=3/2$. The orange, green and the red dot correspond to the largest root of functional $\omega_{1,2,3}(\Delta)$ }
\end{figure}

We start with the question: if the first operator $\Delta_1$ is exactly at  $\Delta_1=2\Delta_\phi+1$, is the theory uniquely fixed? Namely, is the spectrum and their corresponding OPE coefficient unique? Actually from the optimal functional construct in \cite{Mazac:2016qev} and \cite{Mazac:2018mdx}, we see that after the $\Delta=2\Delta_\phi+1$, the functional is definite positive and only have double zeros at $\Delta_{n,odd}=2\Delta_\phi+2n+1$. So act this functional into the sum rule,
\begin{equation}
\omega(0)+c_{2\Delta_\phi+1}^2\omega(2\Delta_\phi+1)+\sum_{\Delta\ge2\Delta_\phi+1}c_\Delta^2\omega(\Delta)=\sum_{\Delta\ge2\Delta_\phi+1}c_\Delta^2\omega(\Delta)=0
\end{equation}
So all the operators must in the position of the double zeros $\Delta_{n,odd}=2\Delta_\phi+2n+1$ in order for this sum rule to hold. But this still does not fix the spectrum, becasue it could be that the second one is not at $2\Delta_\phi+3$, the third one is not at $2\Delta_\phi+5$... We will show that this is impossble. And in order to achieve this we construct a simple family of linear functional,
\begin{align}\label{testfun}
&\omega_1(\Delta)=\langle\mathbf{X},0,\Delta_1,\Delta\rangle\notag\\
&\omega_2(\Delta)=\langle\mathbf{X},0,\Delta_1,\Delta_2,\Delta\rangle\notag\\
&\dots\notag\\
&\omega_n(\Delta)=\langle\mathbf{X},0,\Delta_1,\Delta_2,\dots,\Delta_n,\Delta\rangle,\quad\Delta_n=2\Delta_\phi+2n-1,~n\in\mathrm{odd~number}
\end{align}
These functionals have the proberty that
\begin{equation}
\omega_n(0)=0,~\omega_n(\Delta_k)=0,~for~k\le n
\end{equation}
Take $\Delta_\phi=3/2$ for example, we first use functional $\omega_1(\Delta)$ to bound the location of the second opearator. Setting $\Delta_1=4$ in $\omega_1(\Delta)$. The largest root of the $\omega_1(\Delta)$ is at $\Delta^*\approx7.24$ (plot of this function is the orange curve of Fig.\ref{fig_fer}). Act this functional to the sum rule we get
\begin{equation}
\omega(0)+c_{4}^2\omega(4)+\sum_{\Delta>4}c_\Delta^2\omega(\Delta)=\sum_{\Delta>4}c_\Delta^2\omega(\Delta)=0
\end{equation}
So we have the bound for the second opearator $\Delta_2\le\Delta^*\approx7.24$. And becase it's the second opearator, its scaling dimension must be larger than the first one $\Delta_2>\Delta_1=4$. In the region $4<\Delta_2\le7.24$, there is only one double zero in the extremal functional (plot of extremal function is the blue curve of Fig.\ref{fig_fer}), which is $\Delta=6$. Combined with these two constraint we get, the second opearator must exactly lie in $\Delta_2=6$ ! After fixing the location of the second opearator $\Delta_2$, we can use $\omega_2(\Delta)$ by setting $\Delta_1=4,\Delta_2=6$ to fix the third opearator $\Delta_3$. The largest root of this function is $\Delta^*\approx9.30<10$. So again, by applying this $\omega_2(\Delta)$ to the sum rule, we see that the third opearator must be exactly at $\Delta_3=8$.

One can iterate this arguement to bound that the location n-th opearator if the  (n+1)-th double zero in the extremal functional is smaller than the largest single zero $\Delta^*$ of functional $\omega_n(\Delta)$. Because if it holds true, the n-th opearator $\Delta_n$ is bounded in the region $\Delta_{n-1}<\Delta_n<\Delta^*$. And the number of double zeros in this region is always one, so the location of $\Delta_n$ is completely fixed. We show the data in Table.\ref{table_1dfer}.But in the last column we found that when $n=8$, the largest root of $\omega_8(\Delta)$ is bigger than the 9-th double zero, which naively would means that the iteration stops here.  We introduce a new functional $\tilde{\omega}(\Delta)$,
\begin{align}
&\tilde{\omega}(\Delta)=\langle\mathbf{X},0,\Delta_1,\dots,\Delta_8,\Delta_i,\dot{\Delta}_i,\Delta\rangle\notag\\
&\{\Delta_1,\Delta_2,\dots,\Delta_8\}=4,6,\dots,18,~\Delta_i=25
\end{align}
By introducing a double zero at $\Delta=\Delta_i=25$\footnote{The number 25 is just a valid choice. One can choice another number to get a smaller gap.}, we can construct a functional with single zero at $\Delta=\{4,6,\dots,18\}$ and with smaller single largest single zero, which is $\Delta^*\approx20.96<22$. So using this functional $\tilde{\omega}(\Delta)$ to replace $\omega_8(\Delta)$ we can continute the iteration. And for higher $n$, we expect that with suitable choice of double zero position $\{\Delta_{a_1},\Delta_{a_2},\dots,\Delta_{a_k}\}$, one can always construct a functional $\omega_n(\Delta)=\langle\mathbf{X},0,\Delta_1,\dots,\Delta_n,\Delta_{a_1},\dot{\Delta}_{a_1},\dots,\Delta_{a_k},\dot{\Delta_{a_k}},\Delta\rangle$ which make the iteration go.

\begin{table}
\centering
\begin{tabular}{|c|c|c|c|c|c|c|c|c|}
\hline
Functional& $\omega_1(\Delta)$ & $\omega_2(\Delta)$ & $\omega_3(\Delta)$ & $\omega_4(\Delta)$ & $\omega_5(\Delta)$ & $\omega_6(\Delta)$ & $\omega_7(\Delta)$ & $\omega_8(\Delta)$\\
\hline
Largest single zero & 7.24 & 9.30 & 11.39 & 13.50 & 15.62 & 17.74 & 19.87 & 22.01\\
\hline
(n+1)-th double zero & 8 & 10 & 12 & 14 & 16 & 18 & 20 & 22\\
\hline
\end{tabular}
\caption{\label{table_1dfer} Largest single zero of $\omega_n(\Delta)$ and (n+1)-th double zero of extremal functional when $\Delta_\phi=3/2$}
\end{table}

\section{Conclusion}
In this paper, we studied the optimal functionals of 1D CFT bootstrap in the context of positive geometry. In the derivative expansion scheme, we've identified that the optimal functional corresponds to a degenerate simplex that is one of the faces for the convex hull of block vectors. In particular, for the $2N{+}1$ derivative order with $N=2k{+}1$, the optimal functional is associated with finding $k{+}1$ block vectors such that when projected through the identity and crossing plane, it forms a simplex that encloses the origin. For 1D CFT, this simplex is unique. Taking the continuous limit, we've shown that in the large $\Delta$ limit the functional reproduces the features for the exact functional proposed in~\cite{Mazac:2016qev}. Put in another way, the integral representation for the exact functional is the answer for our geometric problem in the limit when $N\rightarrow\infty$.

We've also extended our analysis to 2D CFT in the diagonal block and modular bootstrap. For 2D CFT, we find that for $\Delta>0.08$, the same degenerate simplex condition also yields the correct optimal functional. The same conclusion was found for spin-less modular bootstrap where we consider rectangular torus. Note that when spins are included for the 2D CFT, we find that the optimal functional still satisfy the degenerate simplex condition, the only difference is that the spin-4  blocks are involved in determining the simplex. 

There are a vast range of straightforward generalizations to be considered. For example the study of the 2D bootstrap away from the diagonal limit as well as the full analysis for the inclusion of spins for the modular bootstrap. Note that the presence of kink in the 2D CFT is reflected in the fact that the optimal functional changes at small $\Delta_\phi$. Since the optimal functional has a clear geometric interpretation for all cases, this implies that there should be a geometric interpretation of the position of the kink. Finally, we can also rephrase the maximization of OPE coefficients as an intersection problem in our positive geometry. Given that the boundaries of the cyclic polytope is known, the range of allowed OPE coefficients should then correspond to the intersection of the line associated with any given block vector, and the boundaries of the cyclic polytope. Identification of the particular boundary should give the bound on the OPE coefficient. We leave this for future exploration.

\section{Acknowledgements}
The authors would like to thank Aninda Sinha, Ahmadullah Zahed for their discussion and collaboration.  We also thank  Song He, Avinash Raju and Somyadip Thakur for discussions during the initial phase of this project. 
 YTH is supported by MoST Grant No. 106-2628-M-002-012-MY3. YTH is also supported by Golden Jade fellowship.

\newpage
\appendix
\section{Equivalence of two functional}\label{prof}
In this we give a rigorous proof of the equivalence of these two kind of functional $\omega(\Delta)=\langle\mathbf{X},0,\Delta_1,\Delta_2,\cdots,\Delta\rangle$ and $\omega(\Delta)=\vec{\alpha}\cdot\vec{\mathbf{F}}_\Delta^{\Delta_\phi}$. First, we multiply function $z^{-2\Delta_\phi}$ to both sides of s-channel OPE expansion
\begin{equation}\label{GL}
\sum_\Delta c_\Delta^2 z^{-2\Delta_\phi}G_\Delta(z)=\sum_\Delta c_\Delta^2\tilde{G}_\Delta(z)=\mathcal{G}(z)z^{-2\Delta_\phi}=\tilde{\mathcal{G}}(z)
\end{equation}
where $\tilde{G}_\Delta(z)=z^{-2\Delta_\phi}G_\Delta(z),\tilde{\mathcal{G}}(z)=z^{-2\Delta_\phi}\mathcal{G}(z)$. Expand bothside at $z=1/2$, we get
\begin{equation}
\sum_\Delta c_\Delta^2\tilde{\mathbf{\mathbf{G}}}_\Delta=\mathbf{\tilde{\mathcal{G}}}\Rightarrow M(\Delta_\phi)\cdot\Big[\sum_\Delta c_\Delta^2\cdot\mathbf{G}_\Delta\Big]=M(\Delta_\phi)\cdot\vec{\mathcal{G}}
\end{equation}
where $M(\Delta_\phi)$ is just a $k\times k$ matrix depending only on $\Delta_\phi$. This matrix corresponds to the $GL(k)$ transformation due to the $z^{-2\Delta_\phi}$ prefactor. An explicit example of  $M(\Delta_\phi)$ in $\mathbb{P}^3$ is,
\begin{equation}
M(\Delta_\phi)=
\left(
\begin{array}{cccc}
 2^{2 \Delta _{\phi }} & 0 & 0
   & 0 \\
 -2^{2 \Delta _{\phi }+2}
   \Delta _{\phi } & 2^{2
   \Delta _{\phi }} & 0 & 0 \\
 2^{2 \Delta _{\phi }+3}
   \Delta _{\phi }^2+2^{2
   \Delta _{\phi }+2} \Delta
   _{\phi } & -2^{2 \Delta
   _{\phi }+2} \Delta _{\phi }
   & 2^{2 \Delta _{\phi }-1} &
   0 \\
 -\frac{1}{3} 2^{2 \Delta
   _{\phi }+5} \Delta _{\phi
   }^3-2^{2 \Delta _{\phi }+4}
   \Delta _{\phi
   }^2-\frac{1}{3} 2^{2 \Delta
   _{\phi }+4} \Delta _{\phi }
   & 2^{2 \Delta _{\phi }+3}
   \Delta _{\phi }^2+2^{2
   \Delta _{\phi }+2} \Delta
   _{\phi } & -2^{2 \Delta
   _{\phi }+1} \Delta _{\phi }
   & \frac{1}{3} 2^{2 \Delta
   _{\phi }-1} \\
\end{array}
\right)
\end{equation}
So easily we get
\begin{equation}\label{app1}
\langle\mathbf{\tilde{X}},\tilde{0},\tilde{\Delta}_1,\tilde{\Delta}_2\cdots,\tilde{\Delta}\rangle=\det[M(\Delta_\phi)]\langle\mathbf{X},0,\Delta_1,\Delta_2,\cdots,\Delta\rangle
\end{equation}
Next we will relate $\langle\mathbf{\tilde{X}},\tilde{0},\tilde{\Delta}_1,\tilde{\Delta}_2\cdots,\tilde{\Delta}\rangle$ to numerical functional $\vec{\alpha}\cdot\vec{\mathbf{F}}_\Delta^{\Delta_\phi}$. Here we use ``$\tilde{\quad}$" to represent vectors that are GL rotated by $M(\Delta_\phi)$. Crossing symmertry implies that $\tilde{\mathcal{G}}(z)$ satisfies
\begin{equation}\label{crossing2}
\tilde{\mathcal{G}}(z)=\tilde{\mathcal{G}}(1-z)
\end{equation}
So the crossing plane $\mathbf{\tilde{X}}$ will be simplied to
\begin{equation}
\mathbf{\tilde{X}}=
\begin{pmatrix}
1&0&0&\cdots&0\\
0&0&0&\cdots&0\\
0&1&0&\cdots&0\\
0&0&0&\cdots&0\\
0&0&1&\cdots&0\\
\vdots&\vdots&\vdots&\vdots&\vdots\\
0&0&0&\cdots&1
\end{pmatrix}
\end{equation}
Also notice that $F_\Delta^{\Delta_\phi}(z)=\tilde{G}_\Delta(z)-\tilde{G}_\Delta(1-z)$, so
\begin{equation}
\frac{\mathrm{d}^{2i+1}}{\mathrm{d}z^{2i+1}}F_\Delta^{\Delta_\phi}(z)\Big|
_{z=1/2}=2\frac{\mathrm{d}^{2i+1}}{\mathrm{d}z^{2i+1}}\tilde{G}_\Delta(z)\Big|
_{z=1/2}
\end{equation}
So the vector $\vec{\mathbf{F}}_\Delta^{\Delta_\phi}$ is just picking out the odd component in of $\tilde{\mathbf{G}}_\Delta$. And the crossing plane $\mathbf{\tilde{X}}$ now exactly cancel the even component in $\tilde{\mathbf{G}}_\Delta$, so we get the relation,
\begin{equation}\label{app2}
\langle\mathbf{\tilde{X}},\tilde{0},\tilde{\Delta}_1,\tilde{\Delta}_2\cdots,\tilde{\Delta}\rangle=\frac{1}{2^N}\langle\vec{\mathbf{F}}_{0}^{\Delta_\phi},\vec{\mathbf{F}}_{\Delta_2}^{\Delta_\phi},\cdots,\vec{\mathbf{F}}_{\Delta}^{\Delta_\phi}\rangle=\alpha\cdot\vec{\mathbf{F}}_{\Delta}^{\Delta_\phi}
\end{equation}
where $\alpha$ is a vector tangent to $\{\vec{\mathbf{F}}_{0}^{\Delta_\phi},\vec{\mathbf{F}}_{\Delta_2}^{\Delta_\phi},\cdots,\vec{\mathbf{F}}_{\Delta_k}^{\Delta_\phi}\}$. Combined (\ref{app1}) with (\ref{app2}) we get the final conclusion,
\begin{equation}\label{appA}
\langle\mathbf{X},0,\Delta_1,\Delta_2,\cdots,\Delta\rangle=\frac{\det[M(\Delta_\phi)]}{2^N}\langle\vec{\mathbf{F}}_{0}^{\Delta_\phi},\vec{\mathbf{F}}_{\Delta_1}^{\Delta_\phi},\cdots,\vec{\mathbf{F}}_{\Delta}^{\Delta_\phi}\rangle=\alpha\cdot\vec{\mathbf{F}}_{\Delta}^{\Delta_\phi}
\end{equation}
\section{Positivity of $\langle\mathbf{X},0,\Delta_{i_1},\Delta_{i_1+1},\dots,\Delta\rangle$}\label{positive}
\begin{figure}
\centering
\includegraphics[width=0.6\textwidth]{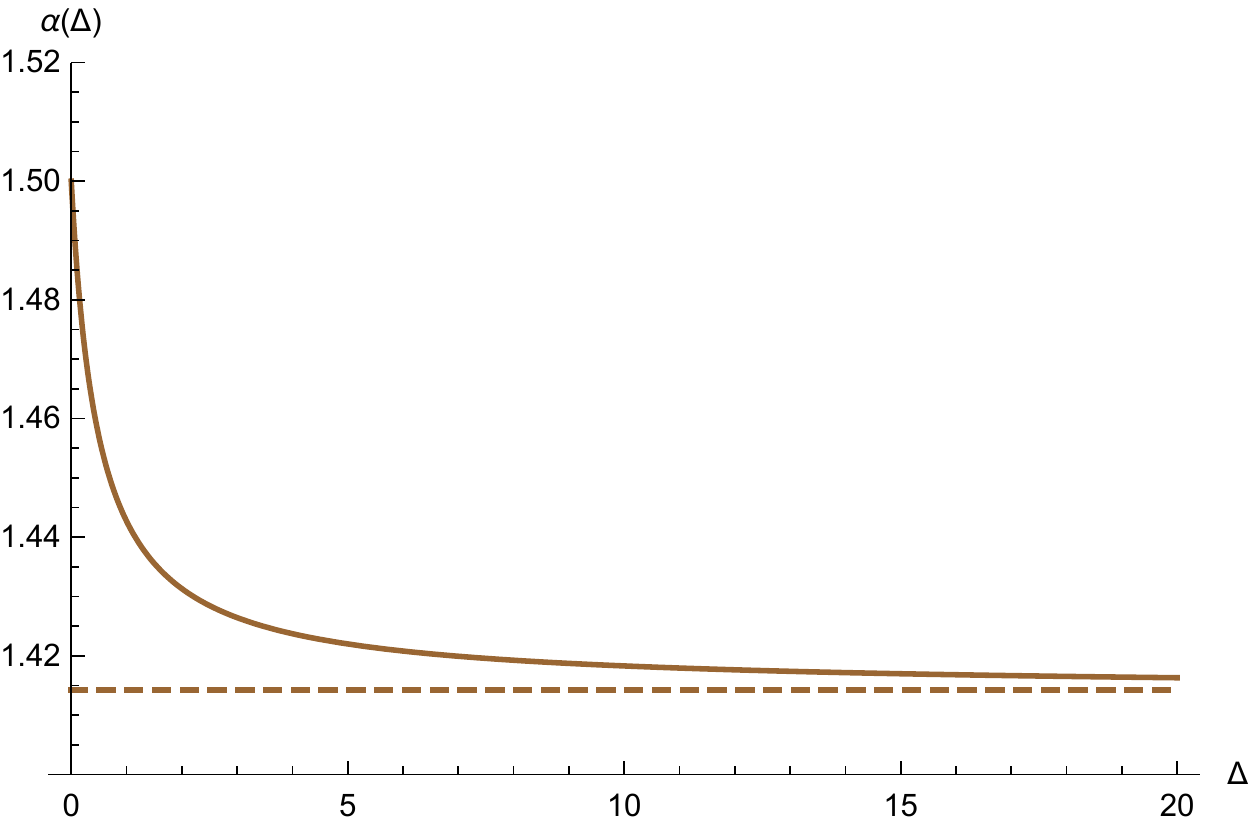}
\caption{\label{fig_alpha} The brown cruve is $\alpha(\Delta)$ defined in (\ref{ini}) as a function of $\Delta$. When $\Delta\to\infty$, the function approaches to constant $\sqrt{2}$ (brown dashed line) }
\end{figure}

In this section, we will show that for large $\Delta$ region $\Delta\gg\Delta_\phi$, functional $\omega(\Delta)=\langle\mathbf{X},0,\Delta_1,\dots,\Delta\rangle$ will be definite positive. Notice that the 1D conformal blocks $G_\Delta(z)$ satisfies the following second order differential equation,
\begin{equation}\label{diffeq}
z^2(1-z)\frac{\mathrm{d}^2}{\mathrm{d}z^2}G_\Delta(z)-z^2\frac{\mathrm{d}}{\mathrm{d}z}G_\Delta(z)-\Delta(\Delta-1)G_\Delta(z)=0
\end{equation}
Let $c_i(\Delta)$ be the Talyor coefficient of $G_\Delta(z)$ expanded around at $z=1/2$ (divided by a positive factor $G_\Delta(1/2)$), we found that first few terms are,
\begin{align}\label{ini}
&c_0(\Delta)=1\notag\\
&c_1(\Delta)=2\Delta\alpha(\Delta)\notag\\
&c_2(\Delta)=2\alpha(\Delta)\Delta+4\Delta(\Delta-1)\notag\\
&c_3(\Delta)=\frac{8}{3}\alpha(\Delta)\Delta(\Delta^2-\Delta+1)\notag\\
&\vdots
\end{align}
where $\alpha(\Delta)$ is the same function defined in \cite{Arkani-Hamed:2018ign}. Its behavior is in Fig.\ref{fig_alpha}

\begin{equation}
\alpha(\Delta)=\frac{{}_2F_1(\Delta,\Delta+1,2\Delta,1/2)}{{}_2F_1(\Delta,\Delta,2\Delta,1/2)}
\end{equation}
The general form of $c_n(\Delta)$ is
\begin{equation}
c_n(\Delta)=P_n(\Delta)+\alpha(\Delta)Q_n(\Delta)
\end{equation}
where $P_n(\Delta)$ and $Q_n(\Delta)$ are polynomial in $\Delta$. Using equation (\ref{diffeq}) we get the recurrence relation for both $P_n(\Delta)$ and $Q_n(\Delta)$,
\begin{equation}
\frac{1}{8}(n+2)(n+1)c_{n+2}+\frac{1}{4}(n^2-1)c_{n+1}-\left[\frac{1}{2}n(n+1)+\Delta(\Delta-1)\right]c_n-(n-1)^2c_{n-1}=0
\end{equation}
From (\ref{ini}) we can get the first two coefficient is $P_0=1,P_1=0$ and $Q_0=0,Q_1=2$. 

Consider the limit $\Delta\gg\Delta_\phi$, from Fig.\ref{fig_alpha} we get $\alpha(\Delta)\to\sqrt{2}$. So in this case $c_n(\Delta)$ is degree $n$ polynomial in $\Delta$. Now we consider the Talyor coefficient $a_n(\Delta,\Delta_\phi)$ of function $F_\Delta^{\Delta_\phi}(z)$ (divided by a positive factor $2^{2\Delta_\phi}G_\Delta(1/2)$), it can be obtained by $c_n(\Delta)$. Below we listed the first two relation,
\begin{align}
&a_1(\Delta,\Delta_\phi)=2c_1(\Delta)-8\Delta_\phi\notag\\
&a_3(\Delta,\Delta_\phi)=48\Delta_\phi(2\Delta_\phi+1)c_1(\Delta)-24\Delta_\phi c_2(\Delta)+2c_3(\Delta)-64\Delta_\phi(\Delta_\phi+1)(2\Delta_\phi+1)\notag\\
&\vdots
\end{align}
So vector $\vec{\mathbf{F}}_\Delta^{\Delta_\phi}$ will be 
\begin{equation}
\vec{\mathbf{F}}_\Delta^{\Delta_\phi}=
\begin{pmatrix}
-8\Delta_\phi+4\Delta\alpha(\Delta)\notag\\
\frac{16}{3}\Big[-2\Delta_\phi(1-3\Delta+3\Delta^2+3\Delta_\phi+2\Delta_\phi^2)+\Delta(1-\Delta+\Delta^2+6\Delta_\phi^2)\alpha(\Delta)\Big]\notag\\
\vdots
\end{pmatrix}
\end{equation}
In large $\Delta$ limit $\Delta\gg\Delta_\phi$, we just substitude $\alpha(\Delta)=\sqrt{2}$ into $\vec{\mathbf{F}}_\Delta^{\Delta_\phi}$. We found that $n$-th component of this vector is degree $2n-1$ polynoimal in $\Delta$,
\begin{equation}
\vec{\mathbf{F}}_{\Delta\gg\Delta_\phi}^{\Delta_\phi}=
\begin{pmatrix}
\#\Delta\\
\#\Delta^3\\
\#\Delta^5\\
\vdots\\
\#\Delta^{2n-1}
\end{pmatrix}
\end{equation}
This is exactly the moment curve, so for large $\Delta_\phi\ll\Delta_1<\Delta_2<\dots<\Delta_k$,
\begin{equation}\label{Fpositive}
\langle\vec{\mathbf{F}}_{\Delta_1}^{\Delta_\phi},\vec{\mathbf{F}}_{\Delta_2}^{\Delta_\phi},\dots,\vec{\mathbf{F}}_{\Delta_k}^{\Delta_\phi}\rangle>0
\end{equation}
Recall that the functional $\langle\mathbf{X},0,\Delta_{i_1},\Delta_{i_1+1}\dots,\Delta\rangle$ is proportional to $\langle\vec{\mathbf{F}}_{0}^{\Delta_\phi},\vec{\mathbf{F}}_{\Delta_{i_1}}^{\Delta_\phi},\vec{\mathbf{F}}_{\Delta_{i_1+1}}^{\Delta_\phi},\dots,\vec{\mathbf{F}}_{\Delta}^{\Delta_\phi}\rangle$. So using the positivity (\ref{Fpositive}) we can conclude that when $\Delta,\Delta_{i_1},\dots,\Delta_{i_k}\gg\Delta_\phi$, functional $\langle\mathbf{X},0,\Delta_{i_1},\Delta_{i_1+1}\dots,\Delta\rangle$ is definite positive

\bibliographystyle{JHEP}
\bibliography{mybib.bib}{}

\providecommand{\href}[2]{#2}\begingroup\raggedright\begin{thebibliography}{10}

\bibitem{Rattazzi:2008pe}
R.~Rattazzi, V.~S. Rychkov, E.~Tonni and A.~Vichi, \emph{{Bounding scalar
  operator dimensions in 4D CFT}},
  \href{http://dx.doi.org/10.1088/1126-6708/2008/12/031}{\emph{JHEP} {\bf 12}
  (2008) 031}, [\href{http://arxiv.org/abs/0807.0004}{{\tt 0807.0004}}].

\bibitem{Hellerman:2009bu}
S.~Hellerman, \emph{{A Universal Inequality for CFT and Quantum Gravity}},
  \href{http://dx.doi.org/10.1007/JHEP08(2011)130}{\emph{JHEP} {\bf 08} (2011)
  130}, [\href{http://arxiv.org/abs/0902.2790}{{\tt 0902.2790}}].

\bibitem{EFT}
N.~Arkani-Hamed, T.-C. Huang and Y.-T. Huang{\emph{{In preparation}} }.

\bibitem{Rychkov:2016iqz}
S.~Rychkov, \emph{{EPFL Lectures on Conformal Field Theory in D>= 3
  Dimensions}}.
\newblock SpringerBriefs in Physics. 2016.
\newblock 10.1007/978-3-319-43626-5.

\bibitem{Simmons-Duffin:2016gjk}
D.~Simmons-Duffin, \emph{{The Conformal Bootstrap}},  in \emph{{Proceedings,
  Theoretical Advanced Study Institute in Elementary Particle Physics: New
  Frontiers in Fields and Strings (TASI 2015): Boulder, CO, USA, June 1-26,
  2015}}, pp.~1--74, 2017.
\newblock \href{http://arxiv.org/abs/1602.07982}{{\tt 1602.07982}}.
\newblock \href{http://dx.doi.org/10.1142/9789813149441_0001}{DOI}.

\bibitem{Poland:2016chs}
D.~Poland and D.~Simmons-Duffin, \emph{{The conformal bootstrap}},
  \href{http://dx.doi.org/10.1038/nphys3761}{\emph{Nature Phys.} {\bf 12}
  (2016) 535--539}.

\bibitem{Poland:2018epd}
D.~Poland, S.~Rychkov and A.~Vichi, \emph{{The Conformal Bootstrap: Theory,
  Numerical Techniques, and Applications}},
  \href{http://dx.doi.org/10.1103/RevModPhys.91.015002}{\emph{Rev. Mod. Phys.}
  {\bf 91} (2019) 015002}, [\href{http://arxiv.org/abs/1805.04405}{{\tt
  1805.04405}}].

\bibitem{Arkani-Hamed:2018ign}
N.~Arkani-Hamed, Y.-T. Huang and S.-H. Shao, \emph{On the positive geometry of
  conformal field theory}, {\emph{JHEP} {\bf 06} (2019) 124}.

\bibitem{Mazac:2016qev}
D.~Mazac, \emph{{Analytic bounds and emergence of AdS$_{2}$ physics from the
  conformal bootstrap}}, {\emph{JHEP} {\bf 04} (2017) 146}.

\bibitem{Mazac:2018mdx}
D.~Mazac and M.~F. Paulos, \emph{{The analytic functional bootstrap. Part I: 1D
  CFTs and 2D S-matrices}},
  \href{http://dx.doi.org/10.1007/JHEP02(2019)162}{\emph{JHEP} {\bf 02} (2019)
  162}, [\href{http://arxiv.org/abs/1803.10233}{{\tt 1803.10233}}].

\bibitem{Hartman:2019pcd}
T.~Hartman, D.~Mazáč and L.~Rastelli, \emph{{Sphere Packing and Quantum
  Gravity}},  \href{http://arxiv.org/abs/1905.01319}{{\tt 1905.01319}}.

\bibitem{Mazac:2019shk}
D.~Mazáč, L.~Rastelli and X.~Zhou, \emph{{A Basis of Analytic Functionals for
  CFTs in General Dimension}},  \href{http://arxiv.org/abs/1910.12855}{{\tt
  1910.12855}}.

\bibitem{Sen:2019lec}
K.~Sen, A.~Sinha and A.~Zahed, \emph{{Positive geometry in the diagonal limit
  of the conformal bootstrap}},
  \href{http://dx.doi.org/10.1007/JHEP11(2019)059}{\emph{JHEP} {\bf 11} (2019)
  059}, [\href{http://arxiv.org/abs/1906.07202}{{\tt 1906.07202}}].

\bibitem{ShuHeng}
S.-H. Shao{\emph{Unpublished notes} }.

\bibitem{Afkhami-Jeddi:2019zci}
N.~Afkhami-Jeddi, T.~Hartman and A.~Tajdini, \emph{{Fast Conformal Bootstrap
  and Constraints on 3d Gravity}},
  \href{http://dx.doi.org/10.1007/JHEP05(2019)087}{\emph{JHEP} {\bf 05} (2019)
  087}, [\href{http://arxiv.org/abs/1903.06272}{{\tt 1903.06272}}].

\bibitem{ElShowk:2012hu}
S.~El-Showk and M.~F. Paulos, \emph{{Bootstrapping Conformal Field Theories
  with the Extremal Functional Method}},
  \href{http://dx.doi.org/10.1103/PhysRevLett.111.241601}{\emph{Phys. Rev.
  Lett.} {\bf 111} (2013) 241601}, [\href{http://arxiv.org/abs/1211.2810}{{\tt
  1211.2810}}].

\bibitem{Paulos:2016fap}
M.~F. Paulos, J.~Penedones, J.~Toledo, B.~C. van Rees and P.~Vieira, \emph{{The
  S-matrix bootstrap. Part I: QFT in AdS}},
  \href{http://dx.doi.org/10.1007/JHEP11(2017)133}{\emph{JHEP} {\bf 11} (2017)
  133}, [\href{http://arxiv.org/abs/1607.06109}{{\tt 1607.06109}}].

\bibitem{ToBePublish}
Y.-T. Huang, W.~Li, A.~Sinha and A.~Zahed{\emph{{In preparation}} }.

\bibitem{Hogervorst:2013kva}
M.~Hogervorst, H.~Osborn and S.~Rychkov, \emph{{Diagonal Limit for Conformal
  Blocks in $d$ Dimensions}},
  \href{http://dx.doi.org/10.1007/JHEP08(2013)014}{\emph{JHEP} {\bf 08} (2013)
  014}, [\href{http://arxiv.org/abs/1305.1321}{{\tt 1305.1321}}].

\end{thebibliography}\endgroup
\nocite{*}

\end{document}